\documentclass[doublespace,a4paper,11pt]{thesis}
\usepackage{amsmath,amsthm,amssymb,euscript,eufrak,mathrsfs,epsf,epsfig}
\usepackage{array,verbatim}
\usepackage[sort&compress]{natbib}
\makeatletter
\makeatletter
\@addtoreset{equation}{section}
\makeatother

\newcounter{multieqs}

\def\a{\alpha}
\def\b{\beta}
\def\c{\chi}
 
\def\dl{\delta}
\def\g{\gamma}
\def\e{\epsilon}

\def\m{\mu}

\def\n{\nu}

\def\r{\rho}

\def\s{\sigma}

\def\l{\lambda}
\def\L{\Lambda}

\def\cb{\ov\chi}
\def\t{\theta}

\newcommand{\Ab}{{\bar{A}}}
\newcommand{\Bb}{{\bar{B}}}
\newcommand{\Cb}{{\bar{C}}}
\newcommand{\Db}{{\bar{D}}}
\newcommand{\Eb}{{\bar{E}}}

\def\ab{{\bar{\alpha}}}

\def\db{{\bar{\delta}}}
\def\dlb{{\bar{\delta}}}

\def\cb{{\bar{\sigma}}}

\def\ar{{\bar{a}}}
\def\br{{\bar{b}}}

\def\dr{{\bar{d}}}
\def\er{{\bar{e}}}

\def\dtp{\frac{d^2p}{(2\pi)^2}}

\newcommand{\cB}{\mathcal B}

\newcommand{\cD}{\mathcal D}

\newcommand{\cG}{\mathcal G}
\newcommand{\cH}{\mathcal H}

\newcommand{\cK}{\mathcal K}
\newcommand{\cL}{\mathcal L}
\newcommand{\cM}{\mathcal M}
\newcommand{\cN}{\mathcal N}
\newcommand{\cO}{\mathcal O}
\newcommand{\cP}{\mathcal P}

\newcommand{\cR}{\mathcal R}
\newcommand{\cS}{\mathcal S}

\newcommand{\cV}{\mathcal V}
\newcommand{\cW}{\mathcal W}

\newcommand{\eqn}[1]{(\ref{#1})}

\def\ha{\frac{1}{2}}
\def\d{\partial}
\def\p{\partial}

\def\dag{\dagger}
\def\del{\partial}

\newcommand{\alp}{\alpha' }

\newcommand{\be}{\begin{equation}}
\newcommand{\bea}{\begin{eqnarray}}

\newcommand{\ee}{\end{equation}}
\newcommand{\eea}{\end{eqnarray}}

\newcommand{\nn}{\nonumber}

\def\Dh{\hat{D}}

\newcommand{\G}[3]{\Gamma^{#1}_{\ #2 #3}}
\newcommand{\Vt}[2]{\V^{#1}_{\ #2}}
\def\beq{\begin{equation}}
\def\eeq{\end{equation}}
\newcommand{\X}{\mathbb{X}}
\def\openone{1 } 
\newcommand{\uc}{{\underline{\sigma}}}
\def\V{\mathcal V}
\def\H{\mathcal H}
\providecommand{\openone}{\leavevmode\hbox{\small1\kern-3.8pt\normalsize1}}

\def\PL{Poisson--Lie T-duality}
\def\tf{{\tilde f}}

\begin{document}

\vspace{1truecm}

\centerline{\LARGE \bf T-duality Invariant Approaches to String Theory\\}
\vspace{1truecm}
\thispagestyle{empty} \centerline{
    {\large \bf Daniel C. Thompson${}^{\dag}$}}

\vspace{.4cm}
\centerline{{\it A thesis submitted for the degree of Doctor of Philosophy}}
\vspace{2cm}

\centerline{{\it ${}^\dag$ Centre for Research in String Theory, Department of Physics}}
\centerline{{ \it Queen Mary University of London}} \centerline{{\it Mile End Road, London E1 4NS, UK}}

\vspace{.4cm}

\topmatter{Abstract}

This thesis investigates the quantum properties of T-duality invariant formalisms of String Theory.  

We introduce and review duality invariant formalisms of String Theory including the Doubled Formalism.  We calculate the background field equations for the Doubled Formalism of Abelian T-duality and show how they are consistent with those of a conventional String Theory description of a toroidal compactification.  We generalise these considerations to the case of Poisson--Lie T-duality and show that the system of renormalisation group equations obtained from the duality invariant parent theory are equivalent to those of either of the T-dual pair of sigma-models. In duality invariant formalisms it is quite common to loose manifest Lorentz invariance at the level of the Lagrangian.  The lack of manifest invariance means that at the quantum level one might anticipate Lorentz anomalies and we show that such anomalies cancel non-trivially.  These represent important and non-trivial consistency checks of the duality invariant approach to String Theory.

\topmatter{Acknowledgements}

The research presented in this thesis was conducted under the support of an STFC studentship. 

 I wish to thank my supervisor David Berman to whom I owe a deep debt of gratitude.  Over the course of my studies he has shared his insight, wisdom and enthusiasm for the subject and has been a tremendous teacher and collaborator.    His advice and guidance were essential in helping me navigate the tortuous  path of research.    
 
 I also wish to thank those others with whom I have collaborated during the course of these studies:  Neil B. Copland; Malcolm J. Perry;  Ergin Sezgin;  Kostas Sfetsos;  Kostas Siampos and Laura C. Tradrowski.  

I thank my fellow colleagues at Queen Mary with whom I have enjoyed many profitable discussions during the course of this work, in particular: Ilya Bakhmatov; Andreas Brandhuber; Tom Brown; Vincenzo Calo; Andrew Low; Moritz McGarrie; Sanjaye Ramgoolam;  Rudolfo Russo; Steve Thomas; David Turton and Gabriele Travaglini. 

Lastly, I thank my family and most of all my wife Jennifer for unflinching belief and unfailing support.

\topmatter{Declaration}

Except where specifically acknowledged in the text the work in this thesis is the original work of the author.   Much of the research presented in this thesis has appeared in the following publications by the author:  \cite{Berman:2007xn} together with D. Berman and N. Copland; \cite{Berman:2007yf} with D. Berman; \cite{Sfetsos:2009vt} and  \cite{KSKSDT2}  with K. Sfetsos and K. Siampos.

Additionally, during the course of the preparation of this thesis the author has published several other articles \cite{Berman:2008be,Berman:2009kj,Berman:2009xd,Berman:2009zz} the results of which are not included in this thesis.

\tableofcontents
 \mainmatter

\chapter{Introduction}

\section{Unification and Duality}
Two important themes in theoretical physics are Unification and Duality. At first these themes seem to be, if not contradictory, at least competing ideas;
Unification is the reduction of multiple theories to a single all-encompassing theory,
whereas Duality is the notion that there may be several distinct and complementary frameworks that describe the same physics.  Nonetheless, both of these ideas have lead to tremendous advances in our theoretical understanding of the universe.

The older idea of Unification, whose roots can be traced at least as far back as Maxwell's theory of Electro-Magnetism, underpins the development of what is perhaps the pinnacle of scientific endeavour -- the Standard Model of particle physics. Time and time again Unification has proven to be a guiding light for theoretical physicists. The ultimate objective of Unification would be a theory that unites Gravity with Quantum Mechanics and with the other forces of nature encapsulated in the Standard Model. To this end, String Theory \cite{Green:1987sp}  is widely accepted as the leading candidate for a Quantum Theory of Gravity and offers the tantalising possibility of Unification of all the forces.

Duality is a comparatively newer concept and not only is it an interesting and deep theoretical property in its own right, it can also help make physical predictions. For instance, a question which may be very hard to answer within the context of one framework may be addressed much more easily by means of a dual description. One of the most important examples of duality, the $AdS-CFT$ correspondence \cite{Maldacena:1997re} does just this; questions asked in a strongly coupled gauge theory can be answered by means of a calculation performed in a weakly coupled gravity theory. The full power of this sort of duality is only now being unlocked and the past few years have seen its application to a range of physical systems including: the Quark Gluon Plasma \cite{Policastro:2001yc,Kovtun:2004de} thought to have been observed at the RHIC experiment; fluid mechanics in general \cite{Bhattacharyya:2008jc}  and condensed matter systems such as high $T_c$ superconductors \cite{Hartnoll:2009sz}.

There are many other fascinating examples of duality, particularly in the context of Gauge Theories \cite{Montonen:1977sn,Seiberg:1994aj,Seiberg:1994rs,Seiberg:1994pq}, but the focus of this thesis is on Duality within the context of String Theory. Nowhere is Duality more pronounced than in String theory and indeed the development of both go hand-in-hand to an extent. Certainly, a fuller understanding of String Theory necessitates a deeper understanding of Duality.

\section{String Theory and its Dualities}

Although born out of an ultimately unsuccessful attempt to formulate a theory of the Strong Nuclear Force,  Superstring Theory was first developed as a Quantum Theory of Gravity during the late 1970's and early 1980's.    In String Theory the fundamental objects are not point particles but string-like extended objects whose characteristic length is of the Planck scale ($1.6 \times 10^{-33}$ cm).   The primary reason for the success of String Theory as a Quantum Theory of Gravity is that the spectrum of closed strings includes a  massless, spin-two excitation corresponding to the graviton.   Not only that, the low energy effective space time action described by String Theory can be recognised as a generalisation of Einstein's theory of gravity.   

However, rather quickly it became apparent that there were some surprising features of Superstring Theory.   Firstly, there are in fact not one but five separate and consistent superstring theories: Type I; Type IIA; Type IIB; Heterotic $SO(32)$ and Heterotic $E_8$.  Given that finding a Quantum Theory of Gravity was such a difficult problem it does seem strange to discover five potential answers; one is reminded of waiting for a London bus only to find that five come along at once!   Additionally, according to the philosophy of unification, one might have hoped to have found a single unique theory.    As we shall see, Duality is the key to understanding that these theories are not as distinct as they first seemed. 

The second rather surprising feature of these Superstring Theories is that their quantum consistency demands that there be exactly ten spacetime dimensions.  To square this with the four-dimensional view of the universe one has to find a way for the six extra predicted dimensions to be rendered unobservable by current experiments.  The typical solution to this is to use the old idea of Kaluza-Klein reduction. In this, the extra dimensions are considered to be compact and small enough that they would only be visible to a very high-energy probe.   This idea of  `compactification' actually transforms what might be considered a fatal flaw into a virtue;  the Kaluza-Klein reduction leads to gauge fields in the lower dimensional theory and thereby gives a possible route to unification of {\it all} the forces of nature.   

It is also through this idea of compactification that we encounter our first `string--string' duality and the topic of this thesis: T-duality.  The extended nature of strings compared with point particles allows them to have an extra degree of freedom when a space time direction is compactified on, say, a circle of radius $R$.  Unlike the point particle,  strings are able to wind or wrap themselves around this compact direction.  This facilitates the astonishing property that the string theory defined on a circle of radius $R$ can actually be equivalent to that defined on a circle of radius $\alp/ R$ ($\alp$ is the square-root of the inverse string tension).    

Returning to the five Superstring Theories, T-duality states an equivalence between the IIA and IIB theories reduced on a circle and also between the two heterotic theories. 
This is just part of a much larger web of dualities which were uncovered during the mid 1990's 
\cite{Hull:1994ys,Witten:1995ex} (see  \cite{Vafa:1997pm} for a review).    In addition to these T-dualities there are conjectured non-perturbative `S-dualities'  which relate a theory at strong coupling to a, possibly different,  theory at weak coupling.  S-duality relates the Type I and the Heterotic $SO(32)$ theory and the IIB theory to itself.    The IIA theory displays a quite different behaviour at strong coupling; its strong coupling limit is an eleven-dimensional theory whose low energy limit is eleven-dimensional supergravity but whose full definition is not yet understood.   This eleven dimensional theory, known as M-theory, can be used to connect all the String Theories through compactifications and dualities. 

 M-theory leads to new duality conjectures, for example, M-theory compactified on a particular four-dimensional Calabi-Yau manifold $K3$ is  dual to heterotic string compactified on a three-torus.   M-theory can also provide explanations of the origin of other dualities, for example  the $SL(2,\mathbb{Z})$ S-duality group of the IIB theory can be mapped to the modular group of a toroidal compactification of M-theory.   Whilst a full understanding of M-theory remains elusive,  during the time this thesis was written several breakthroughs have been achieved in describing the low energy limit of multiple coincident M2 branes (a fundamental object in M-theory) \cite{ Bagger:2006sk, Bagger:2007jr, Bagger:2007vi,Gustavsson:2007vu,Aharony:2008ug}.

\section{Duality Invariance}

In this thesis, we will concentrate on just one aspect of Duality within String Theory and M-theory; T-duality.   Because T-duality is so particular to String theory one might hope that developing a broader understanding of T-duality may eventually provide clues to the underlying nature of String Theory and ultimately M-theory.  

 We will examine ways of reformulating String Theory in which T-duality can be promoted to the role of a manifest symmetry.   Such duality invariant approaches to String Theory have a long history  \cite{Duff:1989tf,Tseytlin:1990nb,Tseytlin:1990va, Rocek:1991ps,Schwarz:1993vs,Maharana:1992my,Klimcik:1995ux,Klimcik:1995dy,Cremmer:1997ct}  and have recently received a great deal of interest following Hull's proposed Doubled Formalism \cite{Hull:2004in,Hull:2006va,Hull:2009sg}.    A common theme of these approaches is to construct a new theory with double the number of fields from which either of the two T-dual theories can be obtained.  One might say that this represents the application of the mindset of Unification to the topic of Duality.   Aside from the rather grand objective of better elucidating String Theory, the Doubled Formalism has a more practical use in describing a novel class of string backgrounds, the so called non-geometric compactifications \cite{Kachru:2003uq,Flournoy:2005kx,Shelton:2005cf,Shelton:2006fd,Hellerman:2004fk,Dabholkar:2005ve}.  

The question that essentially lies at the heart of this thesis is whether such duality invariant approaches to String Theory make sense beyond the classical level?  By a careful quantum treatment we will provide evidence that the answer to this question is yes.  Along the way we will provide perspectives on different notions of T-duality and different approaches to duality invariance in String Theory. 

\section{Structure of Thesis}

In Chapter 2 we provide a detailed introduction to T-duality in the simplest setting - that of radial inversion duality for a bosonic strings with a single compact dimension.  We provide various derivations of the T-duality including the Buscher procedure and also using canonical transformations.   A new result included in this section is a derivation of Fermionic T-duality by means of a canonical transformation.   We also point out that the canonical transformation approach can also be straightforwardly extended to Ramond-Ramond backgrounds. 

Chapter 3 reviews abelian T-duality for the case of many extended dimensions and the associated $O(d,d,\mathbb{Z})$ duality group.  In this chapter we also introduce T-folds - string backgrounds which are constructed by gluing locally geometric patches with T-duality for transition functions.   Finally, we introduce the Doubled Formalism. 

In Chapter 4 we examine the quantum properties of the Doubled Formalism.  We consider the one-loop effective action obtained by performing a background field expansion of an action that corresponds to the Doubled Formalism together with its supplementary constraint.   We calculate the Weyl divergences in the effective action and the associated beta-functionals.  Demanding that these divergences vanish produces a set of equations which the background fields must satisfy.  
After the inclusion of a suitable dilaton field into the theory we show that the background field equations for the Doubled Formalism are consistent with those of a conventional string theory description of a toroidal compactification.   We also consider the possibility of Lorentz anomalies and show that contributions to such anomalies in the effective action cancel out non-trivially.

In Chapter 5 we apply similar techniques to the context of Poisson--Lie T-duality,  an extension of Abelian T-duality.  We calculate the renormalisation of  the Poisson--Lie duality symmetric action. We show that the resultant beta-functions match those of the two T-dual related sigma-models, first by means of explicit examples and then through an algebraic argument that holds in all generality.

Following Chapter 5 we present brief conclusions.  There are several appendices to this material containing useful background information on conventions, notation, the background field method and dimensional reduction.   Also the calculations of various Wick contractions relevant to both chapter 3 and 4 are placed in this appendix.

\chapter{T-duality}
\begin{quote} This chapter serves as an introduction to target space duality (T-duality) in String Theory.  We begin with a historical context and summary of the basic approaches to T-duality.  Additionally we illustrate how the fermionic extension to T-duality can be viewed classically as a canonical transformation in phase space.  
\end{quote}

\section{Twenty-five Years of T-duality}

One of the most profound lessons that String Theory has taught us is that we can not always trust our classical notions of geometry when attempting to understand physics at its most fundamental level.  This is well illustrated by target space duality which heuristically states that strings do not make a distinction between large and small compact spaces in which they propagate.    This is an intrinsically stringy effect not found in conventional field theory and is only made possible by the fact that closed strings have the ability to wind themselves around compact dimensions.  T-duality gives rise to the notion that there is a minimum length scale for String Theory set by the inverse string tension.    T-duality is one of the cornerstones of String Theory and forms an integral part of the intricate web of dualities between different varieties of String Theory.  From a space-time perspective T-duality can be thought of as a solution generating symmetry of the low energy effective theory.  As we shall see, from the string world-sheet point of view it can be though as a non-perturbative symmetry. 

The pre-history of T-duality reaches back almost a quarter of a century to the early studies of toroidal string compactifications in which the first evidence of a symmetry in the one-loop effective potential for the compactification radius was observed.  Using techniques and results from \cite{Green:1982sw},  it was shown in \cite{Sakai:1985cs,Kikkawa:1984cp} that this potential had a minimum when the compactification radius $R= \sqrt{\alp}$  and displayed a symmetry under the inversion $R\rightarrow \alp/R $.  Following the seminal work of Narain et. al.   \cite{Narain:1986am,Narain:1985jj} it was understood \cite{Giveon:1988tt,Shapere:1988zv} that this $\mathbb{Z}_2$ inversion symmetry is part of a larger of group of dualities given by $O(d,d,\mathbb{Z})$ for strings compactified on a $d$-dimensional torus. 

The more modern perspective of String Theory has its roots in the pioneering work of Buscher \cite{Buscher:1987sk,Buscher:1987qj} in which T-duality was derived from a world-sheet perspective as a symmetry of the string path integral.  The relations between the dual geometries that arise as the result of performing the manipulations in  \cite{Buscher:1987sk,Buscher:1987qj} have become known as the Buscher rules.  This world-sheet approach to T-duality was developed further in the early 1990's \cite{Rocek:1991ps,Giveon:1991jj,delaossa:1992vc,Alvarez:1993qi,Alvarez:1994wj} and good reviews  of this period can be found in \cite{Giveon:1994fu,Alvarez:1994dn}.  
 
  With T-duality well understood in the case of toroidal backgrounds with commuting isometries, much attention then turned to cases where the internal space possessed non-abelian isometry \cite{Giveon:1993ai,Quevedo:1993vq,Alvarez:1994wj,Alvarez:1994zr,Alvarez:1994jp}.   An important development in this direction was made by Kilmcik and Severa with the introduction of Poisson-Lie T-duality \cite{Klimcik:1995ux,Klimcik:1995jn,Klimcik:1995dy} which depends on an interesting mathematical structure known as  the Drinfeld double.   Despite much effort, it proved impossible to establish the validity of these non-abelian dualities for all genus of string world-sheet. These non-abelian `dualities' are best thought of as maps between related conformal field theories \cite{Giveon:1993ai}.  Nonetheless, it seems that T-duality should have extensions beyond the case of abelian isometries where Buscher rules can be applied and this remains an area of current research \cite{Dabholkar:2005ve,Hull:2009sg,ReidEdwards:2010vp}.

It has long been known that backgrounds for string compactification need not be smooth geometric manifolds and may be, for example, orbifolds.  An area that has received much attention over recent years has been the idea of using T-duality to construct new `non-geometric' compactifications \cite{Kachru:2003uq,Flournoy:2005kx,Shelton:2005cf,Shelton:2006fd,Hellerman:2004fk,Dabholkar:2005ve}.  From a string perspective, a perfectly good background for compactification is one which is locally geometric but not a manifold since patches are joined together with T-dualities for transition functions.  Roughly speaking, in these compactifications which have become known as T-folds, big circles may be joined to small circles and momentum modes glued to winding modes.  Such non-geometric backgrounds may play a crucial role in moduli stabilisation \cite{Becker:2006ks} and certainly are an important part  of any string landscape.  

Some other remarkable aspects of T-duality include its application to mirror symmetry of Calabi-Yau manifolds \cite{Strominger:1996it}; this is important from both a mathematical point of view but also for string compactifications.  A very exciting recent application of T-duality has been its use in exposing some unexpected properties of $\cN = 4$ supersymmetric Yang-Mills theory; namely, the connection between scattering amplitudes and Wilson loops and related underlying `dual' superconformal symmetry.  By performing successive T-dualities to the gravity dual $AdS_5\times S_5$, Alday and Maldacena \cite{Alday:2007hr} were able to map a configuration describing a scattering amplitude to a configuration describing a Wilson loop.  The T-duality explanation was clarified further by Berkovits and Maldacena \cite{Berkovits:2008ic} who showed that when a  fermionic extension to T-duality  is included the $AdS_5\times S_5$ background is exactly self dual.  

Given the importance of T-duality and its many applications, it is desirable that this symmetry is made manifest in the string sigma model.    There have been various attempts in the past to develop a formalism where T-duality is a symmetry of the action including \cite{Duff:1989tf,Tseytlin:1990nb,Tseytlin:1990va, Rocek:1991ps,Schwarz:1993vs,Maharana:1992my,Klimcik:1995ux,Klimcik:1995dy,Cremmer:1997ct}.   Many of these approaches share a common theme; one considers strings propagating on a target space which is enlarged so as to accommodate both the physical target space and its T-dual partner.  The advent of non-geometric backgrounds has motivated a resurgence of research into duality symmetric String Theory most notably the `Doubled Formalism' championed by Hull \cite{Hull:2004in,Hull:2006va,Hull:2009sg}.

\section{Radial Inversion Duality}

In this section we will illustrate T-duality in its simplest context, that of bosonic strings propagating in a flat target space with a single compact direction. The moduli space of such conformal field theories are parameterised by the radius $R$ of the compact dimension.  From a field theory perspective one might anticipate that there is a one-to-one map between any real value of $R$ and distinct field theories.  However, due to the winding modes of closed strings there exists a $\mathbb{Z}_2$ duality group that identifies the theory at radius $R$ with the theory at radius $\alp/R$. 

We shall provide four different perspectives demonstrating this duality: firstly, the path integral derivation of Buscher  \cite{Buscher:1987sk,Buscher:1987qj} complete with some of the more subtle global aspects as discussed by  \cite{Rocek:1991ps,Alvarez:1993qi}; secondly, a proof that the partition function is invariant under T-duality which also clarifies the transformation of the dilaton; thirdly a demonstration of the duality in canonical quantisation and finally we present how T-duality can be realised as a canonical transformation of phase space variables.

\subsection{Path Integral (Buscher) Approach}
We consider String Theory whose target space is $d$ dimensional flat space with coordinates $X^I$ but with one direction  periodic and so we may write  $X^I = (x^i, \theta)$.  We choose to work with dimensionless radii and coordinates and with conventions outlined in Appendix A. With these conventions $\theta \sim  \theta + 2 \pi$ and $G_{\theta \theta}= R^2$.  Furthermore, let us assume that conformal gauge has been adopted on the world-sheet so that the string sigma-model is given by
\bea
\label{ungauged}
S_0 
  &=&\frac{1}{4\pi} \int d^2 \sigma \, (G + B)_{IJ}\partial_{+}X^I \partial_{-}X^J \nn \\
 &=& \frac{1}{4\pi} \int d^2 \sigma \,  R^2 \partial_+ \theta \partial_- \theta + E_{i\theta}\partial_+x^i \partial_- \theta + E_{\theta i}\partial_+ \theta \partial_- x^i  +     E_{ij}\partial_{+}x^i \partial_{-}x^j \, , \quad
\eea  
in which we have adopted light cone coordinates $\sigma^\pm = \frac{1}{2}(\tau \pm \sigma)$ and defined a generalised metric $E= G + B$.  To begin with we shall consider just a classical procedure when the world-sheet is of genus zero.  

To exhibit T-duality we first notice that the sigma model (\ref{ungauged}) possesses a global $U(1)$ symmetry whose action is $\delta \theta = \omega$.  It is therefore natural to consider gauging this symmetry which we can do by introducing a $U(1)$ valued gauge field $A = A_+ d\sigma^+ + A_-d\sigma^-$ which has a gauge transformation rule $\delta A = -d \omega$.  The covariant derivative is
\be
D \theta = d \theta + A  \, .
\ee 
We constrain the gauge connection to be flat using a Lagrange multiplier.  With this constraint the gauged sigma model
\bea
\label{gauged}
S_{1} &=& \frac{1}{4\pi} \int d^2 \sigma \,  R^2 D_+ \theta D_- \theta + E_{i\theta}\partial_+x^i D_- \theta + E_{\theta i}D_+ \theta \partial_- x^i \nonumber \\ 
&& \quad \quad \quad \quad \quad+     E_{ij}\partial_{+}x^i \partial_{-}x^j + \lambda F_{+-}  \,, 
\eea
is actually equivalent to the ungauged sigma model (\ref{ungauged}). This can be seen by invoking the constraint $F_{+-} = 0$ which can be solved locally by
\bea
A_+ = \p_+ \phi \ , \quad A_- = \p_- \phi\, .
\eea
There are of course some global concerns that we must treat carefully on an arbitrary genus world-sheet to which we shall shortly return.  Substituting this pure gauge connection back into (\ref{gauged}) yields 
\bea
S_{1} &=& \frac{1}{4\pi} \int d^2 \sigma\,  R^2 \partial_+ (\theta + \phi) \partial_-  (\theta + \phi) + E_{i\theta}\partial_+x^i \partial_-  (\theta + \phi) \nn \\
&& \, \quad \quad \quad \quad  + E_{\theta i}\partial_+  (\theta + \phi) \partial_- x^i  +     E_{ij}\partial_{+}x^i \partial_{-}x^j
\eea  
which is equivalent to (\ref{ungauged}) after a trivial field redefinition.  

However, there is an alternative way in which one can proceed; by performing integration by parts one may express the gauged action as 
 \bea
S_{1}&= &\frac{1}{4\pi} \int d^2 \sigma \, R^2 D_+ \theta D_- \theta + E_{i\theta}\partial_+x^i D_- \theta + E_{\theta i}D_+ \theta \partial_- x^i \nonumber \\ && \qquad \qquad +  \partial_- \lambda A_+ - \partial_+ \lambda A_-   +     E_{ij}\partial_{+}x^i \partial_{-}x^j\,.
\eea
In this form the gauge fields are auxiliary (they have an algebraic equation of motion) and can be eliminated via their equations of motion. This procedure is made simplest by fixing the gauge in which $\theta = 0$  where the relevant equations of motion become
\bea
\nn 0 &=& R^2 A_+ +  E_{ i\theta} \partial_- x^i - \partial_+ \lambda \ ,\\
0 &=& R^2 A_- + E_{\theta i} \partial_- x^i + \partial_- \lambda \, .
\eea  
Replacing $A_\pm$ by these expressions results in an action given by
\be
\label{dual}
S_{2}= \frac{1}{4 \pi} \int d^2 \sigma \,  \frac{1}{R^2}  \partial_+ \lambda \partial_- \lambda - \frac{1}{R^2} E_{i\theta}\partial_+x^i \partial_- \lambda +\frac{1}{R^2} E_{\theta i}\partial_+ \lambda \partial_- x^i  +     \left( E_{ij} - \frac{E_{i\theta}E_{\theta j}}{R^2} \right)\partial_{+}x^i \partial_{-}x^j \, .
\ee
This dual action is of the same form as the initial sigma model (\ref{ungauged}) but with the following redefinitions
\bea
\label{buscher}
G_{\theta \theta } &\rightarrow& \frac{1}{G_{\theta \theta} }\nn \, , \\
E_{\theta i}  &\rightarrow&  \frac{1}{G_{\theta \theta}} E_{\theta i} \nn\, , \\
E_{i \theta}  &\rightarrow& - E_{i \theta}  \frac{1}{G_{\theta \theta}} \nn \, ,\\
E_{ij} &\rightarrow& E_{ij} - E_{i\theta}  \frac{1}{G_{\theta \theta}}  E_{\theta j}\, .
\eea
This derivation is, in essence, due to Buscher  \cite{Buscher:1987sk,Buscher:1987qj} and the above transformations and their generalisations are known as the Buscher rules.  Since the actions $S_0$ and $S_2$ are both equivalent (so far classically) to a common action $S_1$ we say that these are dual descriptions of the same physics.  It is from this perspective that we can consider this a non-perturbative duality on the world-sheet by regarding $R$ as the coupling constant.   In fact, this derivation is somewhat similar to that used in showing S-duality in the Lagrangian of $\cN=2$ gauge theories in four dimensions wherein a Lagrange multiplier is used to enforce the Bianchi identity for the field strength; if instead of integrating out the Lagrange multiplier one integrates out the gauge field itself one finds the S-dual action \cite{Seiberg:1994aj, Seiberg:1994rs}.

We now move to the path integral version of the duality for an arbitrary genus world-sheet.  The first addition we make is the inclusion of a dilaton field\footnote{Henceforward $\phi$ refers to the dilaton field and not the pure gauge part of the gauge field introduced earlier. }  $\phi$ coupled to the world-sheet curvature by means of a Fradkin-Tseytlin term \cite{Fradkin:1984ai}:
\be
S' = S_0 + \frac{1}{8\pi} \int d^2 \,  \sigma \phi R^{(2)}  \, . 
\ee
One can place this action in a path integral and repeat the above steps.  The only difference is that the gauge fields are now formally integrated out as a Gaussian integral.  Performing this integral results in a determinant factor which can be absorbed by transforming the dilaton field appropriately \cite{Buscher:1987sk,Buscher:1987qj,Schwarz:1993fk}. Hence, at the level of the path integral one must supplement the rules (\ref{buscher}) with a dilaton transformation  
\be
\label{dilshift}
\phi \rightarrow \phi -\ln ( G_{\theta \theta}) \, .
\ee
For a constant dilaton, the Fradkin-Tseytlin term is proportional to the Euler character of the world-sheet.  In the Polyakov path integral for the string we sum contributions of all genus weighted by their Euler character.   Since we are now considering arbitrary genus world-sheets we must return to consider global issues. 

 The key question is whether the gauged sigma-model is globally equivalent to the ungauged sigma model and, in particular, what happens to the topological information contained in the gauge field.  On a world-sheet Riemann surface of genus $g$ there are a set of $2g$ canonical homology one-cycles labelled $(A_i,B_i)$ and one must therefore keep track of the information contained in the holonomies of the connection around these cycle given by the path ordered exponential ${\cal P} \exp i \oint_\gamma A$.  For the two sigma models (\ref{gauged}) and (\ref{ungauged}) to be truly equivalent as conformal field theories we require that such holonomies are trivial.   To see how this works we look at the torus world-sheet and use the Lagrange multiplier term given by
\be
S_{\lambda}  = \frac{1}{4\pi} \int d^2 \sigma \,   \partial_-\lambda A_+ -   \partial_+ \lambda A_- = \frac{1}{2\pi} \int_\Sigma d\lambda \wedge A  \, . 
\ee
In the path integral we should sum over all configurations for $\lambda$ including topological sectors.  We therefore write 
\be
d \lambda = d \hat{\lambda} + 2\pi L  p \alpha +2 \pi L q \beta\,, \quad p,q \in \mathbb{Z}
\ee
where  $\hat{\lambda}$ is single valued, and where $(\alpha, \beta)$ are the Poincar\'e dual 1-forms to the cycles $(A,B)$ obeying $\oint_A \alpha = \oint_B \beta = 1$ and $\oint_A \beta= \oint_B \alpha = 0$.   $L$ is an, as yet undetermined, periodicity for the Lagrange multiplier.  In the winding sector of the path integral we have, making use of Riemann's bilinear identity (see e.g. \cite{Frak}),
\bea
\sum_{p,q} e^{\frac{i2\pi L}{2\pi}\int (p\a + q\b)\wedge A }&=&\sum_{p,q}  e^{\frac{i2\pi L}{2\pi}\left(p \oint_B A - q \oint_A A \right)} \nonumber \\
&=& \sum_{m,n} \delta\left(m - \frac{L}{2\pi} \oint_A A \right)\delta\left(n - \frac{L}{2\pi} \oint_B A \right)\, .  
\eea 
Then when the periodicity of the Lagrange multiplier is tuned such that $L=1$ we find that the holonomies of the gauge field are just the identity element of $U(1)$.  Thus,  integrating out the single valued piece of the Lagrange multiplier puts the gauge field in pure gauge as before and integrating out the winding modes of the gauge field makes the holonomies trivial \cite{Rocek:1991ps}.\footnote{When extending to non-abelian T-duality the Lagrange multiplier no longer constrains the holonomies of the gauge connection -- an immediate obstruction is that in the non-abelian case the holonomies require path ordering which is not captured in the above analysis \cite{Giveon:1993ai}.}
One can readily extend this argument to higher genus by simply adding in a sum over canonical pairs of homology cycles.  In a similar way one can show that the constant mode of the Lagrange multiplier enforces the constraint $\oint_\Sigma F=  0$ and hence restricts the curvature to a trivial class. These arguments ensure the full equivalence of the gauged and ungauged sigma models.\subsection{Partition Function Equivalence}

A feature of any duality is that there exists a precise matching of the spectrum and partition function in both T-dual models.  Let us illustrate this feature by demonstrating the equivalence of the partition function for a single compact boson under T-duality.  We begin by considering a torus world-sheet $\Sigma$ with modulus $\tau = \tau_1 + i \tau_2$ whose partition function is given by
\bea
 Z[\tau, R] =  \int [d X] \, e^{- \frac{R^2}{4\pi} \int_{\Sigma} d X \wedge \ast d X}\, . 
\eea
There are two contributions to this partition sum:  oscillations about globally defined solutions of the equations of motion and instanton/winding contributions. 
The oscillator sector is Gaussian and produces a determinant which can be evaluated by considering a basis of eigenfunctions to yield the standard result  \cite{Ginsparg:1988ui}
\bea
Z_{osc}[R] = R \left( {\rm det}^\prime \Box \right)^{-\ha} = \frac{R}{\sqrt{\tau_2}} \frac{1}{| \eta(\tau) |^2}\, ,
\eea
in which the factor $R$ comes from the normalisation of the zero mode integration in the path integral\footnote{There is only one zero mode since $b_0=1$ for the compact Riemann surfaces we are considering.} and the prime on the determinant indicates that the zero mode contribution is omitted when multiplying eigenvalues.   
We now consider the instanton contribution by considering the cohomological contributions of the form:
\bea
 dX = 2\pi ( n \alpha + m \beta)
\eea
where a suitable basis for the cohomology is given by 
\bea
\a= d\sigma_1 - \frac{\tau_1}{\tau_2} d\sigma_2 \, , & \beta = \frac{1}{\tau_2} d\sigma_2 \, .
\eea
Since Hodge star acts on this basis of forms as\footnote{Since here we are working with a Euclideanised world-sheet $(\ast)^2 =-1$.}
\bea
\ast \alpha = \frac{\tau_1}{\tau_2}\alpha +\frac{|\tau|^2}{\tau_2}\beta \, , & \ast\beta =-\frac{1}{\tau_2}\alpha- \frac{\tau_1}{\tau_2}\beta\, ,
\eea
the instanton sector provides a  contribution to the partition sum is given by
\bea
Z_{inst} = \sum_{n,m \in \mathbb{Z}} \exp [-\frac{R^2 \pi}{\tau_2}(n^2 |\tau|^2  - 2 nm \tau_1 + m^2)]\, .
\eea
We may use the Poisson re-summation formula and Gaussian integral formula
\bea
\sum_n f(nT+t) &=& \frac{1}{T} \sum_k \int dx f(x) e^{-2\pi i x k/ T} e^{2\pi i kt/T}\\ 
\int_\infty^\infty dx  e^{-bx^2  + c x}&= & \sqrt{\frac{\pi}{b}}e^{\frac{c^2}{4b}}
\eea
to recast this expression in a more illuminating way.  After completing the square and then re-summing over $m$ we obtain 
\bea
Z_{inst} &=& \sum_{n,s} \sqrt{\frac{\tau_2}{R^2}} \exp \left( -\pi \tau_2 \left( R^2n^2 + \frac{s^2}{R^2} + 2i \frac{ \tau_1}{\tau_2}ns \right) \right)\\
&=& \sum_{n,s}\sqrt{\frac{\tau_2}{R^2}} \exp \left( \pi i \tau p_R^2  - \pi i  \bar{\tau} p_L^2 \right)
\eea
where we have introduced 
\be
\label{momenta}
p_L = \frac{1}{\sqrt{2}} \left( \frac{s}{R} + R n  \right), \quad  p_R =  \frac{1}{\sqrt{2}} \left( \frac{s}{R} - R n  \right)\, . 
\ee
It is important to note that there is a factor of $R$ obtained from performing a Gaussian integral after re-summation.  For a higher genus $g$ world-sheet we need to include the $2g$ dimensional basis of canonical cycles. The manipulation is basically the same, however one should remember that the modulus $\tau$ is now a $g\times g$ period matrix $\tau_{ij}$ with $i = 1 \dots g$.\footnote{If $\omega_i$ are a standard basis of of (1,0)-forms of $H_{(1,0)}(\Sigma, \mathbb{Z})$ the period matrix is defined by $\oint_{A_i} \omega_j = \delta_{ij}$ and $\oint_{B_i}\omega_j = \tau_{ij}$ (see \cite{Frak} for more details).  Note that for the genus one torus $\omega = \alpha + \tau \beta = d \sigma_1 + i d \sigma_2 = d z $.  } Thus, winding and momenta numbers are now vectors in $\mathbb{Z}^g$.  Upon doing the Gaussian integral a determinant factor is produced which gives the result for genus $g$
\be
Z_{inst}[R] = \sum_{n,s} \det(\tau_2)^{\frac{1}{2} }R^{-g}  \exp \left( \pi i  p_R \tau p_R  - \pi i p_L \bar{\tau} p_L \right)\,.
\ee
All told, at genus $g$ the full partition sum $Z_{g} [R] = Z_{osc}[R] Z_{inst}[R]$ has R dependance of $R^{1-g}$.  

To consider the T-dual partition function we replace $R \rightarrow \frac{1}{R}$.  For the a genus $g$ contribution we have
\bea
Z_g[R^{-1}] = R^{2g-2} Z_g[R] = R^{-\chi} Z_g[R]
\eea
where $\chi= 2- 2g $ is the Euler character of the world-sheet.  In the Polyakov path integral we sum over all genus of world-sheet weighted by their Euler character\footnote{Note that in our conventions, chosen to be in accord with the literature on T-duality e.g.\cite{Giveon:1994fu}, the dilaton comes with a factor of a half in the wieghting of the Polyakov sum; this corresponds to the the normalisation of the Fradkin Tseytlin term.   Whilst this may differ from other places in the literature note that it is only a naming convention, the important unambiguous quantity is the string coupling constant defined so that each loop in closed string perturbation theory comes with a factor  $g_c^2$,   in our convention as  $g_c = \langle e^{\frac{\phi}{2}  }\rangle$ .}
\bea
Z[\phi, R] = \sum_g e^{-\frac{ \phi}{2} \chi} Z_g[R]\, .
\eea
Then we conclude that the full partition function is invariant providing the dilaton shifts as 
\bea
Z[\phi -  2\ln R,    R^{-1}] = Z[\phi, R]\, ,
\eea
which is in accordance with the stated result (\ref{dilshift}).
\subsection{Spectral Approach to T-duality}  

T-duality is, perhaps, most easily observed by examining the spectrum of the bosonic string obtained from a simple canonical quantisation. We consider a single periodic boson and revert to the more traditional conventions in which the boson has dimension and periodicity $X \simeq X + 2\pi R$.   The equations of motion arising from the action 
\bea
S=  \frac{1}{4\pi \alp}\int d^2 \sigma \p_\a X \p^\a X\, , 
\eea
are simply
\bea
\p_+\p_- X = 0 \, .
\eea
These have the solution 
\bea
X = X_L(\sigma^+)  + X_R(\sigma^-)\, ,
\eea
which can be expressed in Fourier modes by 
\bea
 X_L(\sigma^+) = \frac{x}{2} +   \sqrt{ \frac{\alp}{2}} p_L (\tau + \sigma)  + i\sqrt{ \frac{\alp}{2}} \sum_{n \neq 0} \frac{\a_n}{n} e^{-i n \sigma^+}\, , \\
   X_R(\sigma^-) = \frac{x}{2} + \sqrt{ \frac{\alp}{2}} p_R (\tau - \sigma) + i\sqrt{ \frac{\alp}{2}} \sum_{n \neq 0} \frac{\a_n}{n} e^{-i n \sigma^-}\, .
\eea
The periodic identification in target space allows us to consider the general closed string boundary condition  $X(\tau, \sigma) = X(\tau, \sigma + 2\pi) + 2\pi n R$ where $n \in \mathbb{Z}$ is the winding number of the string.  Since
\bea
X(\tau, \sigma + 2\pi)- X(\tau, \sigma) = 2\pi \sqrt{ \frac{\alp}{2}} ( p_L  - p_R)
\eea
we require a quantisation condition
\be
 p_L  - p_R =  R n \sqrt{ \frac{2}{\alp} }\, , \quad n \in \mathbb{Z}\, .
\ee
A further constraint arises since the total centre of mass momenta on a periodic direction must be quantised.\footnote{A nice way of thinking about this is that the operator which corresponds to a complete circulation around the compact direction i.e. $\exp(i p 2 \pi R)$, should be realised as the identity on states.} This total momenta is given by the integral over the length of the string of the conjugate momenta to $X$.  One finds that 
\be
p =\frac{1}{\sqrt{2\alp}} \left( p_L  + p_R \right) =  \frac{s}{R}\, , \quad  s \in \mathbb{Z}  \, .
\ee
Hence we find that
\be
p_L = \frac{1}{\sqrt{2}} \left( s \frac{\sqrt{\alp}}{R} + n \frac{R}{\sqrt{\alp}} \right)\, , \quad  p_R = \frac{1}{\sqrt{2}} \left( s \frac{\sqrt{\alp}}{R} - n \frac{R}{\sqrt{\alp}} \right)\, , 
\ee
which is in accordance with (\ref{momenta}).

To obtain a critical string theory we supplement this compact direction with a further 25 non-compact directions.  Using the twenty-five dimensional mass shell $m^2 = - p_\mu  p^\mu$ (where $\mu = 0 \dots 24$) condition and the Virasoro conditions on physical states  we find that the mass of a closed string state is given by 
\be
m^2 =\frac{s^2}{R^2} + \frac{n^2 R^2}{\alpha^{\prime2}} + \frac{2}{\alp} \left( N + \tilde{N} - 2 \right)\,, 
\ee
where $N$ and $\tilde{N}$ are the standard number operators.  This is invariant under
\bea
s \leftrightarrow n &  \frac{\sqrt{\alp}}{R} \leftrightarrow \frac{R}{\sqrt{\alp}}\,,
\eea
which is the action of T-duality.\footnote{The oscillators are also transformed as $\alpha_n \leftrightarrow \alpha_n$ and $\tilde{\alpha}_n \leftrightarrow -\tilde{\alpha}_n$ and so from a world-sheet perspective T-duality has the effect of a reflection under which left movers are even and right movers are odd.}     In the limit where the $R\rightarrow \infty$ it is clear that the winding modes become heavy and the momentum modes tend to a continuum,  as would be expected for a non compact direction.  In the opposite limit  $R\rightarrow 0$, we find that the momentum modes become heavy and the winding modes become a continuum.   The limits $R\rightarrow \infty$ and $R\rightarrow 0$ are physically identical with the role of momenta and winding switched.

\subsection{T-duality as a Canonical Transformation} 
An alternative way to think about T-duality, albeit classical in nature, is as a canonical transformation of phase space variables, first demonstrated in \cite{Alvarez:1994wj,Curtright:1994be}.  This approach has a certain elegance since it does not require the introduction of any structure such as the gauge fields introduced in the Buscher derivation.  We briefly review this calculation in the simplest setting before considering its extensions. 

We begin with the bosonic sigma-model Lagrangian
\be
{\cal L} = \frac{1}{2} (G+B)_{IJ}\partial_+ X^I \partial_- X^I \, ,
\ee
and, as before, demand that the background fields are independent of some coordinate $x_1$. We define
\be
\label{Jpm}
J_+ = \frac{1}{2} E_{i 1} \partial_+ x^i\, , \quad J_- = \frac{1}{2} E_{1 i} \partial_i x^i\, ,
\quad V = - \frac{1}{2} E_{ij} \partial_+x^i \partial_- x^j
\ee
such that the momenta conjugate to $x_1$ is given by
\be
p = \frac{\delta {\cal L}} {\delta \dot x_1} = R^2 \dot x_1 + J_+ + J_-\, .
\ee
The `Hamiltonian' obtained by performing the Legendre transform only on the active field $x_1$ is then
\bea
{\cal H} &= & \frac{1}{2 R^{2}  } p^2 - \frac{1}{R^{2}} p\ ( J_+ + J_-) + (J_+ - J_- )\ x_0^\prime +
\frac{ R^2}{2} x_0^\prime{}^2 \nonumber\\
&& \quad + \frac{1}{2 R^2} (J_+ + J_-)^2  + V \, .
\eea
We now propose the following canonical transformation (a map $(p,x_1) \rightarrow (\tilde{p}, \tilde{x}_1)$
to a new set of phase space variables which preserves the symplectic structure):
\be
p = \tilde{x}_1^\prime \, , \quad x_1^\prime = \tilde{p} \, .
\ee
One can readily verify that these transformations can be obtained from a generating functional that is independent of time and is given by 
\be
{\cal F} = \frac{1}{2} \int d \sigma \,  x_1 \tilde{x}_1^\prime -   x_1^\prime \tilde{x}_1 \, , 
\ee
such that
\be
p =  \frac{\delta \cal F}{\delta x_1}\, , \quad \tilde{p} = \frac{\delta \cal F}{\delta \tilde{x}_1}\, . 
\ee

Since the generating function is time independent we have the standard result $H(p,x_1) = \tilde{H}(\tilde{p}, \tilde{x}_1)$.

What is both remarkable, and the crux of the arguement, is that the dual Hamiltonian can be cast into exactly
the same form as the original Hamiltonian after a redefinition of the background fields.  After some simple algbera one establishes that the required redefinition of the background fields corresponds exactly to Buscher T-duality rules for $E_{IJ}$.  Hence T-duality related sigma
models are canonically equivalent.   Of course, since this approach is classical in nature it does not yield the dilaton transformation rule although it has been suggested that this can be recovered by considering the normalisation of the functional measure in phase space \cite{Lozano:1996sc}.  This canonical approach has been extended to the Ramond-Neveu-Schwarz (RNS) form of the superstring \cite{Hassan:1995je} and also to non-abelian and Poisson-Lie T-duality \cite{Lozano:1995jx,Sfetsos:1996pm, Lozano:1996sc,Sfetsos:1997pi}.        

\section{Canonical Transformations in Ramond Backgrounds}

Thus far we have only considered the bosonic string, however, when considering the superstring the Buscher rules as we have presented them are incomplete.  
The massless background fields which can couple to the Type II Superstring, and which correspond to the fields of Type II Supergravity,   include not only fermionic fields but also additional bosonic fields namely the Ramond-Ramond (RR) fluxes.  It is important to understand the transformation rules of these fields since many very important string backgrounds have non zero RR flux, most notably $AdS_5\times S_5$. The purpose of this section is two-fold, firstly to understand T-duality in backgrounds with non-trivial RR flux and secondly to point out that the canonical approach to T-duality is also applicable in this case. 

The transformation rules for the RR sector were first established in \cite{Bergshoeff:1995as} by comparing the dimensional reduction of IIA supergravity to nine-dimensions with that of the IIB theory.  This approach is limited in the sense that it is a first approximation in $\alpha^\prime$ and lacks the non-perturbative robustness of a world-sheet derivation {\it \`a la} Buscher.  Such a derivation was provided using the Green Schwarz (GS) form  \cite{Cvetic:1999zs, Kulik:2000nr}  and recently a somewhat more compact derivation was provided \cite{Benichou:2008it} using Berkovits' Pure Spinor formalism of the superstring.

The Pure Spinor approach to the superstring allows both manifest space time supersymmetry and covariant quantisation thereby combining the strengths of the RNS and GS approaches.   For the purpose of this thesis, however, the virtue of the Pure Spinor formalism is that it readily describes strings in backgrounds with non zero RR fluxes.  It would be somewhat lengthy, and lead us too far astray from the theme of this thesis, to fully review this formalism.\footnote{Comprehensive introductions to this can be found in the lecture notes \cite{Berkovits:2007wz,Bedoya:2009np}.}  Instead we immediately present the form of the Pure Spinor string when coupled to background fields \cite{Berkovits:2001ue}.

 The action in a curved background is given by
\begin{eqnarray} \label{actionPS}
S &=& \frac{1}{2 \pi \alpha '} \int d^2 z \left[
\frac{1}{2}(G_{MN}(Z)+B_{MN}(Z))\partial Z^M \bar{\partial} Z^N +  P^{\alpha \hat{\beta}}(Z) d_{\alpha} \hat{d}_{\hat{\beta}}\right. \nonumber \\
  & & +\ E^{\alpha}_M (Z) d_{\alpha} \bar{\partial}Z^M +
E^{\hat{\alpha}}_M (Z) \hat{d}_{\hat{\alpha}} \partial Z^M + \Omega_{M \alpha}\,^{\beta}(Z) \lambda^{\alpha} w_{\beta} \bar{\partial}Z^M +
\hat{\Omega}_{M \hat{\alpha}}\,^{\hat{\beta}}(Z) \hat{\lambda}^{\hat{\alpha}}
\hat{w}_{\hat{\beta}} \partial Z^M \nonumber \\
  & & \left. +\ C_{\alpha}^{\beta \hat{\gamma}}(Z) \lambda^{\alpha} w_{\beta}
  \hat{d}_{\hat{\gamma}} + \hat{C}_{\hat{\alpha}}^{\hat{\beta} \gamma}(Z)
  \hat{\lambda}^{\hat{\alpha}} \hat{w}_{\hat{\beta}} d_{\gamma} + S_{\alpha \hat{\gamma}}^{\beta \hat{\delta}}(Z) \lambda^{\alpha} w_{\beta}
  \hat{\lambda}^{\hat{\gamma}} \hat{w}_{\hat{\delta}} \right]  \nonumber \\
  &+ &\frac{1}{4 \pi} \int d^2 z (\Phi(Z) {\cal R}^{(2)}) + S_{\lambda} +
    \hat{S}_{\hat{\lambda}}\, . 
\end{eqnarray}

This rather complicated action requires some explanation.  The fields $Z^M$ describe mapping of the world-sheet into a superspace  and can be broken up into bosonic and fermionic parts $Z^M = (Z^m, \theta^\alpha , \hat{\theta}^{\hat{\alpha}})$.  In the type IIA theory $\theta$ and $\hat{\theta}$ have opposing chiralities whereas in the IIB theory they have the same chirality.

The remaining fields $\omega_\a$ and $\lambda^\a$ are commuting spinors of the ghost sector.  $\lambda^\a$ is a pure spinor obeying a constraint $\lambda^\a \gamma^m_{\a \b} \lambda^\beta = 0 $.    The field $\omega_\a$ is the conjugate momenta to $\lambda^\a$.    The term in the action  $S_{\lambda}$ is a kinetic term for the ghost sector which plays no r\^ole in this discussion.    In flat space $d_\a$  is given as the constraint that the momenta obey   and does not enter into the action but it is used to build the nilpotent BRST operator.   However, in the curved space action, we view $d_\a$ as  an independent variable and demanding that the BRST operator is nilpotent and holomorphic requires that the background fields obey the equations of motion of type II Supergravity.

We now give a brief description of the various background superfields entering into the action (\ref{actionPS}).  The superfields $G_{MN}$ and $B_{MN}$ contain as their lowest component bosonic parts the NS sector metric and two form and obey a graded symmetrization:
\begin{equation}
G_{MN}= (-)^{MN}G_{NM}\,  , \quad B_{MN}= -(-)^{MN}B_{NM}\, ,
\end{equation}
in which $(-)^{MN}$ is equal to $+1$ unless both $M$ and $N$ are spinorial indices in which case it is equal to $-1$. We combine these fields by defining $L= G+ B$.\footnote{Here notation $L$ has been used for this combination rather than $E$ to distinguish it from the vielbein.}

The field $P^{\a \hat{\b}}$ contains the information about the RR sector fluxes and has lowest components\footnote{This expression has been modified in comparison to \cite{Berkovits:2008ic,Berkovits:2001ue} so as to keep with the definition of the dilaton field used already in thesis.  Our conventions are such that $\phi_{here}= 2 \phi_{there}$.} 
\begin{equation}
P^{\a \hat{\b}}|_{\theta=\hat{\theta} =0}  = -{ i \over 4}  e^{\phi / 2} F^{\a\hat{\b}}
\end{equation}
where  $F^{\a\hat{\b}}$
is the Ramond-Ramond field strength in bispinor notation (the sum of $p$-form fluxes contracted with the antisymmetric product of $p$ gamma matrices).

The field $E^{\alpha}_M$ is part of the super-vielbein superfield and when the index $M$ corresponds to bosonic coordinate ($M= m$) the lowest component is the gravitino. The fields $\hat{\Omega}_{m \hat{\alpha}}\,^{\hat{\beta}}$ contain as their lowest spin connection modified with torsion generated by the field strength of $B_{mn}$.    The field $C_{\a}^{\b \ \hat{\gamma}}$ is related to the gravitino field strength and  $ S_{\alpha \hat{\gamma}}^{\beta \hat{\delta}}$  to the Riemann tensor (generalised to include torsion).    

For this action a Buscher procedure was used in \cite{Benichou:2008it}  to derive the T-duality rules. It is  also straightforward, if a little tedious, to verify that the canonical transformation approach can also be applied to this general form of the action.   We follow exactly the same steps as in the simple scenario making the replacement
\be
\partial \rightarrow \partial_+ \, , \quad \bar{\partial} \rightarrow \partial_- \, ,
\ee
with the $J_\pm$ now more complicated than the expressions in (\ref{Jpm}), for example, 
\be
J_+ = \frac{1}{2} L_{I 1} \partial_+ Z^I  + E_1^\a d_\a + \Omega_{1 \a}^{\ \ \b} \lambda^\a \omega_\b   \, ,
\ee
with a similar expression for $J_-$ involving $\p_-$ derivatives and hatted fermionic variables.   The potential $V$ is also suitably modified to a rather lengthy expression which can be easily read from the action.  

We use exactly the same canonical transformation as for the simple case i.e.
\be
p = \tilde{x}_1^\prime \, , \quad x_1^\prime = \tilde{p} \, .
\ee
and supplement this with a transformation of hatted fermions which acts by multiplication of the gamma matrix $\gamma_{\bar{1}}$  (the overbar indicating that this is the flat space gamma matrix which squares to the identity rather than the curved space gamma matrix).   The supplementary transformations of fermions are trivial from a canonical point of view but are required since T-duality maps the type IIA and type IIB theories to each other.     

Performing these transformations results in a Hamiltonian of the same form as the initial Hamiltonian (albeit with different chiralities of fields) with the background fields being redefined according to T-duality rules.  For the metric sector (now promoted to superfields) these are the obvious generalisation of the Buscher Rules:
\be
G_{11} \rightarrow  \frac{1}{G_{11}} \, ,  \quad L_{1 I} \rightarrow \frac{L_{1 I}}{G_{11}} \, ,   \quad L_{ I 1} \rightarrow - \frac{ L_{I 1 }}{G_{11}} \, ,  \quad
L_{IJ} \rightarrow L_{IJ} - (-)^{IJ} \frac{ L_{ I 1 } L_{1 J} }{G_{11}} \, . 
\ee
The RR fields transform according to 
\be
\label{RRtrans}
P^{\a \hat{\b} } \rightarrow \left( P^{\a \hat{\delta} }   +  2  \frac{ E_{ 1 }^{\ \a}\hat{E}_{1}^{\ \hat{\delta} }}{G_{11}}     \right)  \left( \gamma_{\bar{1}} \right)_{\hat{\delta}}^{\ \hat{\beta}}   \, . 
\ee 
In addition there are transformation rules for the spin connection, vielbeins and curvature which we omit for brevity and which can be found in \cite{Benichou:2008it}.   As before, the canonical transformation does not immediately tell us about the transformation rule for the dilaton but the path integral calculation reproduces the expected transformation law (\ref{dilshift}). 

 To better understand the transformation law for the RR fields we `unpack' the expression   (\ref{RRtrans}) for a simple case, namely that of a bosonic background (so the gravitino component $E_1^{\ \a}$ is set to zero) with no B field and no off-diagonal metric pieces. Starting from IIB and going to IIA, with tildes indicating IIA T-dual fields, we have from  (\ref{RRtrans}) that:
\bea
\label{RR1}
&&e^{\tilde{\phi}/2} \left(  \frac{1}{2!} \tilde{\gamma}^{m_1m_2} \tilde{F}^{[2]} _{m_1m_2} +    \frac{1}{4!} \tilde{\gamma}^{m_1m_2 m_3 m_4} \tilde{F}^{[4]}_{m_1m_2 m_3 m_4}   \right) \\
&& \quad \quad =  e^{\phi / 2} \left(   \gamma^m F^{[1]}_m + { 1 \over 3 !}
\gamma^{ m_1 m_2 m_3} F_{m_1 m_2 m_3}^{[3] }  + { 1 \over 2} { 1 \over 5!}
 \gamma^{ m_1\cdots m_5} F_{m_1 \cdots  m_5}^{[5]}   \right) \gamma_{\bar{1}}  \nonumber \, . 
\eea
in which we have suppressed the spinor indices.   Notice that the gamma matrices on both sides actually differ; they are in curved space and so are dressed with the vielbeins of the two dual geometries.   For the simple metric ansatz we have that $\tilde{\gamma}^i=\gamma^{i} $ for $i\neq 1$ and $\tilde{\gamma}^1 =G_{11} \gamma^1$, and  $\gamma_{\bar{1}} = \sqrt{G_{11} } \gamma^1$.   Then we can see that the left hand side of (\ref{RR1}) becomes
\be
e^{\phi/2}\frac{1}{ \sqrt{G_{11} }   }  \left(  \frac{1}{2} \gamma^{ij} \tilde{F}^{[2]} _{ij} +  G_{11} \gamma^{i} \gamma^1 \tilde{F}^{[2]} _{i1}  +   \dots \right)
\ee
 whilst the right hand side becomes 
\bea
e^{\phi/2} \left( \gamma^1 F^{[1]}_1 +\gamma^i F^{[1]}_i    + { 1 \over 2 !}
\gamma^{ ij} \gamma^1 F_{ij 1}^{[3] }  + \dots  \right) \gamma^1\sqrt{G_{11}}
\eea
where ellipses indicate the higher form contributions.   Since $F^{[1]}_1 = \partial_1 A^{[0]}=0  $ by virtue of the fact that the background should be independent of $x_1$ we find that the RR fluxes are related by
\be
\tilde{F}^{[2]}_{ij} = F^{[3]}_{ij1}\, ,  \quad  \tilde{F}^{[2]}_{i1} = F^{[1]}_{i} \, , 
\ee
and similarly that
\be
\tilde{F}^{[4]}_{ijkl} = F^{[5]}_{ijkl1}\, ,  \quad  \tilde{F}^{[4]}_{ijk1} = F^{[3]}_{ijk} \, . 
\ee
These T-duality rules have a natural interpretation in terms of D-branes which source the RR fluxes \cite{Polchinski:1995mt}.  Performing  a T-duality along the world volume of a Dirichlet $p$-brane results in a $(p-1)$-brane whereas the dual in a direction orthogonal to the brane's world volume yields a $(p+1)$-brane.   This interpretation is natural in the context of open strings which may end on D-branes; performing a T-duality in a certain direction has the effect of switching Neumann and Dirichlet boundaries conditions in that direction and thus the dimensionality of the Brane changes.

\section{Fermionic T-duality as a Canonical Transformation} 
As a final application of the canonical approach to T-duality we now  consider the fermionic T-duality proposed by Berkovits and Maldacena \cite{Berkovits:2008ic}.  Essentially one repeats the Buscher procedure when there is a shift symmetry in (a non-compact) fermionic direction.  This interesting new development has been been employed to explain the connection between Wilson loops and scattering amplitudes in ${\cal N}=4$ supersymmetric gauge theory (see also \cite{Alday:2007hr} and \cite{Ricci:2007eq,Beisert:2008iq} for more about the role T-duality plays in this story).   Whilst this fermionic T-duality only holds, at least in its present form, at tree level in string perturbation it represents a new solution generating symmetry of Supergravity \cite{Bakhmatov:2009be} .

We consider the Lagrangian density
\be
{\cal L } = \frac{1}{2} L_{MN} \partial_+ Z^M \partial_- Z^N
\ee
where as before $Z^M = (x^m, \theta^\a)$ are coordinates on a superspace so that the $\theta$ variables are anti-commuting fermions.  One may either view this as a warm-up before considering the pure spinor action or in its own right as the Green Schwarz action in curved space.  The superfield $L_{MN}= G_{MN} + B_{MN}$ is the same as that of the previous section with lowest components corresponding to   target space metric and B-field.

We assume the action is invariant under a shift symmetry in one of the fermionic directions which we
denote as $\theta$ and that the background superfield is independent of this coordinate (this is much
the same as working in adapted coordinates for regular bosonic T-duality).  We define $Z^{\hat{M}}$ as
running over all bosonic and fermionic directions except $\theta$.  

We wish to establish whether fermionic T-duality can be understood as a canonical transformation. Our strategy is to assume the T-duality rules presented in \cite{Berkovits:2008ic} and to search for a suitable transformation. 

With the defintions
\begin{eqnarray}
B_{\theta \theta} = B  \nonumber  \, , & \quad &
V =   -\frac{1}{2} L_{\hat{M}\hat{N}} \partial Z^{\hat{M}} \partial Z^{\hat{N}} \, , \nonumber \\
J_+ = \frac{1}{2} L_{ \hat{N} \theta }\partial_+ Z^{\hat{N}} \, , &\quad& 
J_- = - (-)^{\hat{N}} \frac{1}{2} L_{ \theta \hat{N} }\partial_- Z^{\hat{N}} \, , 
\end{eqnarray}
the sigma model Lagrangian can be written as
\begin{equation}
{\cal L } =  -B \dot{\theta} \theta^\prime + ( \dot{\theta} + \theta^\prime) J_{-} + J_{+} ( \dot{\theta} - \theta^\prime) - V\,.
\end{equation}

The canonical momenta conjugate to $\theta$ is given by
\be
\Pi = \frac{\delta L} { \delta \dot{\theta}} = - B \theta^\prime - J_+ + J_-
\ee
and obeys the Poisson Bracket
\be
\{\theta(\s), \Pi(\s^\prime) \}_{PB} = - \delta(\s - \s^\prime)\, .
\ee
Note that the sign convention in this equation is a consequence of the fermionic nature of $\theta$ and the fact that derivatives act from the left.

Since the Lagrangian is first order in time derivatives, the velocities can not be solved in terms of momenta and
instead we have an anti-commuting second class constraint:
\be
\label{consteqn}
f = \Pi  + B \theta^\prime + J_+ - J_- \approx 0
\ee
The naive Hamiltonian is given as
\be
{\cal H} =\dot{\theta} \Pi -  {\cal L} =  - \theta^\prime  (J_{+} + J_{-})  + V  \, ,
\ee
however, this should be amended to take account of the constraint.
We follow the Dirac procedure\footnote{See \cite{henneauxbook} for a detailed treatment of constrained systems.} by first modifying the Hamiltonian with an,
as yet unknown, local function which resembles an anti-commuting Lagrange multiplier:
\be
H_1 = \int d\sigma \,  [{\cal H}(\sigma)   + \lambda(\sigma) f(\sigma)]\, .
\ee
We now need to check whether any secondary constraints are produced by considering
 the time evolution of the constraint and demanding that
\be
\dot{f} =  \{ f , H_1 \}_{P.B.} \approx 0\, .
\ee

In order to calculate this time evolution it will be necessary to know
\bea
\{ f(\sigma) , f(\sigma^\prime) \}_{P.B.} &=&  \{ \Pi(\s) + B \theta^\prime(\s)  , \Pi(\s^\prime) +  B(\s^\prime) \theta^\prime(\s^\prime)   \}_{P.B.} \nonumber \\
&=& B(\s^\prime)  \{ \Pi(\s) , \theta^\prime(\s^\prime)   \}_{P.B.} +   B(\s) \{  \theta^\prime(\s)  , \Pi(\s^\prime)  \}_{P.B.}\nonumber \\
&=& (  B(\s^\prime)  - B(\s) ) \frac{ \partial}{\partial \sigma}  \delta( \sigma - \sigma^\prime)\nonumber  \\
&=& -B^\prime ( \s^\prime) ( \s - \s^\prime   ) \frac{ \partial}{\partial \sigma}  \delta( \sigma - \sigma^\prime) + \dots \nonumber \\
&=&   B^\prime ( \s^\prime)   \delta ( \s - \s^\prime)\, , 
\eea
where the dots indicate higher terms in the Taylor expansion which yield no contribution upon making use of the identification $x \delta (x) = 0$.  Note that since the Poisson bracket of these constraints is non-zero they are second class constraints; to consider the quantisation of the theory one should upgrade Poisson brackets to Dirac brackets.

We also need that
\bea
\{  f(\sigma) ,  {\cal H}    \} &=& - \{ \Pi , \theta^\prime (J_+ + J_-) \} \\
&=&(J_+ + J_-)   \frac{ \partial }{ \partial \s^\prime}  \delta ( \s - \s^\prime)  \, .
\eea

Then the time evolution is given by
\bea
\dot{f}(\sigma) &=&  \{ f(\sigma) , H_1 \}_{P.B.} =  \int d\sigma^\prime \, (J_+ + J_-)
   \frac{ \partial }{ \partial \s^\prime}  \delta ( \s - \s^\prime)       -
\lambda(\s^\prime)  B^\prime ( \s^\prime)   \delta ( \s - \s^\prime)\nonumber   \\
&=&   - (J_+ + J_-)^\prime       - \lambda(\s)  B^\prime ( \s)        \, ,
\eea
note that the minus sign in the second factor is due to the fact that $\lambda(\s)$ is anti-commuting.
Demanding that $ \dot{f}(\sigma) \approx 0$ does not produce a new constraint but instead fixes the
Lagrange multiplier function as
\be
\lambda(\s) = - \frac{(J_+ + J_-)^\prime  }{ B^\prime ( \s) }
\ee


Thus, the total Hamiltonian is given by
\be
{\cal H}_T =    - \theta^\prime  (J_{+} + J_{-})  + V     -  \frac{(J_+ + J_-)^\prime  }{ B^\prime }  (\Pi  + B \theta^\prime + J_+ - J_-  )\, .
\ee

Since the T-dual theory has the same structure as the initial theory the dual total Hamiltonian is also of the form
\be
{\cal \tilde{H} }_T=    - \tilde{\theta}^\prime  (\tilde{J}_{+} + \tilde{J}_{-})  + \tilde{V}
-  \frac{(\tilde{J}_+ + \tilde{J}_-)^\prime  }{ \tilde{B}^\prime }  (\tilde{\Pi}  +\tilde{ B} \tilde{\theta}^\prime + \tilde{J}_+ - \tilde{J}_-  )  \, ,
\ee
where the dual background fields are related according to the fermionic T-duality rules of Berkovits and Maldacena  \cite{Berkovits:2008ic}
\bea
\tilde{B} = - \frac{1}{B}\, , \quad \tilde{J_{\pm}} = \frac{J_{\pm}}{B} \, , \quad \tilde{L}_{\hat{M}\hat{N}} = L_{MN} - \frac{1}{B} L_{ \theta \hat{N}} L_{\hat{M} \theta }  \, .
\eea

We now ask whether there exists a transformation $(\theta , \Pi) \rightarrow (\tilde{\theta}, \tilde{\Pi})$ such that
\be
{\cal H}_T =  {\cal \tilde{H} }_T\, ,
\ee
and moreover whether such a transformation is canonical.
This implies relationships between the background fields in the two canonically equivalent models which take the form of T-duality rules.

Writing the dual Hamiltonian in terms of the original background fields yields
\bea
{\cal \tilde{H} }_T&=&    - \tilde{\theta}^\prime  \frac{({J}_{+} + {J}_{-})}{B}  + V +
\frac{1}{2B} L_{ \theta \hat{N} } L_{\hat{M} \theta } \partial_+ Z^{\hat{M}} \partial_- Z^{\hat{N}}
     \nonumber \\ && \quad   - \left( \frac{(J_+ + J_-)^\prime B}{B^\prime } - (J_+ + J_-)  \right)
(\tilde{\Pi}  -\frac{ 1}{B} \tilde{\theta}^\prime + \frac{ J_+ - J_-} {B}  )  \\
&=& -\tilde{\theta}^\prime  \frac{ (J_+ + J_-)^\prime  }{B^\prime } \nonumber
 + \tilde{\Pi} \left( \frac{(J_+ + J_-)^\prime B}{B^\prime } - (J_+ + J_-)  \right) + V  \\
 && \quad   +   \frac{J_+ - J_- }{ B^\prime }( J_+ + J_-)^\prime \, ,
\eea
in this we have used that
\be
J_+J_- = - \frac{1}{4} (-)^{\hat{N}} L_{\hat{M}\theta} \partial_+ Z^{\hat{M}} L_{\theta \hat{N}} \partial_- Z^{\hat{N}} =  \frac{1}{4}  L_{\theta \hat{N}}  L_{\hat{M}\theta}   \partial_+ Z^{\hat{M}} \partial_- Z^{\hat{N}} \, .
\ee

We now compare this against the initial hamiltonian, which we rewrite as
\bea
{\cal H}_T&=&\Pi\frac{ (J_+ + J_-)^\prime  }{B^\prime } \nonumber
 + \theta^\prime \left( \frac{(J_+ + J_-)^\prime B}{B^\prime } - (J_+ + J_-)  \right) + V  \\
 && \quad   +   \frac{J_+ - J_- }{ B^\prime }( J_+ + J_-)^\prime  \, .
\eea
Then we deduce that the transformation
\be
 \tilde{\theta}^\prime = - \Pi  \,  , \quad \tilde{\Pi} = \theta^\prime
\ee
ensures that ${\cal H} = {\cal \tilde{H}}$ provided that the backgrounds are related by fermionic T-duality.

One should still check whether this is indeed canonical in the sense that it preserves Poisson bracket. We have that
\be
\{\Pi(\s), \theta(\s^\prime) \} = - \delta(\s - \s^\prime) \,, \quad \{\Pi(\s), \theta^\prime(\s^\prime)\} = + \frac{\partial}{\partial \sigma} \delta(\sigma- \s^\prime)\,.
\ee
We now consider the action of the  transformation:
\be
\frac{\partial}{\partial \sigma} \delta(\sigma- \s^\prime)  = \{\Pi(\s), \theta^\prime(\s^\prime)\} = -\{\tilde{\theta}^\prime(\s), \tilde{\Pi} (\s^\prime)\} = - \frac{\partial}{\partial \sigma} \{\tilde{\theta}(\s), \tilde{\Pi} (\s^\prime)\}
\ee
We thus conclude that
\be
\{\tilde{\theta}(\s), \tilde{\Pi} (\s^\prime)\}  = - \delta(\sigma- \s^\prime) + const
\ee
which establishes that the proposed transformation is indeed canonical up to a zero mode ambiguity which is present also in the bosonic case.  We remark that these transformations are the same as the bosonic case up to a crucial sign swap due to the fermionic nature of the variables.  It seems likely that this result is a natural consequence of supersymmetry.  In fact, in order that the transformations be canonical it is sufficient that they may be obtained from a generating function (of the first kind) which is  given by 
\be
{\cal F} = \int d\sigma \theta^\prime \tilde{\theta}
\ee
so that
\be
\frac{\delta {\cal F} } {\delta \theta} = \Pi \, , \quad \frac{\delta {\cal F} } {\delta \tilde{\theta}} = - \tilde{\Pi} \, . 
\ee

We note that the total Hamiltonians presented are formally equivalent but may display some singular behaviour when the function $B^\prime(\sigma)$ has zeros.   This is somewhat similar to the fact that in bosonic T-duality the Hamiltonians may have singular behaviour due to fixed points of the Killing vector action.  Of course, when $B$ is constant things seem particularly divergent. In this case however, the term in the Lagrangian $B \theta^\prime \dot{\theta}$   is purely topological in nature.\footnote{Related to this is the fact that in the limit that $B^\prime(\sigma)=0$ the constraint \eqn{consteqn} becomes first class and the dynamics are trivialised. }  Since the fermionic T-duality only holds for trivial topology anyway we may safely ignore such a contribution.   It would be interesting to better understand what $B^\prime(\sigma)=0$ means in the context of the supergravity. For instance, in the full pure spinor formalism supergravity torsion constraints mean that the derivative of $B$ can be related to a certain Killing spinor bilinear \cite{Berkovits:2008ic}. One would thus expect that the singular behaviour of the Hamiltonian can be traced to some special behaviour of the Killing spinors.   This represents the topic of ongoing research which the author hopes to be able to report on soon.  

A final comment on this derivation was that the construction of the total Hamiltonian was essential in finding the duality transformations; the naive Hamiltonian was independent of momenta whereas the correct canonical transformation switches momenta and position.  
\section{Further Perspectives}
Before continuing to the more general setting of several compact directions it is worth briefly mentioning at this stage a number of important further perspectives on the topic.

\begin{itemize}
\item Because of T-duality the moduli space of the theory of a single compact boson is $\mathbb{R}/\mathbb{Z}_2$.  This space has singularities at the fixed points of the duality group given by $R=1 (=\alp)$.  At these fixed points the behaviour of the string is special; it experiences enhanced gauged symmetry due to the presence of extra massless states in the spectrum giving rise to an $SU(2)\times SU(2) $ current algebra.
\item In the preceeding analysis we assumed that the target spaces are exactly flat and that $E=G+B$ have no coordinate dependance at all. However the heuristics suggest that a sufficient condition to perform a T-duality is only that there exists a $U(1)$ isometry in the target space. In \cite{Rocek:1991ps} it was shown that when there is an invariance  generated by some vector $\delta X^I = \epsilon K^I$ such that
\be
\cL_K G = \cL_K H = \cL_K \phi =  0
\ee
then one can find adapted coordinates $X^I = ( x^i ,\theta)$ so that $K = \partial_\theta$, all the background fields are invariant of $\theta$ and the Buscher duality rules apply.  Notice that it is the field strength that enters in these conditions, $B$ need not be invariant but can instead vary to a total derivative.  Although the background is not trivial the String Theory will be conformal provided the Weyl anomaly vanishes.\footnote{See appendix.}  If the String Theory is conformal the dual theory will also be conformal (at least to one-loop in the $\alp$ expansion).  One can express the Buscher rules in a more covariant way in terms of the Killing vector $K$.  
\item T-duality may produce topology change in the target space. A striking and important example of this phenomenon is Witten's two-dimensional black hole \cite{Witten:1991yr} obtained by gauging a $U(1)$ subgroup in an $SL(2,R)$ Wess-Zumino-Witten (WZW) model.  The resultant metric has the form
\be
\label{cigar}
ds^2 = dr^2 + \tanh^2 r \, d \theta^2\,, 
\ee  
which has a $U(1)$ isometry (the B-field is pure gauge). After T-duality one obtains
\be
\label{trumpet}
ds^2 = dr^2 + \coth^2 r\,  d \theta^2 \, . 
\ee
These geometries have different fundamental groups (as can be easily seen from a visual inspection the former has $\pi_1 = 0$ and the latter $\pi_1 = \mathbb{Z}$). 
\item In fact, the example above gives the key to Rocek and Verlinde's \cite{Rocek:1991ps}  explanation of T-duality in terms of Abelian currents.  Both the geometries (\ref{cigar}) and (\ref{trumpet}) can be obtained from gauging the $SL(2,R)$ WZW model in two different ways.  

In general, for a WZW model on a group G there is a semi-local invariance $g(z, \bar{z}) \rightarrow h_L ( z) g h^{-1}_R(\bar{z})$ for the group elements $g$ and $h$.  One can promote this invariance to a fully local invariance for $h_{L/R} (z,\bar{z})$ in some subgroup $H \subset G$ by introducing appropriate gauge fields which are non-propagating.  The result of integrating out the gauge fields (and gauge fixing) is that one ends up with a sigma-model whose target space is dimension $\dim(G) - \dim (H)$.    Anomalies restrict the choice of subgroups that can be gauged and for a $U(1)$ subgroup there are exactly two possibilities  $h_L = h_R$ (known as vector gauging) and $h_L = - h_R$ (known as axial gauging).    In the $SL(2,R)$ case an axial gauging results in the first geometry (\ref{cigar}) and the vector gauging results in the second geometry (\ref{trumpet}).   In \cite{Rocek:1991ps,Giveon:1991jj} it was suggested that one can always think of abelian T-duality for $d$ commuting $U(1)$ isometries as being the interchange of vector and axial gaugings of a parent sigma model defined in $d$ extra dimensions.   In some senses this parent theory with extra dimensions is prototypical of the duality invariant models we shall study in this thesis. 
\item The interplay between T-duality and supersymmetry is somewhat subtle and has been investigated in \cite{Bakas:1994ba,Bergshoeff:1994cb,Bakas:1995hc,Alvarez:1995zr}.  From a world-sheet point of view $\cN=2$ world-sheet supersymmetry is always manifestly preserved under T-duality.\footnote{Complications arise with more world-sheet supersymmetry: $\cN= 4$ world sheet supersymmetry constrains the target space to be hyper-kahler and equipped with a triplet of covariantly constant complex structures obeying an $SU(2)$ algebra, however upon performing an abelian T-duality this hyper-kahler structure need not always be preserved and may result in a dual model with only manifest $\cN = 2$ worldsheet supersymmetry with the remaining supersymmetry realised non-locally \cite{Bakas:1995hc} .}  However, from a space time perspective, an application of the Buscher rules can apparently destroy or even create supersymmetry.  The resolution to this puzzle is that in some cases one must go beyond the supergravity approximation which is lowest order in $\alp$.  Lost symmetries can be recovered as non-local symmetries.  Similar phenomenon where naive T-duality needs to be corrected occurs when considering the dual of NS5 branes \cite{Gregory:1997te,Tong:2002rq,Harvey:2005ab}.

\end{itemize}

\chapter{ $O(d,d,\mathbb{Z})$, T-folds and the Doubled Formalism} 

 \begin{quote} 
We review the generalisation of  T-duality for toroidal backgrounds and the associated duality group $O(d,d,\mathbb{Z})$.   We introduce the concepts of T-folds and non-geometric backgrounds.  Finally, we provide a detailed introduction to the T-duality invariant Doubled Formalism of String theory.
 \end{quote}  

\section{Toroidal Compactification and $O(d,d)$ Duality Group}
A full understanding of toroidal compactification defined by $d$ compact bosonic directions $X^i \simeq X^i + 2 \pi$ and the internal metric data $E_{ij} = G_{ij} + B_{ij}$ was first given by Narain et al.     \cite{Narain:1985jj,Narain:1986am}.\footnote{For simplicity we ignore spectator fields in this presentation.}   It is straightforward to deduce, by using an ansatz for the zero mode of $X^i= \sigma n^i + q^i(\tau)$,   that the zero mode momenta must have the following form
\bea
p_{L \bar{i}}& =&  p_L^i G_{ij} e^i_{ \bar{i}} = (n_i + E_{ji}m^j) e^i_{ \bar{i}} \, , \\
p_{R \bar{i}}& =&  p_R^i G_{ij} e^i_{ \bar{i}} = (n_i + E_{ij}m^j) e^i_{ \bar{i}} \, ,
\eea
with $n_i$ and $m^i$ integer valued and in which we have introduced vielbeins defined such that
\be
G_{ij}  = e_i^{ \bar{i} } e_{j}^{\bar{j}} \delta_{\bar{i}\bar{j}}\, . 
\ee
 These momenta $(p_L, p_R)$ define a lattice $\Gamma_{d,d} \subset R^{2d}$ , and moreover, since 
\be
p_L^2 - p_R^2 = 2 m^in_i \in 2\mathbb{Z}
\ee 
this lattice is said to be even.  A more detailed investigation of partition functions shows that the lattice must also be self dual to ensure modular invariance \cite{Narain:1985jj}.   The contribution of these momenta to the Hamiltonian is given by 
\be
H_0 = \frac{1}{2}(p_L^2 + p_R^2)  = \frac{1}{2} \left(\begin{array}{c} m^i \\ n_j \end{array}\right) \left(\begin{array}{cc} (G - BG^{-1}B)_{ij}  & (BG^{-1})_{i}^{\ j} \\ - (G^{-1}B)^{i}_{\ j} & G^{ij} \end{array}\right)      \left(\begin{array}{cc} m^j & n_j \end{array}\right) \, .
\ee
These last expressions illustrate underlying moduli space structure of toroidal compactification which is \cite{Narain:1985jj}, 
\be
\label{modspace}
\cM_d =O(d,\mathbb{R} )\times O(d, \mathbb{R})/O(d,d, \mathbb{R})\backslash {\rm T-dualities}\, ,
\ee  
where we have indicated a duality group which we will shortly identify.  The dimension of this moduli space  is $d^2$ in accordance with the number of parameters in $E_{ij}$. The fact that the momentum reside in an even self dual lattice  gives rise to the numerator in (\ref{modspace}) and the invariance of Hamiltonian under  separate $O(d,\mathbb{R})$ rotations of $p_L$ and $p_R$ gives rise to the $O(d) \times O(d)$ quotient. 

$O(d,d,\mathbb{R})$ is the group which consists of real matrices $M$ that respect an inner product with $d$ positive and $d$ negative eigenvalues given, in a conventional frame, by  
\be
\label{oddmetric}
 L_{IJ} = \left(\begin{array}{cc} 0 & 1 \\ 1 & 0 \end{array}\right) 
\ee
such that
\be
M^t L  M = L\, . 
\ee
Note that the condition that the lattice $\Gamma_{d,d}$ is even is defined with respect to this inner product.  In what follows we shall frequently make use of the $O(d,d,\mathbb{R})/\left( O(d, \mathbb{R} ) \times  O(d, \mathbb{R} )  \right)$ coset form of the $d^2$ moduli fields :
\be
\cH_{IJ} = \left(\begin{array}{cc} (G - BG^{-1}B)_{ij}  & (BG^{-1})_i^{\ j} \\ - (G^{-1}B)^{i}_{\ j} & G^{ij} \end{array}\right)  \, .  
\ee 
The action of an $O(d,d,  \mathbb{R})$ element
\be
\cO =  \left(\begin{array}{cc} a  & b \\ c & d \end{array}\right)\,  , \quad \cO^t L \cO = L\, ,
\ee
is transparent on the coset form
\be
\label{Hacts}
\cH^\prime = \cO^t \cH \cO\, ,
\ee
but the equivalent action on the generalised metric $E=G+B$ is more complicated and is given by the fractional linear transformation 
\be
\label{Eacts}
E^\prime = (a E + b)(c E + d )^{-1}\, . 
\ee

For an $S^1$ compactification we saw that the moduli space was further acted on by a discrete duality group (in that case $\mathbb{Z}_2$).  We may ask what is the duality group for toroidal compactification, or what subgroup of $O(d,d, \mathbb{R})$ transformation leave the physics completely invariant.  

There are essentially three types of contributions, the first are large diffeomorphisms of the compact torus which preserve periodicities and correspond to the action of a $GL(d, \mathbb{Z})$ basis change. We refer to such transformations as being in the geometric subgroup of the duality group.   These correspond to $O(d,d)$ matrices of the form
\be
\cO_{A} = \left(\begin{array}{cc} A^T & 0 \\ 0 & A^{-1} \end{array} \right)\, , \quad A \in GL(d, \mathbb{Z})\, , \quad  \det A = \pm 1\, .
\ee  

The second transformation arises by considering constant shifts in the  $B$ field.  Such a shift in the $B$ field by an integer results in a shift in the action by $2 \pi  \mathbb{Z}$ and produces no change in the path integral.  For $d$ compact directions this allows us to consider the shift $B_{ij} \rightarrow B_{ij}+ \Omega_{ij}$ where $\Omega$ antisymmetric with integer entries.   The $O(d,d)$ form of this transformation is 
\be
\cO_{\Omega} = \left(\begin{array}{cc} 1 & \Omega \\ 0 & 1 \end{array} \right)\,,  \quad \Omega_{ij}=-\Omega_{ji} \in \mathbb{Z} \, .
\ee  

The final set of dualities consist of things more akin to the radial inversion and have the form \cite{Giveon:1994fu}
\be
\cO_{T} =  \left(\begin{array}{cc} 1 -e_i & e_i \\ e_i & 1-e_i \end{array} \right)\,, 
\ee  
where $(e_i)_{jk} = \delta_{ij}\delta_{ik}$.   All together these three transformations generate the duality group $O(d,d, \mathbb{Z})$ which henceforward shall be known as the T-duality group. These results above were first fully established in \cite{Giveon:1988tt,Shapere:1988zv}.  To calculate a T-dual geometry one simply performs the action (\ref{Hacts}) or (\ref{Eacts}) using an   $O(d,d, \mathbb{Z})$ transformation. The generalisation of the dilaton transformation (\ref{dilshift}) for the toroidal case becomes
\be
\label{dilshift2}
\phi^\prime = \phi + \frac{1}{2} \ln \frac{\det G^\prime}{\det G}   \, .
\ee
The calculation of this shift is rather subtle involving careful regularisation of determinants and is detailed in \cite{Tseytlin:1990va, Schwarz:1993fk}. 

\subsection{Example of $O(2,2, \mathbb{Z})$}
An illustrative example of this $O(d,d,\mathbb{Z})$ can be found by considering String Theory whose target is a $T^2$ and this will be useful for our discussion of T-folds. The duality symmetry is now quite rich since the theory with a torus target space has four moduli encoded in the three components of $G_{IJ}$ and the single component of $B_{IJ}$. In general we may write the metric of a torus as 
\be
ds^2 = \frac{A}{\tau_2} | dx + \tau dy  |^2
\ee
for coordinates $(x,y)\simeq(x+1,y) \simeq(x, y+1)$.  The three metric moduli are neatly encoded in the volume modulus $A$ and the complex structure modulus $\tau$.  The volume modulus and integrated B field can be combined to form a complexified K\"ahler modulus $\rho = B_{xy} + i A$. 

The geometric subgroup of $O(2,2,\mathbb{Z})$ is $GL(2,  \mathbb{Z})$ whose volume preserving subgroup  $SL(2,\mathbb{Z})$ acts as $\cH^\prime = \cO^t \cH \cO$ with generators
\be
\cO_T= \left(\begin{array}{cccc}
1&0&0&0\\
1&1&0&0\\
0&0&1&-1\\
0&0&0&1\\
\end{array}\right)\, , \quad \cO_S= \left(\begin{array}{cccc}
0&-1&0&0\\
1&0&0&0\\
0&0&0&-1\\
0&0&1&0\\
\end{array}\right)\, . \ee
These act to transform the complex structure  $\tau$ by the standard modular transformations
\be
\label{geosub}
T: \tau \rightarrow \tau + 1 \, \quad S: \tau \rightarrow - \frac{1}{\tau} \, . 
\ee    

The part of the duality group which is the generalisation of radial inversion swaps K\"ahler and complex structure  $\tau \leftrightarrow \rho$ and is generated by transformations of the following type
\be
\cO_R= \left(\begin{array}{cccc}
0&0&1&0\\
0&1&0&0\\
1&0&0&0\\
0&0&0&1\\
\end{array}\right)\,. 
\ee
For the case where $B=0$ and $G=diag(R_1^2 , R_2^2) $ this simply acts by inversion of the radius $R_1 \rightarrow 1/R_1$. Whereas the $SL(2,\mathbb{Z})$ identifications are geometric, this duality is fundamentally stringy in its nature.   

In summary, the duality group  $O(2,2, \mathbb{Z})$ can be identified with  $(SL(2,\mathbb{Z})\times SL(2,\mathbb{Z}))/\mathbb{Z}_2 $.\footnote{A further discrete $\mathbb{Z}_2$ identification correspond to world-sheet parity.}

\section{T-folds}
\label{secT-folds}
 
The last few years have seen the discovery \cite{Kachru:2003uq,Flournoy:2005kx,Shelton:2005cf,Shelton:2006fd,Hellerman:2004fk,Dabholkar:2005ve} of a new class of string compactifications that are known variously as `non-geometric backgrounds', `T-folds', `monodrofolds' and `twisted tori'. 

An old idea in compactification theory is that of twisted reductions introduced by Scherk and Schwarz \cite{Scherk:1979zr}.  In this generalisation of Kaluza-Klien reduction one only requires fields to be periodic in the compact direction up to the action of some element of a global symmetry group.  This twisting creates a potential, known as the Scherk-Schwarz potential, for the fields.   Such a mechanism is useful for compactification since one of the difficulties in String Theory is the presence of a large number of moduli fields describing the geometry of the internal space which have no potential.  Finding a way to `stabilise' all moduli is a long standing challenge.

 In \cite{Dabholkar:2005ve} it was suggested that that the $O(d,d , \mathbb{Z})$ T-duality group  provides a suitable symmetry for performing a twisted Scherk-Schwarz compactification.    The resultant background is known as a T-fold.  A T-fold is a rather strange object since in a T-fold locally geometric patches are glued together using T-duality as a transition function.   Similar backgrounds can also be obtained from T-dualising a background with H-flux (i.e. a background for which the field strength for the Kalb-Ramond 2-form has non zero values around some cycles of the compact space). 

 These idea can be well illustrated using the explicit example of a $T^3$ which has been studied in some detail  \cite{Kachru:2003uq,Flournoy:2005kx,Shelton:2005cf,Shelton:2006fd,Hellerman:2004fk,Dabholkar:2005ve}.   We begin with a $T^3$ equipped with flux described by 
\be
\label{H-flux}
ds^2 = dx^2 + dy^2 + dz^2 \ , \quad  B_{xy}=  N z \, , \, N\in \mathbb{Z}\, .
\ee
This background may be thought of as a $T^2$ in the $(x, y)$ directions fibred over an $S^1$.  After performing a T-duality in $x$ direction we find a slight surprise -- there is no longer any B-field and the resultant metric is given as
\be
\label{Geoflux}
ds^2 = (dx - N z dy)^2 + dy^2 + dz^2\, . 
\ee
This background is still a torus fibration over the $S^1$, however, it is non-trivial and topologically distinct from the starting $T^3$ since the complex structure  given by 
\be
\tau = -Nz, 
\ee
has a monodromy under a circulation of the base $S^1$
\be
\label{mon}
\tau \rightarrow \tau - N \,. 
\ee
This geometry is known as a twisted torus (also sometimes called a nillmanifold) and is a T-fold since the monondromy is one of the $O(2,2, \mathbb{Z})$ transformations.  However things are still reasonably geometric; the element of the duality group corresponding to the transformation (\ref{mon}) is contained in the geometric subgroup (\ref{geosub}).  This background (\ref{Geoflux}) is often described as having `geometric-flux'  in analogy to the H-flux of the background (\ref{H-flux}).

Things get even more exciting upon performing a second T-duality along the $y$ direction.  Buscher rules give the resultant geometry as 
\be
\label{F-flux}
ds^2 = \frac{1}{1+ N^2 z^2} (dx^2 + dy^2) +  dz^2\, , \quad B_{xy} =\frac{Nz}{1+ N^2 z^2}\, . 
\ee  
In this case it is the Kahler modulus 
\be
\rho = \frac{Nz}{1+N^2 z^2} + i  \frac{1}{1+ N^2 z^2}
\ee
that has monodromy under $z\rightarrow z+1$
\be
\rho \rightarrow \rho - \frac{i N}{N - i } \,. 
\ee
In terms of $O(2,2, \mathbb{Z})$ this transformation mixes metric and B-field and hence does not lie in the geometric subgroup and is generated by the following $O(2,2)$ matrix
\be
\cO =  \left(
\begin{array}{llll}
 1 & 0 & 0 & 0 \\
 0 & 1 & 0 & 0 \\
 0 & N & 1 & 0 \\
 - N  & 0 & 0 & 1
\end{array}
\right)\, . 
\ee
This background is locally geometric but clearly no-longer globally geometric.  In the terminology of \cite{Shelton:2005cf,Wecht:2007wu} this background possesses non-geometric flux or  `Q-flux'.   It should be pointed out at this stage that the above example is really only a toy demonstration; it does not represent a complete string background satisfying the conformal invariance criteria of vanishing beta functionals.\footnote{These beta functionals are presented in the appendix.}  It is somewhat tricky to find an explict full string background with a twisted torus in its geometry.

One may even go a step further and postulate that there should also be a notional third T-dual background obtained by T-dualising the $S^1$ corresponding to the $z$ coordinate.  However, at this point the Buscher procedure ceases to be of use since the background is not isometric in the $z$ direction.   Furthermore, it is expected that such a background may not even have a locally geometric interpretation since the description of physics may depend locally on both the physical coordinates and the T-dual coordinates \cite{Dabholkar:2005ve}.   Such backgrounds are often described as having `R-flux' and have been explored in \cite{Dabholkar:2005ve,DallAgata:2007fk,Hull:2009sg}.   

A significant amount of work has been done in understanding the construction of these non-geometric backgrounds and their implications for string compactifications.  One reason that such backgrounds have seen such interest is because they allow for moduli stablisation  \cite{Becker:2006ks,Flournoy:2005kx,Wecht:2007wu}.  There has been some suggestion in the literature that these non-geometric backgrounds may be as numerous as conventional backgrounds \cite{McOrist:2010jw}.  The idea of T-folds can be naturally extended to compactifications with U-duality twists (U-folds) \cite{Hull:2004in,Hull:2006va,Cederwall:2007je} and, given the understanding of mirror symmetry as T-duality \cite{Strominger:1996it}, may allow mirror-folds for Calabi-Yau compactifications \cite{Kawai:2007uq}.
   
An interesting recent proposal \cite{Boer:2010fk}  is that `exotic branes' (obtained by dualising co-dimension two objects in a dimensionally reduced string or M-theory and for which there is no obvious higher dimensional origin) can be identified with T-folds (or more generally with U-folds).   In \cite{Boer:2010fk} it is further suggested that such T-fold geometries might play an important role in black hole micro-state counting.

\section{The Doubled Formalism}

The T-fold backgrounds introduced above are examples where String Theory disregards familiar notions of geometry.  Understanding String Theory on such backgrounds is an important question for two main reasons. Firstly, T-folds compactifications represent an interesting and novel corner of any string landscape and so should be better understood in their own right.  The second reason is that by considering scenarios where the picture of a string propagating in a geometric target space no longer makes sense, we may hope to uncover some clues as to the deeper nature of String Theory. 

Parallel to the need to understand non-geometric string backgrounds is a long standing desire to make T-duality a more manifest symmetry of the string sigma model.   This idea of a duality symmetric  formulation of String Theory for abelian T-duality has a long history with previous approaches including  the formulation of Tseytlin \cite{Tseytlin:1990va, Tseytlin:1990nb}, the work of Schwarz and Maharana \cite{Schwarz:1993vs, Maharana:1992my} and the earlier work by Duff \cite{Duff:1989tf}.   In this section we introduce the duality symmetric theory known as the Doubled Formalism recently championed by  Hull in \cite{Hull:2004in, Hull:2006va} and which has its roots in the early approaches of \cite{Schwarz:1993vs, Maharana:1992my, Duff:1989tf} and the subsequent work of \cite{Cremmer:1997ct}.

The Doubled Formalism \cite{Hull:2004in,Hull:2006va} is an alternate description of String Theory on target spaces that are locally $T^n$ bundles over a base $N$.   For a geometric background local patches are glued together with transition functions which include  $GL(n, \mathbb{Z})$ valued large diffeomorphisms of the fibre.\footnote{A more general definition of a geometric background is one where transition functions can also include integer shifts in the B-field and are valued in $GL(n, \mathbb{Z}) \ltimes \mathbb{Z}^{n(n-1)/2} \subset O(n,n, \mathbb{Z})  $.}  As discussed above the T-duality group for this background is $O(n,n, \mathbb{Z})$ and a non-geometric background is formed by gluing together local patches with transition functions that take values not just in   $GL(n, \mathbb{Z})$ but in the whole duality group  $O(n,n, \mathbb{Z})$.

The essence of the Doubled Formalism is to consider adding an additional $n$ coordinates so that the fibre is doubled to be a $T^{2n}$. The Doubled Formalism recasts the string sigma model for the target space which is locally a $T^n$ bundle as a sigma model whose target space is the doubled torus.  Because the number of coordinates have been doubled it is necessary to supplement the sigma model with a  constraint to achieve the correct degree of freedom count.   This constraint takes the form of a self duality (or chirality) constraint.  The final step of the Doubled Formalism is to define a patch-wise splitting $T^{2n} \rightarrow  T^n \oplus \tilde{T}^n$ which specifics a physical $T^n$ subspace and its dual torus $\tilde{T}^n$.    
 
  In this doubled fibration the T-duality group $O(n,n, \mathbb{Z})$ appears naturally as a subgroup of the $GL(2n, \mathbb{Z})$ large diffeomorphisms of the doubled torus.  Because of this, geometric and non-geometric backgrounds are given equal footing in the Doubled Formalism.  They are distinguished only according to whether the splitting (also known as polarisation)  can be globally extended or not. 
  
 \subsection{Lagrangian and Constraint}
 
 To describe the sigma model we first introduce some local coordinates on the $T^{2n}$ doubled torus which we denote by $\X^I$ and coordinates on the base $N$ denoted $Y^a$.   The world-sheet of the sigma model is mapped into the doubled torus by $\X^I(\sigma)$.  This allows us to define a world-sheet 1-form\footnote{Following  \cite{Hull:2004in,Hull:2006va} we will work with world-sheet forms on a Lorentzian world-sheet and will slightly abuse notation by not indicating the pull back to the world-sheet explicitly.}
 \be
{ \cal P}^I = d\X^I \, . 
 \ee
 Additionally we introduce a connection in the bundle given by the  1-form 
\be
{\cal A}^I ={ \cal A}^I_a dY^a\, , 
\ee
and covariant momenta
\be
\hat{{ \cal P}}^I = d\X^I  + {\cal A}^I  \, . 
\ee
 As before, the $n^2$ moduli fields on the fibre are packaged into the coset form 
 \be
 \label{fibremetric}
 \cH_{IJ}(Y) = \left(\begin{array}{cc} (G - BG^{-1}B)_{ij}  & (BG^{-1})_{i}^{\ j} \\ - (G^{-1}B)^{i}_{\ j} & G^{ij} \end{array}\right) \, ,
\ee
but note that they are allowed to depend on the coordinates of the base.    

The starting point for the doubled sigma model is the Lagrangian 
\be
\label{doubledlag}
\mathcal{L} = \frac{1}{4} \H_{IJ}(Y)\hat{{ \cal P}}^I\wedge\ast \hat{{ \cal P}}^J - \frac{1}{2} L_{IJ}{ \cal P}^I \wedge {\cal A}^J + \mathcal{L}(Y) + \mathcal{L}_{top}(\X)\, , 
\ee
where  $\mathcal{L}(Y)$ is a standard sigma model on the base $N$ given (in these conventions) by 
\be
\label{baselag}
 \mathcal{L}(Y) = \frac{1}{2}g_{ab}dY^a \wedge \ast dY^b +  \frac{1}{2}b_{ab}dY^a \wedge  dY^b \, .
 \ee 
The final term in (\ref{doubledlag}) is a topological term which we will examine in a little more detail when considering the equivalence with the standard formulation of String Theory.    Note that the kinetic term of (\ref{doubledlag}) is a factor of a half down compared with the Lagrangian on the base (\ref{baselag}) and that an overall factor of $2\pi$ in the normalisations of these Lagrangians is omitted for convenience. 

This Lagrangian is to be supplemented by the constraint 
\be
\label{constraint}
\hat{{ \cal P}}^I = L^{IK}\H_{KJ} \ast \hat{{ \cal P}}^J\, ,
\ee
where $L_{IJ}$ is the invariant $O(d,d)$ metric introduced in (\ref{oddmetric}) which we repeat for convenience here
\be
\label{oddmetric2}
L_{IJ}  = \left(\begin{array}{cc}  0 & 1 \\ 1 & 0 \end{array}\right)\, .
\ee
Note that for this constraint to be consistent we must require that 
\be
 L^{IK}\H_{KJ}L^{JL}\H_{LM} = \delta_{M}^I
\ee
which is indeed true for the $O(d,d)$ coset form of the doubled fibre metric (\ref{fibremetric}).  

For the remainder of this thesis we shall make two simplifying assumptions to aid our presentation, first that the fibration is trivial in the sense that the connection $\cal{A}^I$ can be set to zero.  In terms of the physical $T^n$ fibration over $N$ this assumption corresponds to demanding that the off-diagonal components of the background with one index taking values in the torus and the other in the base are set to zero (i.e. $E_{ai}=E_{ia}=0$).  The second is that we shall assume that on the base the 2-form $b_{ab}$ vanishes.   With these assumptions the Lagrangian is simply 
\be
\label{doubledlag2}
\mathcal{L} = \frac{1}{4} \H_{IJ}(Y)d\X^I\wedge\ast d\X^J + \mathcal{L}(Y) + \mathcal{L}_{top}(\X) \, , 
\ee
 and the constraint is 
 \be
 \label{constraint2}
 d\X^I = L^{IJ}\H_{JK}\ast d \X^K\, .
 \ee
  To understand this constraint it is helpful to introduce a vielbein to allow a change to a chiral frame (denoted by over-bars on indices) where:
\bea
\H_{\Ab\Bb}(y) = \left( \begin{array}{cc}
\openone &0\\
0 & \openone
\end{array}\right), &  L_{\Ab\Bb} = \left( \begin{array}{cc}
\openone &0\\
0 & -\openone
\end{array}\right).
\eea
In this frame the constraint (\ref{constraint2}) is a chirality
constraint ensuring that half the $\X^\Ab$ are chiral Bosons and half
are anti-chiral Bosons.  Such a frame can be reached by introducing a vielbein for the fibre metric $\H_{IJ}$ given by (\ref{fibremetric}) 
\bea
\H_{IJ} =  V^{\hat{I}}_I \delta_{\hat{I}\hat{J}} V^{\hat{J}}_{ J}\, .
\eea
with 
\bea
V^{\hat{A}}_A = \left( \begin{array}{cc}
e^{t} & 0 \\ -e^{-1}B& e^{-1}
\end{array} \right)\, ,
\eea 
and $G_{ij} =e_i^{\ \hat{i}} e_j^{\ \hat{j}} \delta_{\hat{i}\hat{j}}$.   
This allows the definition of an intermediate frame in which 
\bea
L_{\hat{I}\hat{J}} = \left(\begin{array}{cc}
	0 & 1 \\ 1 & 0 
\end{array} \right)\, , &  \H_{\hat{I}\hat{I}} = \left(\begin{array}{cc}
	1 & 0 \\ 0 & 1
\end{array} \right)\,
\eea 
Supplementing this with a basis transformation defined by
\bea
O_{\bar{A}}^{\, \hat{I}} = \frac{1}{\sqrt{2}} \left(\begin{array}{cc}
	1 & 1 \\ 1 & -1 
\end{array} \right)
\eea
brings $\H$ and $L$ into the required form.   In this chiral basis the one-forms ${\cal P}^{\bar{A}}= (P^{\bar{i}}, Q_{\bar{i}})$ obey simple chirality constraints 
\bea
 P^{\bar{i} }_+ = 0\, ,  &  Q_{\bar{i} - }=0 \, .
\eea

\subsection{Action of $O(n,n,\mathbb{Z})$ and Polarisation}

The Lagrangian (\ref{doubledlag2}) has a global symmetry of $GL(2n, \mathbb{R})$ acting as 
\be
\cH^\prime  = \cO^t \cH \cO\, ,  \quad  \X^\prime = \cO^{-1} \X \, , \quad \cO \in GL(2n, \mathbb{R})\,.
\ee
However,  to preserve the periodicity of the coordinates $\X$ this symmetry is broken down to the discrete subgroup  $GL(2n, \mathbb{Z})$.  Under these transformations the constraint (\ref{constraint2}) becomes
\be
\cO^{-1} d\X^I = L^{-1} \cO^{t} \H \ast d \X  \, .
\ee
Hence, to preserve the constraint we require that 
\be
\cO^t L \cO = L 
\ee 
and therefore that the global symmetry is reduced to $O(n,n,\mathbb{Z}) \subset GL(2n ,\mathbb{R})$.  It is now clear that the T-duality group has been promoted to the role of a manifest symmetry.   

To complete the description of the Doubled Formalism one needs to specify the physical subspace which defines a splitting  of the coordinates $\X^I = (X^i, \tilde{X}_i)$.  Formally this is done by introducing projectors $\Pi$ and $\tilde{\Pi}$ such that $X^i = \Pi^i_{\ I}\X^I$    and   $\tilde{X}_i = \tilde{\Pi}_{i I}\X^I$ .   These projectors are required to pick out subspaces that are maximally isotropic in the sense that $\Pi L^{-1} \Pi^t = 0$.    These projectors are sometimes combined into a  `polarisation vielbein' given by $\Theta = (\Pi , \tilde{\Pi})$.   

This is a somewhat formal way of saying that we are picking out a preferred basis choice for the sigma model but has use when describing T-duality.    Indeed, the formulas for the coset matrix $\H_{IJ}$ and the invariant metric $L_{IJ}$  given by (\ref{fibremetric}) and (\ref{oddmetric2}) should be understood to be in the basis defined by this splitting.  The components of $\H$ are given by acting with projectors so that for example
\be
G^{ij} = \Pi^{i I} \Pi^{j J} \cH_{IJ} \, ,
\ee   
in which we have raised the indices on the projectors with the invariant metric $L^{IJ}$.  

There are now two ways to view T-duality. The first is known as active whereby the polarisation is fixed but the geometry transforms according to the usual transformation law
\be 
\cH^\prime = \cO^t \cH \cO \, . 
\ee  
The second approach is to consider the geometry as fixed but that the polarisation vielbein changes according to
\be
\Theta^\prime = \Theta \cO \, . 
\ee
In this view T-duality amounts to picking a different choice of polarisation. 

\subsection{Recovering the standard sigma model}

We now indicate two approaches to recover the standard sigma model from the doubled Lagrangian. The first approach is to solve the constraint and show that the equations of motion subject to the constraint are equivalent to those arising from the standard string sigma model.  The second approach is more sophisticated and works by recognising the constraint can be interpreted in terms of the vanishing of a current.  If this current can be successfully gauged, a judicial choice of gauge fixing invokes the constraint. 
\begin{itemize}
\item Solving the constraint
\end{itemize}
The equations of motion arising from the doubled Lagrangian (\ref{doubledlag2}) are, for the fibre coordinates,
\be
\label{eqm1}
0= \H_{IJ} \partial^2 \X^J + \partial_a \H_{IJ} \partial_\alpha Y^a \partial^\a \X^J \, ,
\ee
and for the base coordinates
\be
\label{eqm2}
0= \frac{1}{4} \partial_c \H_{IJ} \partial_\a \X^I \partial^\a \X^J - g_{ac}\left( \partial^2 Y^a +\hat{\Gamma}^a_{de} \partial_\a Y^d \partial^\a Y^e  \right) \, , 
\ee
where $ \hat{\Gamma}^a_{de}$ is the Christoffel symbol constructed from the base metric. 

The way to proceed is to notice that, after choosing a polarisation, the constraint can be written as
\be
 \partial^\a \tilde{X}_j = G_{jk} \e^{\a\b} \partial_\beta X^k +B_{jk}\partial^\a X^k \, , 
\ee
and that this can be used at the level of equations of motion to replace all occurrences of  the dual coordinates $\tilde{X}_i$ with the physical coordinates $X^i$.    After expanding out in the $(X,\tilde{X})$ basis one finds that the equation of motion (\ref{eqm1}) can be written as
\be
0 = g_{ij} \partial^2 X^j + \partial_a g_{ij} \partial_\a Y^a \partial^\a X^j + \partial_a b_{ij} \e^{\a\b} \partial_\a Y^a \partial_\b X^j 
\ee
and (\ref{eqm2}) as 
\be
0=   \frac{1}{2}  \partial_a g_{ij} \partial_\a X^i \partial^\a X^j + \frac{1}{2} \partial_a b_{ij} \e^{\a\b} \partial_\a X^i \partial_\b X^j 
 -      g_{ac}\left( \partial^2 Y^a +\hat{\Gamma}^a_{de} \partial_\a Y^d \partial^\a Y^e  \right)\, . 
\ee
The latter two equations correspond to the equations of motion obtained from the standard sigma model of the form 
\be
\label{standardsigma}
\cL = \frac{1}{2} g_{ij}(Y)  d X^i \wedge \ast dX^j + \frac{1}{2} b_{ij}(Y) dX^i\wedge dX^j + \frac{1}{2} g_{ab}(Y) dY^a \wedge dY^b \, . 
\ee
This demonstrates that the doubled sigma model together with the constraint is equivalent, at the classical level of equations of motion, to the standard string sigma model. 
\begin{itemize}
\item Gauging the Current
\end{itemize}

To go further one would like to establish the equivalence of the two sigma models at the quantum level.  The principal difficulty with this approach is that the constraint needs to be implemented in a consistent way.   In terms of the Dirac constraint classification the chirality constraint is second class and may not be implemented by means of a simple Lagrange multiplier method.\footnote{See appendix for more details concerning chiral bosons and treating the chirality constraint.}     However an approach to this is given in \cite{Hull:2006va} whereby one notices that the conserved current associated to shift symmetries in the fibre together with the trivial Bianchi identity for fibre momenta can be used write down a conserved current
\be
J_{I} = \H_{IJ} d\X^J - L_{IJ} \ast d\X^J \, . 
\ee 
The constraint equation (\ref{constraint2}) is equivalent to demanding that the polarisation projected component $J^i = 0$.    The symmetry giving rise to this current can be consistently gauged by introducing a world-sheet gauge field however to ensure that the action is completely gauge invariant (including large gauge transformations) it was found in  \cite{Hull:2006va}  that a topological term 
\be
\cL_{top} = \frac{1}{2} d\tilde{X}_i \wedge dX^i \, ,
\ee 
must also be included to supplement the minimal coupling.   As well as ensuring gauge invariance this topological term also ensures that the winding contributions from the dual coordinate $\tilde{X}$ can   be gauged away.  This gauged theory can be seen to be equivalent to the standard sigma model (\ref{standardsigma}) and upon fixing a gauge equivalent to the Doubled Formalism.

\subsection{Geometric vs. Non-Geometric}
To discuss the difference between geometric and non-geometric backgrounds one must examine topological issues and in particular the nature of the transition functions between patches of the base.  For two patches of the base $U_1$ and $U_2$, on the overlap  $U_1 \cap U_2$ the coordinates are related by a transition function
\be
\X_1 = g_{21} \X_2 \, , \quad  g_{21} \in O(n,n , \mathbb{Z}) \, .
\ee
In the active view point, where polarisation $\Theta$ is fixed but the geometry is transformed,  the condition for this patching to be geometric is that the physical coordinates, obtained from projecting with the polarisation, are glued only to physical coordinates.  More formally we have 
\be
\Theta \X_1 = (X_1, \tilde{X}_1) \, , \quad   \Theta \X_2 = (X_2, \tilde{X}_2)\, , 
\ee
 and therefore 
 \be
 \Theta \X_2 = \Theta g_{12} \X_1= \Theta g_{12} \Theta^{-1} \Theta \X_1 = \hat{g}_{12} \Theta \X_1\, . 
\ee
Then the physical coordinates are glued according to 
\be
X^i_2 =  (\hat{g}_{12})^{i}_{\ j} X_1^j + (\hat{g}_{12})^{ij} \tilde{X}_{1j}  \, . 
\ee
Thus, for the physical coordinates to be glued only to physical coordinates, all transition functions must be of the form
\be
(\hat{g})^I_{\ J} = 
\left(
\begin{array}{cc}
 g^i_{\ j}  & 0     \\
  g_{ij }  &  g_i^{\ j}  
\end{array}
\right)
\ee
which are the $O(n,n, \mathbb{Z})$ elements formed by the semi direct product of $Gl(n, \mathbb{Z})$ large diffeomorphism of the physical torus and integer shifts in the B-field.    Otherwise the background is said to be non-geometric.\footnote{One can tighten the definition of geometric to exclude the B-field shifts. This restriction is equivalent to demanding that the non-physical T-dual coordinates also only get glued amongst themselves.}

\subsection{Extensions to the Doubled Formalism}
We now outline a few advances and extensions to the above formalism. 
\subsubsection{Dilaton}
Alongside the metric and B-field a background in String Theory is equipped with a scalar field, the dilaton.  One is forced to ask  how should this scalar field be included in the Doubled Formalism.  A first guess is to include a scalar field which replicates the same coupling as for the standard string namely
\be
S_{dil}= \frac{1}{8\pi} \int d^2 \sigma \sqrt{h} \Phi(Y)  R^{(2)}
\ee 
where $R^{(2)}$ is the scalar curvature of the world-sheet and, in keeping with the ansatz for metric and B-field, the scalar only depends on the base coordinates.  This `doubled dilaton' is duality invariant under $O(n,n,\mathbb{Z})$ and related to the standard dilaton field $\phi$ by \cite{Hull:2006va}
\be
\Phi = \phi - \frac{1}{2}\ln \det G
\ee
so that from 
\be 
\Phi^\prime = \phi^\prime - \frac{1}{2}\ln \det G^\prime = \phi - \frac{1}{2}\ln \det G
\ee
we recover the dilaton transformation rule (\ref{dilshift2}). 
\subsubsection{Branes}
In the original work of Hull \cite{Hull:2004in}, the extension of the Doubled Formalism to open strings and their  D-branes interpretation was suggested.  Since T-duality swaps  boundary conditions,  of the $\X$ coordinates exactly half should obey Neumann boundary conditions and half Dirichlet.   This splitting is, in general, different from the splitting induced by polarisation choice.   If, for a given polarisation, exactly N of the physical coordinates obey Neumann conditions then there is a Brane wrapping N of the compact dimensions.  The dimensionality of the brane then changes according to the usual  T-duality rules under a change in polarisation.  The projectors that specify Neumann and Dirichlet conditions are required to obey some consistency conditions detailed in   \cite{Lawrence:2006ma,Albertsson:2008gq}.  Further related work on the role of D-branes in non-geometric backgrounds can be found in \cite{Kawai:2007qd}.
\subsubsection{Supersymmetry}
A world-sheet $\cN=1$ supersymmetric extension of the formalism has been considered in  
\cite{HackettJones:2006bp,Hull:2006va}.  To achieve this one simply promotes coordinates into superfields and world-sheet derivatives into super-covariant derivatives.  The constraint   (\ref{constraint2}) then is generalised to a supersymmetric version which includes a chirality constraint on fermions. 
\subsubsection{Quantum Effects}
Some progress has been made on understanding the quantum aspects of the Doubled Formalism.  In  \cite{Hull:2006va} it was explained how to implement the constraint at the quantum level through gauging.   In \cite{HackettJones:2006bp} a canonical quantisation was performed, with Dirac brackets used to invoke the (second class) constraint.     In \cite{Berman:2007vi} the 1-loop partition function of the doubled string was calculated and shown to be equivalent to the standard string after a holomorphic factorisation.  This was generalised to the supersymmetric partition function in \cite{Chowdhury:2007ba}.
An interesting direction of research has been  to take some of the ideas of the Doubled Formalism and to apply them to the realm of String Field Theory \cite{Hull:2009mi,Hull:2009zb,Hohm:2010jy}.   
\subsubsection{Doubled Twisted Torus}
For the example of section (\ref{secT-folds}) with a non-geometric $T^2$ over a base $S^1$ it has been suggested \cite{Dabholkar:2005ve}  that a further duality around the base may be possible even though isometry is not present.  This situation has been considered from the context of a generalised Doubled Formalism in which the base $S^1$ is also doubled.   In this case that doubled space is no longer a doubled torus but has a more general description as the  quotient of a group manifold by a discrete subgroup and is known as a doubled twisted torus.   In general this means considering a doubled fibre metric that has internal coordinate dependance.  The doubled sigma model description of this has been studied in    \cite{Hull:2007jy,ReidEdwards:2009nu,ReidEdwards:2010vp}.  Essentially, the momenta entering into the action are replaced with the Maurer-Cartan left-invariant one-forms on the group and a supplementary term is included into the action which is a Wess-Zumino term given by the pull back of the canonical three form on the group.    There is some similarity between such sigma-models and Poisson-Lie T-duality invariant sigma-models \cite{Klimcik:1995jn,Klimcik:1995dy} that we will consider in later chapters.

\chapter{Quantum Aspects of Doubled String Theory}
\begin{quote}
In this chapter we examine the Doubled Formalism at the quantum level.  We calculate the beta-functionals of the string in the Doubled Formalism so as to determine the background-field equations for the doubled space.  We relate these to the equations that arise for the standard, non-doubled, string.  The results presented in this chapter have appeared in the author's previous publications \cite{Berman:2007xn,Berman:2007yf}.
\end{quote}

\section{Overview}
The duality invariant formalism described in the preceding chapter successfully promotes T-duality to a manifest symmetry of the string sigma model action.  Whilst it is straightforward to demonstrate the classical equivalence of this formalism to the usual sigma model,  showing quantum equivalence is less trivial.  In addition to the arguments presented in  \cite{Hull:2006va}, progress in this direction has been made by showing the equivalence of the partition function  for both the bosonic and supersymmetric extension of the Doubled Formalism \cite{Berman:2007vi,Chowdhury:2007ba} and in   \cite{HackettJones:2006bp} the canonical quantisation of the Doubled Formalism was investigated. 

In this chapter we ask: what are the beta functionals for the couplings in the Doubled Formalism?  Demanding that the conformal invariance of String Theory holds at the quantum level requires the vanishing of such beta functionals.  This, in turn, restricts the classes of backgrounds on which the string can propagate to those that (to first order in $\alp$) obey equations \cite{Callan:1985ia} analogous to Einstein's equations.\footnote{See appendix.}    

This study was prompted by various questions. The most important was to determine if the background-field equations arising from the one-loop beta-functional for the Doubled Formalism were the same as those for the usual string. A priori, this did not have to be the case.  
In fact, one may imagine that they will be different since the doubled
formalism naturally incorporates the string winding modes which could
in principle correct the usual beta-function. We know that
worldsheet instantons correct T-duality \cite{Harvey:2005ab, Tong:2002rq} and 
since the doubled space contains the naive T-dual one may think of all sorts of
possibilities that could arise for corrections to the doubled geometry. 

The outline of this chapter is as follows:  we begin by studying the constraint that supplements the doubled action and describe how the PST procedure \cite{Pasti:1996vs,Pasti:1995us,Pasti:1997mj} can be used to implement the constraint in a Lorentz covariant manner. We then show how gauge fixing the residual PST symmetry produces an action that closely resembles the duality invariant formalism of Tseytlin \cite{Tseytlin:1990nb,Tseytlin:1990va}. This is followed by a calculation of the beta functionals by means of a background field expansion.   We relate these to the analogous equations of the standard string and conclude by providing a direct linkage between the beta functions and the field equations obtained from a toroidally compactified gravity theory.

\section{Implementing the Constraint}
We recall that the doubled string has an action 
\be
\label{doubleact}
\mathcal{L} = \frac{1}{4} \cH_{IJ}(Y)d\X^I\wedge\ast d\X^J  +
\mathcal{L}(Y) + \mathcal{L}_{top}(\X)\,,
\ee
where $\mathcal{L}(Y)$ is the standard Lagrangian for a string on the
base and $\H(Y)$ is a metric on the fibre.  The topological term plays no role in the rest of this chapter and can be safely ignored for the moment.
This action is supplemented by a constraint 
\be
\label{constrainthew}
d\X^I = L^{IJ}\cH_{JK}\ast d\X^K \, ,
\ee
which is essentially a chirality constraint on the bosons. 

To understand this constraint more clearly we consider the simplest example, the one-dimensional target space of a circle, with
constant radius $R$. The doubled action on the fibre is
\be
S_{d}=\frac{1}{4} R^2 \int dX\wedge \ast dX + \frac{1}{4} R^{-2} \int d\tilde X \wedge \ast d\tilde X .
\ee
We first change to a basis in which the fields are chiral by defining:
\bea
\X_+= RX+R^{-1}\tilde{X}, &  \d_-\X_+ =0\, , \\ \X_-= RX-R^{-1}\tilde{X}, & \d_+\X_- =0\, . 
\eea
Note that because the radius is constant in this case the chirality constraint is simple. 
In this basis the action becomes
\be
S_{d}= \frac{1}{8}\int d\X_+\wedge \ast d\X_+ + \frac{1}{8} \int d  \X_- \wedge \ast d \X_-\, .
\ee
One may then incorporate the constraints into the action using the
method of Pasti, Sorokin and Tonin \cite{Pasti:1996vs,Pasti:1995us,Pasti:1997mj}.  We define one-forms
\bea
\mathcal{P}=d\X_+-\ast d\X_+ ,& \mathcal{Q}=d\X_-+\ast d\X_-\, ,
\eea
which vanish on the constraint.  These allow us to incorporate the constraint into the action via the introduction of two auxiliary closed one-forms $u$ and $v$ as follows: 
\be
S_{PST}= \frac{1}{8}\int d\X_+\wedge \ast d\X_+ + \frac{1}{8} \int d  \X_- \wedge \ast d \X_-   -\frac{1}{8}\int d^2\sigma\left( \frac{(\mathcal{P}_m u^m)^2}{u^2} +\frac{(\mathcal{Q}_m v^m)^2}{v^2}\right).
\ee

As explained in the appendix, the PST action works by essentially introducing a new gauge symmetry, the PST symmetry, that allows the gauging away of fields that do not
obey the chiral constraints. Thus only the fields obeying the chiral
constraints are physical.

There are now two ways to proceed. One may either gauge fix the
PST-style action immediately 
which will break manifest Lorentz invariance
or try to quantise whilst maintaining covariance and introduce ghosts to deal
with the PST gauge symmetry. In this paper we choose the non-covariant
option and immediately gauge fix to give a Floreanini-Jackiw \cite{Floreanini:1987as}
style action. Picking the auxiliary PST fields ($u$ and $v$) to be
time-like produces two copies of the FJ action (one chiral and one anti-chiral)
\be\label{FJL}
S = \frac{1}{4}\int d^2\s ( \d_1\X_+\d_{-}\X_+ -  \d_{1}\X_- \d_+\X_-).
\ee
If we re-expand this in the non-chiral basis we find an action
\be
\label{eqTseyaction}
 S= \frac{1}{2}\int d^2\sigma\left[ -(R\d_1 X)^2 - ( R^{-1}\d_1 \tilde{X})^2 + \partial_0X\partial_1\tilde{X} + \partial_1 X \partial_0 \tilde{X} \right] \, ,
\ee
which may be recognised as Tseytlin's  duality symmetric formulation \cite{Tseytlin:1990va,Tseytlin:1990nb}.  Notice that the unusual normalisation of (\ref{doubleact}) was exactly what was needed for this form of the action to have the correct normalisation.
The constraints 
\be
\d_0 \tilde{X} = R^2 \d_1 X \, ,  \qquad \d_0 X = R^{-2} \d_1 \tilde{X} \, ,
\ee
 then follow after integrating the equations of motion and the string wave equation for the physical coordinate $X$ is implied by combining the constraint equations.\footnote{We fix
   the arbitrary function of $\tau$ introduced by integration by
   observing that (\ref{eqTseyaction}) has $\dl{X} = f(\tau)$ gauge
   invariance as shown in the appendix.}
 
Returning to the general case, the PST procedure yields an action
\bea\label{Daction}
S= \frac{1}{2} \int d^2\sigma\left[ -\cM_{\a\b} \partial_1 X^\a \partial_1 X^\b + \cL_{\a\b} \partial_1 X^\a \partial_0 X^\b + \cK_{\a\b} \partial_0 X^\a \partial_0 X^\b\right]\,   ,
\eea
where $X^\a = (\X^A, Y^a)=(X^i, \tilde{X}_j, Y^a)$ and\footnote{In our notation we reserve the Greek characters $\mu$ and $\nu$ to denote worldsheet indices.}
\be
\cM=
\left(
\begin{array}{cc}
\cH  &   0 \\
 0 &    g
\end{array}
\right), \quad 
\cL=
\left(
\begin{array}{cc}
L &   0 \\
 0 &    0
\end{array}
\right), \quad
\cK=
\left(
\begin{array}{cc}
0  &   0 \\
 0 &    g
\end{array}
\right).
\ee
For the fibre coordinates we have the equation of motion 
\bea
\d_1 \left( \H \d_1 \X \right) = L \d_1\d_0 \X,
\eea
which integrates to give the constraint (\ref{constrainthew}).  
This form of the action will be our starting point for calculating the beta-functional of the theory.

\section{The Background Field Expansion}

To perturbatively study ultra-violet divergences in the doubled
formalism we expand  quantum fluctuations around a classical
background.  By writing the fields $X^\a$ as the sum of a classical piece
$X^\a_{cl}$ and a quantum fluctuation $\pi^\a$ we may define the effective action
\be
e^{iS_{eff}} = \int [d\pi] e^{iS[X_{cl} + \pi]} \, .
\ee
In what follows we shall calculate the 1-loop contribution to this effective action.  

 We make a choice of coordinates that leaves the general
coordinate invariance of the action manifest in the perturbative
expansion \cite{Braaten:1985is, AlvarezGaume:1981hn}.  Instead of expanding in $\pi^\a$, which would not be covariant, one expands in the tangent vector $\xi^\a$ to the geodesic between  $X^\a_{cl}$ and  $X^\a_{cl}+\pi^\a$. The vector $\xi^\a$ is understood to be evaluated at  $X^\a_{cl}$ and has length equal to that of the
geodesic. Since it is a covariant object an expansion in terms of $\xi^\a$ will result in expansion coefficients which are tensorial.  More details of this procedure can be found in the appendix.   To evaluate the one-loop contribution to the action it is sufficient to expand upto quadratic order in fluctuations.  The linear term $S^{(1)}$ provides no contributions since it is, in general, proportional to the equations of motions which we assume the classical background obeys.  The quadratic term $S^{(2)}$ provides the kinetic term for fluctuations as well as certain interactions $S^{(2)}_{int}$.   In taking the exponential above there are one-loop contributions that arise from the connected diagrams formed by considering
\be
S_{1-loop\ eff}= \langle S^{(2)}_{int} \rangle  + \frac{1}{2}\langle  S^{(2)}_{int}   S^{(2)}_{int}   \rangle \, . 
\ee

Note that in this procedure one must specify the background metric with which to perform such an expansion.  Normally, the appropriate target space metric is completely obvious from the form of the sigma model.  In the case of the Doubled Formalism there is some ambiguity; whilst the metric on the base is clearly defined as $g_{ab}$ there is a choice as to whether we consider the metric on the fibre to be $L_{IJ}$ or $\cH_{IJ}$.  From the form action (\ref{Daction}) it certainly seems most sensible to consider the metric of the total target space to be given by $\cM_{\a\b}$.  A different choice may be possible, and one should remember that this is a calculative tool that should not alter the end result.  

We will use the algorithmic method of calculating the background-field expansion developed in \cite{Mukhi:1985vy}. To obtain the $n$th order background-field expanded action we simply act on the Lagrangian with the operator
\beq
\int d^2\s \xi^\a(\uc )D^\s_\a
\eeq
$n$ times and divide by $n!$ ($D^\s_\a$ is the covariant functional derivative with respect to $X^\a(\uc)$). The action of this operator can be summarised as
\bea
\int d^2\s \,\xi^\a(\uc)D^\s_\a \xi^\b(\uc^\prime)&=&0\, ,\\
\int d^2\s \,\xi^\a(\uc)D^\s_\a \d_\m X^\b(\uc^\prime)&=&D_\m\xi^\b(\uc^\prime)\, ,\\
\int d^2\s \,\xi^\a(\uc)D^\s_\a D_\m\xi^\b(\uc^\prime)&=&R^\b_{\phantom{\b}\a\g\dl}\d_\m X^\dl \xi^\a \xi^\g(\uc^\prime)\, ,\\
\int d^2\s \,\xi^\a(\uc)D^\s_\a T_{\a_1\a_2\ldots \a_n}(X(\uc^\prime))&=&D_\b T_{\a_1\a_2\ldots \a_n}\xi^\b(\uc^\prime)\, ,\label{DT}
\eea
where $R^\b_{\phantom{\b}\a\g\dl}$ is the target space Riemann tensor and $T_{\a_1\a_2\ldots \a_n}$ is a rank $n$ tensor and these are understood to be evaluated at the classical value $X_{cl}$. The form of (\ref{DT}) is particularly relevant, leading to simplification when dealing with the metric.

Expanding the first term in (\ref{Daction}) is exactly the same as the
standard sigma model calculation (albeit without worldsheet
covariance), at first order we have\footnote{From now on $X$ refers to the classical field $X_{cl}$.}
\beq
-\cM_{\a\b}\d_1X^{\a}D_1\xi^\b \, , 
\eeq
and at second order
\beq
-\frac{1}{2}\left(\cM_{\a\b}D_1\xi^\a D_1\xi^\b+R_{\g\a\b\dl}\xi^\a\xi^\b\d_1X^{\g}\d_1X^\dl\right),
\eeq
where $R_{\g\a\b\dl}$ is the Riemann tensor constructed from the metric $\cM$. The expansion of the $\cL$ term in the action is more complex giving
\beq
\frac{1}{2}
\left(\cL_{\a\b}\d_0X^{\a}D_1\xi^\b+\cL_{\a\b}D_0\xi^{\a}\d_1X^\b+D_{\a}\cL_{\g\b}\xi^\a\d_0 X^{\g}\d_1X^\b\right)
\eeq
at first order and
\bea
\frac{1}{2}\left(\cL_{\a\b}D_0\xi^{\a}D_1\xi^\b+\frac{1}{2}\left( D_\a D_\b\cL_{\g\dl} + L_{\g\s}R^{\s}_{\ \a\b\dl} +  \cL_{\dl\s}R^{\s}_{\ \a\b\g} \right)\xi^\a\xi^\b\d_0X^{\g}\d_1X^\dl \qquad\right.\nn\\
+D_{\g}\cL_{\a\b}\xi^\g\left(\d_0X^{\a}D_1\xi^\b+D_0\xi^{\a}\d_1X^\b\right)\biggr)\qquad
\eea
at second-order. The  $\cK$ term in (\ref{Daction}) may be expanded in a similar way. 

The first-order terms in $\xi$ vanish as they should (using the
equations of motion of $X_{cl}$) leaving the second-order Lagrangian which is given by
\bea\label{fulag}
2{\mathcal L}_{(2)}&=&-\cM_{\a\b}D_1\xi^\a D_1\xi^\b+\cL_{\a\b}D_0\xi^\a D_1\xi^\b+\cK_{\a\b}D_0\xi^\a D_0\xi^\b\nn\\
&&-R_{\g\a\b\dl}\xi^\a\xi^\b\d_1X^\g\d_1X^\dl+\cL_{\a\b;\g}\xi^\g(D_0\xi^\a\d_1X^\b+\d_0X^\a D_1\xi^\b)\nn\\
&& +\frac{1}{2}D_\a D_\b\cL_{\g\dl}\xi^\a\xi^\b\d_0X^{\g}\d_1X^\dl
+\frac{1}{2}\left( \cL_{\g\s}R^{\s}_{\ \a\b\dl} + \cL_{\dl\s}R^{\s}_{\ \a\b\g}\right)\xi^\a\xi^\b\d_0X^\g\d_1X^\dl\nn\\
&&+2\cK_{\a\b;\g}\xi^\g D_0\xi^\a\d_0X^\b\nn\\
&&  +\frac{1}{2}D_\a D_\b\cK_{\g\dl}\xi^\a\xi^\b\d_0X^{\g}\d_0X^\dl
+\cK_{\g\s}R^{\s}_{\ \a\b\dl}\xi^\a\xi^\b\d_0X^\g\d_0X^\dl \, .
\eea

\section{Simplification Strategy}

The effective action above appears rather complex and it would certainly be intimidating to calculate the correct Wick contraction in this form.  However we have the opportunity to simplify the effective action by using the equations of motion of $X^{\alpha}$ (recall
$X^{\alpha}$ is the classical field configuration which we are expanding
around and so obeys its equation of motion).  This equation of motion is 
\be
D_{1}(\cM_{\a\b}\d_1 X^{\b})=\cL_{\a\b}\d_1\d_0X^\b+\Dh_{0}(\cK_{\a\b}\d_0X^{\b})\, ,
\ee
where $\Dh_{0}$ is a pulled back covariant derivative constructed only from the
base metric $g_{ab}$ ($\Dh_{0}\cK=0$ and $\hat{\Gamma}$ will similarly refer
to the connection constructed from $g_{ab}$). This equation is required to show vanishing of the first-order term in the expansion and we now use it to remove all $\cL$ terms (except the $\cL$
fluctuation `kinetic' term) from the action. Therefore we find,  after performing some integration by parts, that we may use the following substitution:
\bea
&&\cL_{\a\b;\gamma}\xi^{\gamma}\left(\d_0\xi^{\a}\d_1X^{\b}+\d_1\xi^{\a}\d_0X^{\b}\right)+\cL_{\a\b}\G{\a}{\g}{\dl}\xi^{\gamma}\d_1X^{\dl}\d_0\xi^{\b}+\cL_{\a\b}\G{\a}{\g}{\dl}\xi^{\gamma}\d_0X^{\dl}\d_1\xi^{\b}\nn\\
&=&\frac{1}{2}\cL_{\b\dl}\d_{\s}\G{\dl}{\g}{\a}\xi^{\g}\xi^{\a}(\d_1X^{\b}\d_0X^{\s}+\d_0X^{\b}\d_1X^{\s})\nn\\
&&-2\cM_{\dl\b}\G{\dl}{\g}{\a}\xi^{\g}\d_1\xi^{\a}\d_1X^{\b}-\d_{\s}(\cM_{\dl\b}\G{\dl}{\g}{\a})\xi^{\g}\xi^{\a}\d_1X^{\b}\d_1X^{\s}\nn\\
&&+2\cK_{\dl\b}\G{\dl}{\g}{\a}\xi^{\g}\d_0\xi^{\a}\d_0X^{\b}+\d_{\s}(\cK_{\dl\b}\G{\dl}{\g}{\a}\xi^{\g})\xi^{\a}\d_0X^{\b}\d_0X^{\s}\nn\\
&&+\G{\dl}{\g}{\a}\xi^{\g}\xi^{\a}\left(\cM_{\dl\b}\G{\b}{\s}{\tau}\d_1X^{\s}\d_1X^{\tau}-\cK_{\dl\b}\hat{\Gamma}^{\b}_{\  \s\tau}\d_0X^{\s}\d_0X^{\tau}\right)\, .
\eea
This leads to a dramatically simplified Lagrangian given by
\bea\label{Lag}
2{\mathcal L}_{(2)}&=&-\cM_{\a\b}\d_1\xi^\a \d_1\xi^\b+\cL_{\a\b}\d_0\xi^\a \d_1\xi^\b+\cK_{\a\b}\d_0\xi^\a\d_0\xi^\b\nn\\
&&-2\d_\a\cM_{\g\b}\d_1X^\g\xi^\a\d_1\xi^\b-\frac{1}{2}\d_\a\d_\b\cM_{\g\dl}\d_1X^\g\d_1X^\dl\xi^\a\xi^\b\nn\\
&&+2\d_\a g_{\b\g}\xi^\a\d_0\xi^\b\d_0X^\g+\frac{1}{2}\d_\a\d_\b
\cK_{\g\dl}\xi^\a\xi^\b\d_0X^\g\d_0X^\dl  \, \, .
\eea
Note that we have chosen to proceed by expanding covariant derivatives
and simplifying using the equations of motion, rather than leaving
things expressed in terms of covariant derivatives.   This is the somewhat unfortunate price that we have had to pay for having a manageable action.\footnote{ We remark that(\ref{Lag}) could perhaps have been obtained more directly by considering a different metric on the total space.} 

To proceed with the Wick contractions one needs to extract the kinetic term for the propagators.  In (\ref{Lag}) this is not immediately obvious due to the coordinate dependance contained in the matrices $\cM_{\a\b}(Y)$.   One must introduce vielbeins and work in the
chiral frame (which we indicate by placing over-bars on indices) where we know how to find the fluctuation propagators. 

Once vielbeins are introduced there are, of course, terms with derivatives acting on the vielbeins. Normally such terms are accounted for by exchanging the usual connection for the spin connection.  The pull back of the spin connection to the worldsheet transforms as a gauge field.
 There is then a general argument that this gauge field, 
which is minimally coupled, cannot contribute to the Weyl anomaly. 
We have a modified action where the gauge connection is no 
longer minimally coupled and there is no such argument 
(indeed we find contributions from the `gauge' terms). 

The effect of introducing vielbeins is to replace all fluctuations  $\xi^\a$ in the second-order action (\ref{Lag}) by their chiral frame counter parts $\xi^{\bar{\a}}= \cV^\a_{\ \bar{\a}} \xi^\a$  and to add on the following terms created by pulling vielbeins through derivatives:
\bea\label{VLag}
2{\mathcal L}_{\cV}&=&-2\cM_{\ab\b}\d_1\Vt{\b}{\bar{\beta}}\xi^{\bar{\beta}}\d_1\xi^{\ab}-\cM_{\a\b}\d_1\Vt{\b}{\bar{\beta}}\d_1\Vt{\a}{\ab}\xi^{\bar{\beta}}\xi^{\ab}\nn\\
&&+\cL_{\ab\b}\d_1\Vt{\b}{\bar{\beta}}\xi^{\bar{\beta}}\d_0\xi^{\ab}+\cL_{\ab\b}\d_0\Vt{\b}{\bar{\beta}}\xi^{\bar{\beta}}\d_1\xi^{\ab}+\cL_{\a\b}\d_1\Vt{\b}{\bar{\beta}}\d_0\Vt{\a}{\ab}\xi^{\bar{\beta}}\xi^{\ab}\nn\\
&&+2\cK_{\ab\b}\d_0\Vt{\b}{\bar{\beta}}\xi^{\bar{\beta}}\d_0\xi^{\ab}+\cK_{\a\b}\d_0\Vt{\b}{\bar{\beta}}\d_0\Vt{\a}{\ab}\xi^{\bar{\beta}}\xi^{\ab}\nn\\
&&-2\d_{\ab}\cM_{\g\b}\d_1X^\g\d_1\Vt{\b}{\bar{\beta}}\xi^{\ab}\xi^{\bar{\beta}}+2\d_{\ab}
\cK_{\g\b}\d_0X^\g\d_0\Vt{\b}{\bar{\beta}}\xi^{\ab}\xi^{\bar{\beta}} \, \, .
\eea

\section{Wick Contraction}

In this chiral frame, $\cL$ and $\H$ are diagonal and one can calculate the propagators for the fluctuations from the `kinetic terms' in the Lagrangian.  On the fibre these `kinetic terms' are Floreanini-Jackiw style Lagrangians of the form (\ref{FJL}) 
for $n$ chiral and $n$ anti-chiral Bosons in flat space.  The propagators for such chiral Lagrangians have previously been considered by 
Tseytlin \cite{Tseytlin:1990va,Tseytlin:1990nb}.   In the appendix we provide full details and simply quote the result here.   The sum of a chiral and anti-chiral propagator is proportional to a standard Boson propagator $\Delta_0$. The difference of chiral and anti-chiral propagators gives a phase $\theta$.  The general result for our action is that for the coordinates on the fibre
\beq
\label{prop1}
\langle \xi^{\Ab}(\sigma) \xi^{\bar{B}}(\sigma^\prime) \rangle = \Delta_0(\sigma - \sigma^\prime) \H^{\Ab\bar{B}}  +\theta(\sigma - \sigma^\prime)  L^{\Ab\bar{B}}.
\eeq

$\Delta_0$ results in UV divergences as we take $\sigma - \sigma^\prime\rightarrow 0$.  Such divergence need regularisation and renormalisation.   In dimensional regularisation we find a simple pole  $\Delta_0(0) \sim \frac{1}{\epsilon}$.  The coefficients of $\Delta_0(0)$ will thus contribute to the Weyl anomaly and in turn, to the beta-functionals. 

As  $\sigma - \sigma^\prime\rightarrow 0$ the quantity $\theta$ remains finite and does not lead to any UV divergence or contributions to the Weyl anomaly.  Instead, $\theta$ depends on the direction with which $\sigma$ approaches $\sigma^\prime$.  It has been suggested  \cite{Tseytlin:1990va,Tseytlin:1990nb} that since this represents a breakdown of Lorentz invariance  we may think of  $\theta$ as parameterising a
Lorentz anomaly.\footnote{The use of the term anomaly is, perhaps, slightly abusive; the classical Lorentz invariance of the action is not at all obvious from (\ref{Daction}).   However, we understand the lack of manifest Lorentz invariance as being a gauge fixing artefact and in this respect the situation is somewhat similar to working with electromagnetism in say axial gauge.   Also we shall see in the next chapter that, as is the case for chiral bosons in the FJ formalism  \cite{Floreanini:1987as,Sonnenschein:1988ug},  there exist modified Lorentz transformations that leave the action invariant and there is also a notion of `on-shell' Lorentz invariance.}   It seems natural to demand such an anomaly vanishes by setting $\frac{\d  S_{eff}}{\d \theta} = 0$. In principal, this might place additional constraints on the background fields
beyond that of the beta-functionals vanishing. However,  after performing all Wick contractions we find that when the dust settles all occurrences of $\theta$  in the effective action cancel out leaving no Lorentz anomaly.  This seems like an important indicator of the consistency of the formalism, perhaps it is as expected since we have an equal number of Bosons of each
chirality.  In what follows we shall therefore, to avoid needless duplication, we shall only detail the steps required to calculate the contributions to the Weyl anomaly.  

Given (\ref{prop1}) we can deduce the form of more complicated Wick
contractions which are quartic in $\xi$ and contain derivatives of fluctuations. These contractions are relevant to the evaluation of $\langle S^{(2)}_{int} S^{(2)}_{int}  \rangle$ term in the exponential of the effective action and must be included since we need to count all logarithmic divergences.  We again relegate the calculation details to the appendix, and here present the required results: 
\bea
\label{p1}
 \langle \xi^{\ab} \d_1 \xi^{\bar{\beta}} \d_1 \xi^{\cb} \xi^{\db} \rangle &=& \frac{1}{2} \Delta_0\left(\cM \cM -\cL\cL\right)^{(\ab\cb\bar{\beta}\db - \ab\db\bar{\beta}\cb)},\\
\label{p2}
\langle {\xi^\ab \d_1 \xi^{\bar{\beta}} \d_0 \xi^{\cb} \xi^{\dlb}} \rangle &=& - \frac{1}{2}\Delta_0\left(\cM\cL+\cL\cM\right)^{(\leftrightarrow)} -  \t \cL\cL^{(\leftrightarrow)},   \\
\label{p3}
 \langle {\xi^{\ab} \d_0 \xi^{\bar{\beta}} \d_0 \xi^{\cb} \xi^{\dlb}}\rangle &=& - \frac{1}{2}\Delta_0\left(\cM\cM+3\cL\cL\right)^{(\leftrightarrow)} -  \t \left( \cL\cM + \cM\cL\right)^{(\leftrightarrow)},   
\eea
where $\cM \cM^{(\ab\cb\bar{\beta}\db - \ab\db\bar{\beta}\cb)} $ represents $ \cM^{\ab\cb} \cM^{\bar{\beta}\db}  - \cM^{\ab\db} \cM^{\bar{\beta}\cb}$ and  $(\leftrightarrow)$ understood in the same way.\footnote{We will simplify notation by using $\langle \xi^{\Ab} \d_1 \xi^{\bar{B}}\xi^{\Cb} \d_0 \xi^{\Db} \rangle =i\int d^2\s'\langle \xi(\s)^{\Ab} \d_1 \xi(\s)^{\bar{B}} \xi(\s')^{\Cb} \d_0 \xi(\s')^{\Db} \rangle$ since such terms always appear in the expansion of the exponential of the effective action in this form.}

We now evaluate all the contributions to the one-loop effective action.   First we look at the single contraction terms that arise from calculating   $\langle S^{(2)}_{int} \rangle$ and then we consider the double contractions needed to calculate  $\langle S^{(2)}_{int} S^{(2)}_{int}  \rangle$.

\subsection{Single contraction terms for   $\langle S^{(2)}_{int} \rangle$  }

It is helpful to organise our calculations by distinguishing three sorts of contributions:
 `base terms' which contain only the base metric $g$ and 
its derivatives;   `fibre terms' involving only the fibre metric and its derivatives; and `vielbein terms' which come from (\ref{VLag}) and contain derivatives of the vielbein prior to any integrations by parts.

The terms with a single $\left<\xi\xi\right>$ contraction are
\beq
\label{p4}
\frac{1}{2}\d_a\d_b g_{gd}\left<\xi^a\xi^b\right>\d_\mu Y^g\d^\mu Y^d=\frac{1}{2}\d^a\d_a g_{gd}\d_\mu Y^g\d^\mu Y^d\Delta_0 
\eeq
from the base,
\beq
-\frac{1}{2}\d_a\d_b\H_{GD} \d_1\X^G\d_1\X^D\left<\xi^a\xi^b\right>=-\frac{1}{2}\d_a\d^a\H_{GD} \d_1\X^G\d_1\X^D\Delta_0 
\eeq
on the fibre and
\bea
-\cH_{AB}\d_1\Vt{B}{{\bar{B}}}\d_1\Vt{A}{{\Ab}}\langle\xi^{\bar{B}}\xi^{{\Ab}}\rangle&=&\left(-\d_1\Vt{A}{\Ab}\d_1\Vt{\Ab}{A}+\frac{1}{2}\d_1\H_{AB}\d_1\H^{AB}\right)\Delta_0,\\
L_{AB}\d_1\Vt{B}{\bar{B}}\d_0\Vt{A}{\Ab}\langle\xi^{\bar{B}}\xi^{\Ab}\rangle&=&\left(\d_1\Vt{\Ab}{A}\d_0\Vt{A}{\bar{B}}\delta^{\bar{B}\Cb}L_{\Cb\Ab}\right)\Delta_0,\\
-2\d_{\ar}\cH_{BG}\d_1\X^G\d_1\Vt{B}{\bar{B}}\langle\xi^{\ar}\xi^{\bar{B}}\rangle&=&0,\\
g_{ab}\d^\mu\Vt{b}{\br}\d_\mu\Vt{a}{\ar}\langle\xi^{\br}\xi^{\ar}\rangle&=&\left(\d^\mu\Vt{a}{\ar}\d_\mu\Vt{\ar}{a}-\frac{1}{2}\d^\mu g_{ab}\d_\mu g^{ab}\right)\Delta_0,\\
2\d_{\ar}g_{bg}\d^\mu Y^g\d_\mu \Vt{b}{\br}\langle\xi^{\ar}\xi^{\br}\rangle&=&\left(-2\d_ag_{gb}\d_d\Vt{b}{\br}\dl^{\ar\br}\V_{\ar}^{\ a}\right)\Delta_0\, ,
\eea
from the vielbeins.

\subsection{Double contraction terms for $\langle S^{(2)}_{int} S^{(2)}_{int}  \rangle$}

These occur when expanding the exponential of the effective action to
second order.   Although there seem a myriad of possible terms that could contribute,
especially from vielbein terms, many vanish trivially. This is because the block diagonal 
form of the metrics and vielbeins ensure the terms mainly 
separate into base and fibre terms, with a few `cross-terms'. 
We use the propagator contraction (\ref{p1}, \ref{p2}, \ref{p3})  and 
note that these terms are a factor of a half down due to the exponential,
and a further factor of a half down due to the two sitting on the left-hand side of (\ref{Lag}).

On the base we get
\bea
&&-\d_{\ar}g_{g \br}\d^\mu Y^g\d_{\bar{c}}g_{ d \er}\d_\mu Y^{d}\left<\xi^{\ar}\d_1\xi^{\br}\xi^{\bar{c}}\d_1\xi^{\er}\right>\nn\\&&\quad\qquad\qquad\qquad=-\frac{1}{2}\left(\d_ag_{gb}g^{bc}\d^ag_{cd}-\d^ag_{bg}\d^bg_{ad}\right)\d^\mu Y^g\d_\mu Y^d\Delta_0\, ,
\eea
and on the fibre
\bea
&&\d_{\ar}\cH_{G \bar{B}}\d_1\X^G\d_{\bar{c}}\cH_{ D \Eb}\d_1\X^{D}\left<\xi^{\ar}\d_1\xi^{\bar{B}}\xi^{\bar{c}}\d_1\xi^{\Eb}\right>\nn\\&&\quad\qquad\qquad\qquad\qquad=\frac{1}{2}\d_a\H_{GB}\H^{BC}\d^a\H_{CD}\d_1\X^G\d_1\X^D\Delta_0\, .
\eea

Purely from the vielbein piece of the Lagrangian (\ref{VLag}) we have 
\bea
\frac{1}{4}L_{\Ab B}\d_1\Vt{B}{\bar{B}}L_{\Cb
  D}\d_1\Vt{D}{\Db}\left<\xi^{\Ab}\d_0\xi^{\bar{B}}\xi^{\Cb}\d_0\xi^{\Db}\right>&=&\left(-\d_1\Vt{A}{\Ab}\d_1\Vt{\Ab}{A}+\frac{1}{8}\d_1\H_{AB}\d_1\H^{AB}\right)\Delta_0, \nonumber \\
\frac{1}{4}L_{\Ab B}\d_0\Vt{B}{\bar{B}}L_{\Cb
  D}\d_0\Vt{D}{\Db}\left<\xi^{\Ab}\d_1\xi^{\bar{B}}\xi^{\Cb}\d_1\xi^{\Db}\right>&=&\left(-\frac{1}{8}\d_0\H_{AB}\d_0\H^{AB}\right)\Delta_0,\nonumber \\
\frac{1}{2}L_{\Ab B}\d_0\Vt{B}{\bar{B}}L_{\Cb
  D}\d_1\Vt{D}{\Db}\left<\xi^{\Ab}\d_1\xi^{\bar{B}}\xi^{\Cb}\d_0\xi^{\Db}\right>&=&\left(-\d_1\Vt{\Ab}{A}\d_0\Vt{A}{\bar{B}}\hat{\delta}^{\bar{B}}_{\Ab}\right)\Delta_0, \nonumber \\
-\cH_{\Ab B}\d_1\Vt{B}{\bar{B}}L_{\Cb
  D}\d_1\Vt{D}{\Db}\left<\xi^{\Ab}\d_1\xi^{\bar{B}}\xi^{\Cb}\d_0\xi^{\Db}\right>&=&\left(2\d_1\Vt{A}{\Ab}\d_1\Vt{\Ab}{A}-\frac{1}{2}\d_1\H_{AB}\d_1\H^{AB}\right)\Delta_0, \nonumber \\
-\cH_{\Ab B}\d_1\Vt{B}{\bar{B}}L_{\Cb
  D}\d_0\Vt{D}{\Db}\left<\xi^{\Ab}\d_1\xi^{\bar{B}}\xi^{\Cb}\d_1\xi^{\Db}\right>&=&0,\nonumber \\
\cH_{\Ab B}\d_1\Vt{B}{\bar{B}}\cH_{\Cb
  D}\d_1\Vt{D}{\Db}\left<\xi^{\Ab}\d_1\xi^{\bar{B}}\xi^{\Cb}\d_1\xi^{\Db}\right>&=& 0, \nonumber \\
\sum_{\mu=0,1} g_{\ar b}\d^\mu\Vt{b}{\br}g_{\bar{c} d}\d^\mu\Vt{d}{\dr}\left<\xi^{\ar}\d_\mu\xi^{\br}\xi^{\bar{c}}\d_\mu\xi^{\dr}\right>&=& \left(\frac{1}{4}\d^\mu g_{ab}\d_\mu g^{ab}-\d^\mu\Vt{a}{\ar}\d_\mu\Vt{\ar}{a}\right)\Delta_0\, , \nn \\ 
\eea
with one cross-term
\bea
\label{p5}
&&\sum_{\mu=0,1} 2\d_{a}g_{gb}\Vt{b}{\br}\Vt{a}{\ar}\d^\mu Y^g g_{\bar{c} e}\d^\mu\Vt{e}{\dr}\left<\xi^{\ar}\d_\mu\xi^{\br}\xi^{\bar{c}}\d_\mu\xi^{\dr}\right>\nn\\&&\qquad\qquad\qquad=\left(-2\d_ag_{gb}\d_d\Vt{b}{\br}\dl^{\ar\br}\V_{\ar}^{\ a}+\d_ag_{gb}\d_dg^{ba}\right)\d^\mu Y^g\d_\mu Y^d \Delta_0\, .
\eea

\subsection{The Weyl Divergence}

The total Weyl divergence will be given by the coefficient of $\Delta_0$ obtained by summing all contributions from (\ref{p4}) to (\ref{p5}).  We will express this total sum as 
\beq
\label{Sweyl}
S_{Weyl}=\frac{1}{2}\int d^2 \sigma \bigl[ -W_{GD} \d_1 \X^G\d_1 \X^D + W_{gd}\d_\mu Y^g \d^\mu Y^d \bigr]\Delta_0 \, .
\eeq
 On the base the divergence $W_{gd}$ is given by 
\bea\label{Weylb1}
W_{gd}&=&\frac{1}{2}\d^a\d_a g_{gd}-\frac{1}{4}\d_g g_{ab}\d_d g^{ab}-\frac{1}{2}\d_ag_{gb}g^{bc}\d^ag_{cd}\nn\\&&+\frac{1}{2}\d^ag_{bg}\d^bg_{ad}+\d_ag_{gb}\d_dg^{ba}\nn\\
&&-\frac{1}{8}\d_g\H_{AB}\d_d\H^{AB}.
\eea
The divergence on the fibre is 
\beq\label{fibdiv}
W_{GD}=\frac{1}{2}\d^2\H_{GD}-\frac{1}{2}\left((\d_a\H)\H^{-1}(\d^a\H)\right)_{GD}.
\eeq
The divergence on the base, (\ref{Weylb1}), can be rewritten as
\bea\label{Weylb2}
W_{gd}&=&g^{ab}g_{gs}\left(\d_b \G{s}{a}{d}+\G{s}{b}{t}\G{t}{a}{d}\right)\nn\\
&&-\frac{1}{2}\d_g g_{ab}\d_d g^{ab}+\d_ag_{gb}\d_dg^{ba}\nn\\
&&-\frac{1}{8}\d_g\H_{AB}\d_d\H^{AB},
\eea
where we recognise the first two terms as part of the Ricci tensor. We
now, using the base components of the equation of motion for the fields, add zero to the divergence in the form 

\beq
\label{eom0} 
\G{t}{a}{b}g^{ab}\left(\Dh_\mu\left(g_{td}\d^\mu Y^d \right)-\frac{1}{2}\d_t\H_{GD}\d_1 \X^G\d_1\X^D\right)\, . 
\eeq

The base divergence becomes
\bea\label{Weylb3}
W_{gd}&=&-\hat{R}_{gd}-\frac{1}{8}\d_g\H_{AB}\d_d\H^{AB},
\eea
where $\hat{R}_{gd}$ is the Ricci tensor constructed from the base metric $g$ alone. The fibre components of the divergence become
\beq
W_{GD}=\frac{1}{2}\d^2\H_{GD}-\frac{1}{2}\left((\d_a\H)\H^{-1}(\d^a\H)\right)_{GD}-\frac{1}{2}\G{t}{a}{b}g^{ab}\d_t\H_{GD}\,  \label{Weylf}\, .
\eeq

\subsection{Relation to the Doubled Ricci tensor}

If we calculate the Ricci tensor of the doubled space $\cR_{\a \b}$  (the Ricci tensor of $\cM$)\footnote{The Ricci tensor is given as $\cR_{\a\b} = \partial_\gamma \Gamma^\gamma_{\a\b} - \partial_\a \Gamma^\gamma_{\gamma \b} +\Gamma^\tau_{ \a\b}\Gamma^{\gamma}_{\gamma \tau}  - \Gamma^\tau_{\gamma \a} \Gamma^\gamma_{\tau \beta}$\, .  } for comparison, and drop terms proportional to $\H^{AB}\d_d \cH_{AB}=0$,\footnote{This identity only holds due to the $O(d,d)/O(d)\times O(d)$ coset form of $\H_{IJ}$ as can be readily verified by brute-force expansion.}  it also has block diagonal form with
\bea
\cR_{GD}&=&-\frac{1}{2}\d^2\H_{GD}+\frac{1}{2}\left((\d_a\H)\H^{-1}(\d^a\H)\right)_{GD}+\frac{1}{2}\G{t}{a}{b}g^{ab}\d_t\H_{GD}\, ,\label{Ricci}\\
\cR_{gd}&=&\hat{R}_{ab}+\frac{1}{4}\mbox{tr}(\d_g\H\d_d\H^{-1})
\, ,
\eea
for the fibre and base parts respectively. We see that the Weyl divergence is almost equal to minus the Ricci tensor  
except that the term on the base containing the doubled metric $\H$
have an extra factor of $1/2$.   One might try and combine $W_{IJ}$ and $W_{ab}$ into a tensor over the complete doubled bundle:
\be
W_{\a\b} = \left(\begin{array}{cc}
W_{AB} & 0 \\ 0 & W_{ab}
\end{array}\right)\, . 
\ee 
Then we may express
\be
\label{Wdiv}
W_{\a\b} = -  \cR_{\a\b} - \cS_{\a\b}\, ,
\ee
with 
\be
\cS_{\a\b} = \left(\begin{array}{cc}
0 & 0 \\ 0 & -\frac{1}{8}\mbox{tr}(\d_a\H\d_b\H^{-1})\end{array}\right)\, . 
\ee
At this stage it is very unclear how $\cS_{\a\b}$ can be given a geometric interpretation in terms of the curvatures of $\cM_{\a\b}$.  As we shall see shortly, perhaps this is not the best way to think of things. It seems more natural to interpret the above divergences in a geometric way in terms of the metric on the base only.  

We note also that the fibre divergence is contracted with
$\d_1\X\d_1\X$, whereas if we considered an ordinary sigma model with
metric $\cM$ the fibre piece would be contracted with $\d^\mu\X\d_\mu\X$. 
However, we can use the fibre equations of motion to make $W_{GD}$
contract $\d^\mu\X\d_\mu\X$ at the expense of introducing a factor of $1/2$.
Then, comparing $W$ with $R$, all terms containing the doubled metric 
$\H$ would be a factor of $1/2$ down. 
We will see that writing the doubled metric in terms of the 
standard sigma model fields $h$ and $b$ takes care of these extra factors.

\section{Doubled Renormalisation}

One may now proceed directly to regularise and renormalise the divergences coming from $\Delta_0$.  In the standard way one would dimensionally regularise and introduce a mass scale $\mu$  through, say, minimal subtraction and the introduction of counter terms.  We then absorb all scale dependence to define the renormalised couplings
\be
\left\{\cM, \cK, \cL \right\} \rightarrow \left\{\cM^R(\mu), \cK^R(\mu),\cL^R(\mu)\right\}
\ee
producing the renormalised action
\be\label{DactionRen}
S^R= \frac{1}{2} \int d^2\sigma\left[ -\cM^R(\mu)_{\a\b} \partial_1 X^\a \partial_1 X^\b + \cL^R(\mu)_{\a\b} \partial_1 X^\a \partial_0 X^\b + \cK^R(\mu)_{\a\b} \partial_0 X^\a \partial_0 X^\b\right]\,   .
\ee
We can calculate the beta-functions from this by differentiating the renormalised couplings with respect to the log of the mass scale giving
\be
\b[\cM_{\a\b}] = -\left(\begin{array}{cc}
W_{AB} & 0 \\ 0 & W_{ab}
\end{array}\right)\, , \quad  \b[\cK_{\a\b}] = -\left(\begin{array}{cc}
0 & 0 \\ 0 & W_{ab} 
\end{array}\right)\, , \quad  \b[\cL_{\a\b}] = 0 \, .
\ee
 Demanding the vanishing of these beta-functions gives the background field equations. 


\section{Comparison with Standard Sigma Model}
Instead of working directly with these doubled beta-functions we shall make a comparison to the standard sigma model by expanding out the Weyl divergence (\ref{Sweyl}) in terms of the non-doubled metric and B-field.  We will then use the classical equations of motion to eliminate the extra T-dual coordinates from  (\ref{Sweyl}).  The result of this process is then compared against the corresponding Weyl divergence for the standard string.  We will find that the divergences match. 

To aid calculation it is helpful to temporarily simplify matters by assuming a trivial base metric $g_{ab}=\dl_{ab}$. Expanding the Weyl divergence $W_{\a\b}$ using
\be
\cH_{IJ} = \left(\begin{array}{cc}
h-bh^{-1}b &  bh^{-1} \\ -h^{-1}b & h^{-1}
\end{array}\right)\, ,
\ee
we obtain on the fibre
\be
W_{AB} = \frac{1}{2}\left(\begin{array}{cc}\left( r + bh^{-1}r h^{-1}b  - bh^{-1}s  - sh^{-1}b\right)_{ij}  & \left(sh^{-1} -  bh^{-1} r h^{-1}\right)_i{}^j\\ -\left(h^{-1}s-   h^{-1}r h^{-1}b\right)^i{}_j & -\left(h^{-1}rh^{-1}\right)^{ij}\end{array}\right)\, ,
\ee
and on the base
\be
W_{ab} =  \frac{1}{4} t_{ab}\, , 
\ee
where we have defined
\bea
r_{ij} &=& \left(\d^2 h -  \d_a h h^{-1} \d^a h - \d_a b h^{-1} \d^a b\right)_{ij}\, ,\\
s_{ij} &=& \left(\d^2 b -  \d_a b h^{-1} \d^a h - \d_a h h^{-1} \d^a b\right)_{ij}\, ,\\
t_{ab} &=&  \mbox{tr}\left( h^{-1}\d_a h h^{-1}\d_b h - h^{-1}\d_a b h^{-1} \d_b b\right)\, .
\eea
Since the classical fields,  $X^A = \X^A = \left(X^i, \tilde X_i\right)$, must obey as a consequence of their equations of motion
\beq
d \X^A = L^{AB}\cH_{BC}\ast d \X^C ,
\eeq
 we may eliminate the dual coordinates $\tilde{X}_i$ from the Weyl divergence (\ref{Sweyl}) using 
\bea
 \d_1\tilde{X}_i&=& h_{ij}\d_0X^j + b_{ij}\d_1X^j \,.
\eea
We can observe that the right hand side of the above has a sensible interpretation in terms of the standard sigma model; it is proportional to the canonical momentum.  On using the constraint we find that
\beq
W_{AB}\d_1 \X^A\d_1 \X^B = \frac{1}{2}r_{ij}\d_\mu X^i\d^\mu X^j - \frac{1}{2}\e^{\mu\nu}s_{ij}\d_\mu X^i\d_\nu X^j.
\eeq
Thus, we find that prior to renormalisation, the Weyl divergence (\ref{Sweyl}) is given by
\be
\label{Sweyl2}
S_{Weyl} = \frac{1}{4} \int d^2\s \left[ r_{ij}\d_\mu X^i \d^\mu X^j  - s_{ij}\e^{\mu\nu} \d_\mu X^i\d_\nu X^j + \frac{t_{ab}}{2}\d_\mu Y^a \d^\mu Y^b\right]\Delta_0.
\ee
Demanding that this divergence vanishes constrains the background fields to obey $r = s = t =0$.\par
We now wish to compare this result to the standard sigma model in conformal gauge 
\be
S=\frac{1}{2}\int d^2\sigma \left[ G_{IJ}\d_\mu X^I \d^\mu  X^J  - \e^{\mu\nu}B_{IJ}\d_\mu X^I \d_\nu  X^J\right]\, ,
 \ee
with $X^I = (X^i, Y^a)$ and metric and B-field given by 
\be
\label{eqAnsatz}
G_{IJ} = \left(\begin{array}{cc}
h_{ij}(Y) & 0 \\ 0 & \dl_{ab}
\end{array}\right) , \quad  B_{IJ} = \left(\begin{array}{cc}
b_{ij}(Y) & 0 \\ 0 & 0
\end{array}\right).
\ee
The beta-functionals for this sigma model are \cite{Callan:1985ia}\footnote{We have set the non-doubled dilaton to a constant.}
\bea
  \beta[G_{IJ}] &=&  R_{IJ} -\frac{1}{4}H^{MK}_{\phantom{MK}I}H_{MKJ}  \, , \\
 \beta[B_{IJ}]&=&- \frac{1}{2} D^{K} H_{KIJ}\, ,
\eea
where $H_{IJK} = \partial_I B_{JK} + \partial_J B_{KI} +\partial_K B_{IJ} $.  On substitution of our ansatz for $B$ and $G $ we find that the non-vanishing components are
\bea \label{eqSinglebeta1}
\b^{G}_{ij} &=& -\frac{1}{2}\left( r + \frac{1}{2} tr\left(h^{-1}\d_a
h\right)\d^ah\right)_{ij},\label{bGf} \\
\b^{G}_{ab} &=&  -\frac{1}{2}\left( \frac{t_{ab}}{2} + \d_a tr\left(h^
{-1}\d_{b} h \right)  \right) \label{bGb},\\
\b^{B}_{ij} &=& -\frac{1}{2} \left( s+ \frac{1}{2} tr\left(h^{-1}\d_a
h\right) \d^a b\right)_{ij}.\label{bBf}\eea
The Weyl divergent part of the effective action which produces these beta-functions after renormalisation is, 
\be\label{eqSinglebfe}
S_{Weyl} = -\frac{1}{2} \int d^2 \s \bigl[\b^{G}_{ij}\d_\mu X^i \d^\mu X^j + \b^{G}_{ab}  \d_\mu Y^a \d^\mu Y^b - \e^{\mu\nu}\b^{B}_{ij}\d_\mu X^i \d_\nu X^j  \bigr]\Delta_0.
\ee
The equation of motion for the base coordinate $Y$ is 
\bea
2\d^2 Y^a = \d^a h_{ij} \d_\mu X^i \d^\mu X^j  - \d^a b_{ij}\e^{\mu\nu}\d_\mu X^i \d_\nu  X^j , \eea
so upon multiplying both sides $tr\left(h^{-1}\d_a h\right)$ and integrating by parts we have
\bea
&&{}\,\frac{1}{2}\int d^2\s \,\mbox{tr}\left(h^{-1}\d_a h\right)\left(\d^ah_{ij}\dl^{\mu\nu} - \d^ab_{ij}\e^{\mu\nu}\right)\d_\mu X^i\d_\nu X^j  \\
&=&\int d^2\s\,  \mbox{tr}\left(h^{-1}\d_a h\right)\d^2 Y^a = -\int d^2\s \,  \d_a \mbox{tr}\left(h^{-1}\d_{b} h \right)  \d_\mu Y^a\d^\mu Y^b
\eea
so that (\ref{eqSinglebfe}) reduces to
\bea
\label{eqSinglebfe2}
S_{Weyl} =\frac{1}{4} \int \bigl[ r_{ij}\d_\mu X^i \d^\mu X^j -
s_{ij}\e^{\mu\nu} \d_\mu X^i\d_\nu X^j + \frac{t_{ab}}{2}\d_\mu Y^a
\d^\mu Y^b\bigr]\Delta_0 \, .
\eea
This agrees with what we found previously from the Doubled Formalism in (\ref{Sweyl2}). Thus after integrating out the dual coordinate the Doubled Formalism gives {\it exactly} the same divergent terms as the standard string sigma model.   In a general T-fold (at least one of the sort that this Double Formalism can cater for) there may be no global choice of polarisation and one may only be able to integrate out the dual coordinate on a patchwise basis.  In that case we may say that on each patch the background fields need to obey the same equations as the standard string. 

The construction can be extended to include a non-trivial base metric $g(Y)$ though the manipulations are a little more involved. In this case the results for the beta-functionals  (\ref{bGf}),
(\ref{bGb}) and (\ref{bBf}) are respectively supplemented by the following extra terms:
\bea
&&\frac{1}{2}\G{t}{a}{b}g^{ab}\d_t h_{ij}\, ,\\
&&\hat{R}_{ab}+\frac{1}{4}\mbox{tr}(h^{-1}\d^th)\d_tg_{ab}-\frac{1}{2}
\mbox{tr}(h^{-1}\d^th)\d_ag_{tb}\, ,\\
&&\frac{1}{2}\G{t}{a}{b}g^{ab}\d_t b_{ij}\, .
\eea
These terms do indeed follow from the doubled geometry Weyl divergences (\ref{Weylb3}) and (\ref{Weylf}) after application of the equations of motion.  For instance,  one can immediately see $\hat{R}_{ab}$, the Ricci tensor of the base metric $g$ appearing in (\ref{Weylb3}).

\section{A Brief Recapitulation}

Let us, at this point, take stock of the situation. We have been able to calculate the one-loop Weyl divergence of the duality symmetric string which upon renormalisation gives rise 
to the beta-functionals for the doubled geometry metric couplings.  
For the fibre coordinates these are the obvious geometric quantity,
the doubled target space Ricci tensor.  
For base coordinates, the terms in the Ricci tensor that contain the
fibre metric $\H$ pick up an extra factor of a half.

We were then able to show that upon elimination of the background T-dual coordinate by means of the classical  equation of motion, the Weyl divergences for the standard string match those of the doubled string.   In fact, we actually had to do more than just eliminate the T-dual coordinate, we also had to use the equation of motion for the base coordinates to match these divergences.  This may feel somewhat strange since in the construction of  the Doubled Formalism we haven't actually changed the base of the fibration whatsoever. 

There are a few puzzling questions that still remain. Firstly, can we relate the standard string to the doubled string without resorting to the elimination of the classical T-dual fields according to their equations of motion?  Second, what is the underlying geometric picture that the doubled beta functions provide?  Finally, how does the inclusion of an appropriate dilaton field alter these results?  As we shall see, the answers to these questions are interrelated. 

\section{Inclusion of the Dilaton}
The doubled dilaton $\Phi$ is related to the standard dilaton $\phi$ by
\bea
\label{bigDlittled}
 \Phi(Y) = \phi(Y) - \frac{1}{2}\ln \det h
\eea
and is included into the string action (even in the Doubled Formalism) with a standard Fradkin-Tseytlin term\footnote{Note the normalisation of $\frac{1}{8\pi}$ which is for later convenience. A different normalisation would require amending the relation between $\Phi$ and $\phi$. Also, note that in this action $\eta$ refers to a general worldsheet metric.}
\bea
\label{Sdil}
S_{dil} = \frac{1}{8\pi}\int d^2\sigma \sqrt{|\eta|}  \Phi(Y) R^{(2)}\, .
\eea
We emphasise that this term is an order of $\alpha^\prime$ up on the rest of the doubled string action.  We should remark that in the conformal gauge (which we have used to do the background field expansion) the Dilaton decouples from the theory.

\subsection{Using a constant Doubled dilaton}
An important point is that because of the relationship (\ref{bigDlittled}) it is inconsistent in general to simultaneously consider both the doubled dilaton, $\Phi$, and the standard dilaton, $\phi$, to be constant.   In the calculations above we assumed that the only contribution to the Weyl divergence is given by (\ref{Wdiv}) and hence that $\Phi$ is constant.  It therefore makes the most sense to compare the divergence arising in the Doubled Formalism with the non-doubled string with a non constant dilaton 
\be
\phi = \frac{1}{2} \ln \det h + constant \, .
\ee
Therefore in assessing the metric divergences we should really compare the divergences in $S_{weyl}$ to those that give rise to the completed beta functions
 \bea
  \beta[G_{IJ}] & =&\alp\left( R_{IJ}-\frac{1}{4}H^{MK}_{\phantom{MK}I}H_{MKJ} + D_{I}D_{J}\phi\right) \,  , \\
 \beta[B_{IJ}]&=& - \frac{\alp}{2}\left(  D^{K} H_{KIJ} - D^K \phi H_{KIJ} \right) \, .
\eea
When evaluating these one quickly sees that the dilaton contribution exactly cancels the extra terms on the right hand side of   (\ref{bGf}),
(\ref{bGb}) and (\ref{bBf})  to leave 
\be
\b^{G}_{ij} = -\frac{1}{2}r_{ij}\, ,  \quad
\b^{G}_{ab}=  -\frac{1}{4}t_{ab}\, , \quad
\b^{B}_{ij} = -\frac{1}{2}  s_{ij}.\label{bBf2}\ee
Then it is immediately clear that these beta functions correspond to the divergences in (\ref{Sweyl2}) without the need for further manipulations involving the base equation of motion.  

\subsection{Dilaton contributions to existing beta functions}

Let us now consider how a non constant doubled dilaton modifies the doubled beta function for $\cM_{\a\b}$.  We shall use a slightly indirect method to evaluate these contributions by making use of the generalised conformal invariance technique used in \cite{Callan:1986jb}.  Consider starting with a sigma model with no dilaton term.  For the sigma model to be Weyl anomaly free we don't actually need that Weyl divergence, given by 
\be
 -\cW_{\a\b} = \cR_{\a\b} + \cS_{\a\b} \,  ,
\ee
be identically zero.  Since the sigma model should not be affected by on-shell target space field transformation induced by $X^\a \rightarrow X^\a + V^\a$ we only need that the Weyl divergence is zero up to such redefinition.  Therefore we define a full beta function as
\bea
\b_{\a\b} =- \cW_{\a\b} - \cD_{\a} V_{\b} = 0 \, .
\eea
As explained in \cite{Callan:1986jb,Hull:1985rc} we should restrict $ V_{\b}$ to be the derivative of a scalar which we are led to associate with the dilaton. Hence we have, introducing a constant $\lambda$ to account for possible normalisation issues,
\bea
\b_{\a\b} = \cR_{\a\b} + \cS_{\a\b} + \lambda \cD_\a \cD_\b \Phi = 0\, .
\eea
We express this on the base and fibre and obtain:
\bea
\label{basebeta}
\b_{ab}&=&\hat{R}_{ab} + \frac{1}{8}\hat{\nabla}_a\H_{AB}\hat{\nabla}_b\H^{AB} +\lambda   \hat{\nabla}_a \hat{\nabla}_b \Phi \,  , \\
\label{fibrebeta}
\b_{AB}&=&-\frac{1}{2}\hat{\nabla}^2\H_{AB}+\frac{1}{2}\left((\hat{\nabla}_a\H)\H^{-1}(\hat{\nabla}^a\H)\right)_{AB}+\lambda \frac{1}{2} \hat{\nabla}_a\H_{AB}  \hat{\nabla}^a \Phi\, .
\eea
We have chosen to express these with hatted covariant derivatives constructed out of the base metric $g_{ab}$ and the reader should bear in mind that these objects are actually blind to indices on the fibre e.g.  
\be
\hat{\nabla}_a\H_{AB} = \d_a \H_{AB}\,  , \quad  \hat{\nabla}^2\H_{AB}= \d_a\d^a\H_{AB} +\hat{\Gamma}^{a}_{ab}\d^b \H_{AB}\, .
\ee
 With this notation we are tempted to think of fibre indices as not really labelling target space coordinates but labelling moduli fields contained in $\H$.  In other words it seems like these beta-functionals may have a geometric interpretation in a target space that has been dimensionally reduced so that only the base coordinates remain.  We will see later that this is indeed a sensible interpretation.
 
\subsection{Dilaton Beta Function}
Now let us consider the beta function for the dilaton itself.  The leading order part will be the familiar
\bea
\b^\Phi = \frac{26-D}{6} + O(\alpha^\prime)
\eea
where $D$ is the dimension of the non-doubled target space.  The reason for this is seen quite clearly by counting the central charge $c$ of the doubled theory; if we double $d$ coordinates we have $D+d$ bosons however $2d$ of them are chiral and only count for half in the central charge.  We thus expect the central charge to remain equal to the standard target space dimension.  The factor of 26 arises in the standard way from considering the integration over worldsheet metrics in the path integral \cite{Polyakov1981}. 

The next order in $\alpha^\prime$  contribution is more complicated.  There are
two sources of contributions, namely those which are one-loop and arise from the background field expansion of $S_{dil}$ to quadratic order and those which are two-loops and arise from the expansion of the action (\ref{Daction}). The reason that two-loop contributions appear is that the dilaton term is order $\alp$ up on the metric terms in the action.    

Let us first evaluate the one-loop contribution.  Since the Dilaton action is the same as for the standard string we can read off the result that
\bea
\b^\Phi &=& -\frac{1}{2} \alpha^\prime \left( \cD_\alpha \cD^\alpha \Phi - \frac{1}{2}(\cD\Phi)^2 +  \mbox{ 2-loops}\right)\\
&=& -\frac{1}{2} \alpha^\prime \left( \hat{\nabla}_a \hat{\nabla}^a \Phi - \frac{1}{2}(\hat{\nabla}\Phi)^2 +  \mbox{ 2-loops}\right) \, ,
\eea
in which we have used that $\Gamma^I_{Ia} = \mbox{tr} \H^{-1}\d_{a} \H  = 0$ to get to the second line.

To evaluate the two-loop contribution we need to background field expand  (\ref{Daction}) to third and fourth order.  The relevant terms are those that do not directly couple to the classical fields $\d X^\a$ and whose contractions have a power counting of $p^{-2}$.  At third order we find
\bea
S^{(3)} = \frac{1}{2} \int \cL_{\a\b;\g} \xi^\g D_0\xi^\a D_1\xi^\b +\cK_{\a\b;\g} \xi^\g D_0\xi^\a D_0\xi^\b  + \dots\, ,
\eea
and at fourth order
\bea
\nn  S^{(4)} &=&-\frac{1}{6} \int  \cR_{\a\c\dl\b} D_1\xi^\a D_1 \xi^\b \xi^\c \xi^\dl\\ 
\nn  && +\frac{1}{48}\int \left[ 12\cL_{\a\b;\c\dl} + 4\left( \cL_{\a\r}\cR^\r_{\, \c\dl \b} +   \cL_{\b\r}\cR^\r_{\, \c\dl \a} \right)\right] D_0\xi^\a D_1\xi^\b\xi^\c \xi^\dl \\
&& +\frac{1}{48}\int \left[12\cK_{\a\b;\c\dl} + 8\cK_{\a\r}\cR^\r_{\, \c\dl \b}   \right] D_0\xi^\a D_0\xi^\b\xi^\c \xi^\dl + \dots\, ,
\eea
where the ellipsis signify contributions not relevant for the dilaton beta-function.

To calculate the quantum effective action we will need to calculate the Wick contractions of the exponential of the action.  That is, we will have to consider $\langle (S^{(3)})^2\rangle$ and $\langle S^{(4)}\rangle$. One could find the correct propagator contractions by considerations similar to those in the appendix of this thesis.  It would, of course, be a matter of some considerable detail to calculate all the diagrams needed.  However we can quickly see by taking a few test contractions of $\cR_{\a\b\c\dl}$ with $\cM^{\a\b}, \cL^{\a\b}$ and $\cK^{\a\b}$ that the result, which must be a scalar, will be given in terms of two underlying objects; the base Ricci scalar $\hat{R}$ and two-derivative contractions of the fibre coset metric $\cH$.

So with constants $A$ and $B$ undetermined we will have
\be
\label{bphiansatz1} \b^\Phi = -\frac{ \alpha^\prime}{2} \left( \hat{\nabla}_a \hat{\nabla}^a \Phi -  \frac{1}{2}(\hat{\nabla}\Phi)^2 + A \hat{R} + B \left(\hat{\nabla}_a \H^{-1} \hat{\nabla}^a \H \right)\right)\, .
\ee
We shall use an integrability condition as an indirect and efficient way to determine these coefficients. For the standard string the vanishing of the metric beta function and the Bianchi identity imply that the divergence of the metric beta function is equal to the gradient of some scalar \cite{Callan:1986jb}. This scalar is identified as the dilaton beta function.  We shall see that in the doubled case the same is true, that is we have
\bea
\label{intcond}
\cD^\a \b(\cM)_{\a\b} = \Lambda \cD_\b \b(\Phi)
\eea
when $\b(\cM)_{\a\b}=0$ for a constant coefficient $\Lambda$ which we fix in a short moment.   This 
is sufficient to fix all the unknown coefficients $A,B, \lambda, \Lambda$.   Using the Bianchi identity for the Ricci tensor one can show that when $\beta[\cM]=0$, 
\bea
\nn \cD^\a \b(\cM)_{\a\b} &=& \cD^\a \left( \cR_{\a\b} +\cS_{\a\b} + \lambda \cD_\b \cD_\a \Phi\right)\\
&=& \frac{1}{2} \cD_\beta \left(  \cR +  2\lambda  \cD^2 \Phi - \lambda^2(\cD \Phi)^2 \right) - \lambda\cS_{\a\b}\cD^\a \Phi +  \cD^\a \cS_{\a\b}. \quad
\eea
It is also easy to see that on the fibre (when the free index $\b=B$) both sides of (\ref{intcond}) are zero by the assumption that no fields depend on the fibre coordinates and the fact that $\Gamma^d_{cB}=0$. So we will need to evaluate this expression on the base (when the free index $\b=b$).  We find
\bea
\label{intcond2}
\nn \cD^\a \b_{\a\b} \bigg|_{\b =b} &=&  \hat{\nabla}_b \left( \frac{\hat{R} }{2}  + \frac{1}{8} \mbox{tr} (\hat{\nabla}_a \H \hat{\nabla}^a \H^{-1})   +\lambda \hat{\nabla}^2 \Phi - \frac{\lambda^2}{2}(\hat{\nabla}\Phi)^2 \right)\\ &&  + \left( \cD^\a \cS_{\a\b} -\lambda \cS_{\a\b}D^\a \Phi  \right)\bigg|_{\b =b}.
\eea
To proceed we are going to need a suitable Bianchi-like identity for $\cS_{\a\b}$ to allow us to pull out a total derivative. 
To do this it suffices to see that  
\bea
\label{intcond3}
\nn \frac{1}{2}\hat{\nabla}_b \mbox{tr} (\hat{\nabla}_a \H^{-1} \hat{\nabla}^a \H ) &=& \mbox{tr} (\hat{\nabla}_a \H \H^{-1} \hat{\nabla}_b \H  \H^{-1} \hat{\nabla}^a \H  \H^{-1}) \\ &&- \mbox{tr} (\hat{\nabla}^a \H \H^{-1} \hat{\nabla}_a\hat{\nabla}_b \H  \H^{-1})\,.
\eea
Then for the for the second of the $\cS$ terms in (\ref{intcond2}) we have
\bea
\cD^\a \cS_{\a\b} \bigg|_{\b =b} &=& -\frac{1}{8} \hat{\nabla}^a  \mbox{tr} (\hat{\nabla}_a \H \hat{\nabla}_b \H^{-1})\nn \\
 \nn &=&  -\frac{1}{8}   \mbox{tr} (\hat{\nabla}^2 \H \hat{\nabla}_b \H^{-1}) - \frac{1}{4}   \mbox{tr} (\hat{\nabla}_a \H  \H^{-1} \hat{\nabla}_b \H  \H^{-1} \hat{\nabla}_a \H\H^{-1})
 \\ &&+ \frac{1}{8}   \mbox{tr} (\hat{\nabla}^a \H  \H^{-1} \hat{\nabla}_a \hat{\nabla}_b \H  \H^{-1})\,.
\eea
For the first we find
\bea
\label{xx1}
-\lambda \cS_{\a\b}\cD^\a \Phi  \bigg|_{\b=b} = \frac{\lambda}{8}  \mbox{tr} (\hat{\nabla}_b \H^{-1} \hat{\nabla}_a \H )\hat{\nabla}^a \Phi = \frac{\lambda}{8}(\hat{\nabla}_b \H^{IJ})\hat{\nabla}_a \H_{IJ}\hat{\nabla}^a \Phi\, .
\eea
We can now use the vanishing of the fibre components of the doubled metric beta function (\ref{fibrebeta}) to swap the occurrence of the dilaton in (\ref{xx1}) for some more terms involving $\H$.  We find
\be
-\lambda \cS_{\a\b}\cD^\a \Phi  \bigg|_{\b = b}
= \frac{1}{8}  \mbox{tr} (\hat{\nabla}_b \H^{-1} \hat{\nabla}^2 \H) + \frac{1}{8} \mbox{tr} (\hat{\nabla}_a \H  \H^{-1} \hat{\nabla}_b \H  \H^{-1} \hat{\nabla}_a \H \H^{-1}) \, .
\ee
Upon invoking the identity (\ref{intcond3}) we find that
\bea
\left( \cD^\a \cS_{\a\b} -\lambda \cS_{\a\b}\cD^\a \Phi \right) \bigg|_{\b =b} =  -\frac{1}{16} \hat{\nabla}^b \mbox{tr} (\hat{\nabla}_a \H \hat{\nabla}^a \H^{-1}).
\eea
Thus from the integrability condition (\ref{intcond}) we find the dilaton beta function to be
\bea
\label{dilbeta}
 \nn \Lambda \b^\Phi &= & \left( \lambda \hat{\nabla}_a \hat{\nabla}^a \Phi- \frac{\lambda^2 }{2}(\hat{\nabla}\Phi)^2 + \frac{1}{2} \hat{R} + \frac{1}{16} \left(\hat{\nabla}_a \H^{-1} \hat{\nabla}^a \H \right)\right)\\
&=& -\Lambda \frac{ \alpha^\prime}{2} \left( \hat{\nabla}_a \hat{\nabla}^a \Phi -  \frac{1}{2}(\hat{\nabla}\Phi)^2 + A \hat{R} + B \left(\hat{\nabla}_a \H^{-1} \hat{\nabla}^a \H \right)\right)\, .
\eea
This then fixes the coefficients $\lambda=1, \, \Lambda= -2/\alp, \, A= \frac{1}{2},\, B=\frac{1}{16}$.   Thus we find the full set of beta functions for the doubled string:
\be
\label{betafinal}
\boxed{ \begin{array}{r l}
\beta[\cM_{\a\b}] & = \cR_{\a\b} + \cS_{\a\b} + \cD_\a \cD_\b \Phi 
 \\ \beta[\cH_{IJ}] & = -\frac{1}{2} \hat{\nabla}^2 \cH_{IJ} + \frac{1}{2} \hat{\nabla}_a \cH_{IL} \cH^{LM}\hat{\nabla}^a \H_{MJ} +\frac{1}{2} \hat{\nabla}_a \cH_{IJ} \hat{\nabla}^a \Phi 
  \\ \beta[g_{ab}] & = \hat{R}_{ab} + \frac{1}{8}\mbox{tr}  (\hat{\nabla}_b \H^{-1} \hat{\nabla}_a \H ) + \hat{\nabla}_a \hat{\nabla}_b \Phi 
  \\ \beta[\Phi] & = -\frac{\alp}{2} \left( \hat{\nabla}^2 \Phi - \frac{1}{2} (\hat{\nabla}\Phi)^2 + \frac{1}{2} \hat{R} + \frac{1}{16}  \mbox{tr}  (\hat{\nabla}_b \H^{-1} \hat{\nabla}^a\H ) \right)
  \end{array}} 
\ee
\section{ Target Space Interpretation} 

For the standard string, a result of fundamental importance is that the beta functions can be connected to the field equations of a target space theory \cite{Callan:1985ia}.  The identification is schematically given by
\bea
\label{betaid1}\b^G + \frac{2}{\a^\prime}G\b^\phi &\sim& \dl_GS \, ,\\
\label{betaid2}\frac{1}{\a^\prime}\b^\phi &\sim& \dl_\phi S \, ,\\
\label{betaid3} \b^{matter} &\sim& \dl_{matter}  S \, ,
\eea
where $S$ is the action for the target space theory. Given that we have seen that the beta functions appear geometric in a dimensionally reduced target space it is natural to guess that the correct target space theory is the dimensional reduction of the standard bosonic string target space theory. We now show that this is indeed the case.

Starting with the $d=26$  bosonic spacetime action,
\bea
S_{26}= \frac{1}{2\kappa^2}\int d^{26}x \sqrt{-G}e^{-\phi}\bigl( R(G) +  \left(\nabla \phi \right)^2  - \frac{1}{12}H^2\bigr)
\eea
 we reduce on $T^d$ with the relevant ansatz (in this case a simplified diagonal reduction) given by
\bea
G=
\left(
\begin{array}{cc}
h(Y)  &   0 \\
 0 &    g(Y)
\end{array}
\right) &
B=
\left(
\begin{array}{cc}
b(Y)  &   0 \\
 0 &    0
\end{array}
\right) &
 \phi(Y) = \Phi(Y) - \frac{1}{2}\ln \det h\,.\quad 
\eea
With this ansatz we see that the T-dual invariant (i.e. doubled) dilaton $\Phi$ emerges in the reduced action since
\bea
\sqrt{-G}e^{-\phi} = \sqrt{-g}\sqrt{h}e^{-\phi} = \sqrt{-g}e^{-\phi + \frac{1}{2}\ln \det h} = \sqrt{-g}e^{-\Phi}\, .
\eea
The standard result is a low energy effective action which displays manifest $O(d,d)$ symmetry \cite{MyersOdd, Maharana:1992my}. To see this, note that the 
only non-vanishing Christofel symbols and components of field strength are 
\be
\Gamma^i_{ja} = \frac{1}{2} (h^{-1} \partial_a h )^i_{\ j}\, , \quad \Gamma^a_{ij} = -\frac{1}{2}\partial^a h_{ij}\, , \quad \Gamma^a_{bc} = \hat{\Gamma}^a_{bc} \, , \quad H_{aij} = \partial_a b_{ij} \, . 
\ee
Using this one finds 
\bea
R_{ij} &=& -\frac{1}{2}\hat{\nabla}^2 h_{ij} -  \frac{1}{4} \Lambda_a \partial^a h_{ij} + \frac{1}{2}( \partial_a h  h^{-1} \partial^a h )_{ij} \, , \\
R_{ab} &=& \hat{R}_{ab} -\frac{1}{2}  \hat{\nabla}_b \Lambda_a - \frac{1}{4} {\rm Tr}(h^{-1} \partial_a h  h^{-1} \partial_b h h^{-1} ) \, ,\\
R &=& \hat{R} -  \hat{\nabla}_a \Lambda^a + \frac{1}{4} \Lambda_a \Lambda^a - \frac{1}{4} {\rm Tr}(h^{-1} \partial_a h  h^{-1} \partial^a h h^{-1})\, ,
\eea
in which $\Lambda_a = {\rm Tr} (h^{-1} \partial_a h )$.  Recalling that the fibre moduli fields are naturally packaged in the $O(d,d)/O(d)\times O(d)$ coset form $\cH_{AB}$ we have the identity 
\be
{\rm Tr}\left( \partial_a \cH^{-1} \partial^a \cH  \right) = 2 {\rm Tr}\left( \partial_a h^{-1} \partial^a h  + h^{-1} \partial_ab  h^{-1} \partial^a b \right) \, . 
\ee
Armed with the above it is easy to see that the dimensional reduction produces the action
\be
\label{effa}
S_{26-d}= \frac{ vol(T^d)}{2 \kappa^2}  \int d^{26-d}y \sqrt{-g} e^{-\Phi}\bigl\{ \hat{R}(g) + \left(\hat{\nabla} \Phi\right)^2
 + \frac{1}{8}\mbox{tr} \left(L \hat{\nabla}_a \H L \hat{\nabla}^a \H\right) \bigr\}\, .
\ee
The fields arising from the internal components of the metric and B-field are thought of as moduli in that they parametrise the vacuum of the dimensionally reduced theory.  

We can see that the reduced action doesn't `remember' which T-dual compactifaction it arose from.  In this sense it is natural to expect a clear linkage with the Doubled Formalism.  In performing this dimensional reduction we assumed that the fields had no dependence on the internal coordinates.  This assumption doesn't contradict the general aim of understanding T-folds since a non-trivial fibration over the base is still allowed.

Variation of the action with respect to $\Phi, \cH$ and $g_{ab}$ results in the respective equations of motion: 
\bea
 0& =& \hat{\nabla}^2 \Phi - \frac{1}{2} (\hat{\nabla} \Phi)^2 + \frac{1}{2} \hat{R} + \frac{1}{16}  {\rm Tr}  \left( \partial_a \cH^{-1} \partial^a \cH  \right)\, , \label{dilatoneqm}\\ 
 0&=&-\frac{1}{4} \hat{\nabla}^2 \cH_{IJ} + \frac{1}{4}  \left( \partial_a \cH  \cH^{-1} \partial^a \cH  \right)_{IJ} + \frac{1}{4} \hat{\nabla}^a \Phi \hat{\nabla}_a \cH_{IJ} \, , \label{modulieqm} \\ 
 0 &=& \hat{R}_{ab} + \hat{\nabla}_a\hat{\nabla}_b \Phi + \frac{1}{8}  {\rm Tr}  \left( \partial_a \cH^{-1} \partial^a \cH  \right) \nonumber   \\
 && \quad \quad  - g_{ab} \left(  \hat{\nabla}^2 \Phi - \frac{1}{2} (\hat{\nabla} \Phi)^2 + \frac{1}{2} \hat{R} + \frac{1}{16}  {\rm Tr}  \left( \partial_a \cH^{-1} \partial^a \cH  \right)    \right)\, . 
\eea

Immediately we see that fibre components of the doubled metric beta function $\b[\cH_{IJ}]$, given by (\ref{fibrebeta}), are proportional to the field equations for $\H_{IJ}$ (\ref{modulieqm}), the doubled dilaton beta function $\b[\Phi]$ is proportional to the field equation for $\Phi$ (\ref{dilatoneqm}) and the identification rules  (\ref{betaid1}, \ref{betaid2} ,\ref{betaid3}) are satisfied with the base beta function $\b[g_{ab}]$ playing the role of the $\b^G$ in (\ref{betaid1}) .

\section{Discussion and conclusions}
To summarise, using the technique of background field expansion we have been able to calculate the beta functionals for the doubled string.  Using generalised conformal invariance and an integrability condition we were able to establish the complete beta functionals including the dilaton dependance. 

We have shown that the full beta-functions of the T-duality symmetric string have a most natural target space interpretation as the field equations of  an  $O(d,d)$ invariant dimensionally reduced target space theory. This was perhaps to be expected though the detailed relationship between the string in the doubled geometry and the equations of motion of the background are now made manifest. One hopes this may ultimately lead to a better understanding of how T-folds may work. 

In this calculation there were some notable features that we wish to
draw attention to.  First, the topological term which is crucial in establishing equivalence of
partition functions with the standard sigma model, 
played no role in this calculation.  
Second, the non-covariant structure of the action (\ref{Daction})
meant that unlike the calculation for the standard string, the `gauge' or vielbein
terms do contribute to the divergence and increase the
computational difficulty. Third, the B-field is incorporated into the doubled metric and there is no anti-symmetric term in the action. Fourth, the chiral nature of the fibre coordinates suggests that one should be concerned about any potential Lorentz anomaly. This anomaly
actually vanishes, cancelling between the Bosons of opposite chirality.
Finally, we found that the $\cL$ coupling containing the  $O(n,n)$ fibre metric $L$ does not get renormalised. 

 One interesting question to consider is the convergence properties of the perturbation theory of the doubled formalism.  We should really consider the background field method an expansion in a small dimensionless parameter $\sqrt{\alp}/r$ where $r$ is the characteristic radius of a compact space \cite{Green:1987sp}.   However,  in the Doubled Formalism this presents a puzzle since there are always necessarily T-dual pairs of characteristic radii  $r$ and $\alp/r$ one of which is small when the other is large. One might then be concerned that the perturbation theory breaks  down.   Nonetheless, we have seen that this is not so  --  upon integrating out the dual fields the Weyl divergences exactly match those of the standard string. One might just say that the Doubled Formalism perturbation theory is as exactly as good as the standard string. Furthermore, one might anticipate that the background field expansion of the Doubled Formalism  formally matches that of the standard string (compactified on a torus) to any given loop order regardless of how convergent the perturbation theory actually is.   
 
We conclude by making a few remarks about potential generalisations.  In general, the dimensional reduction above includes a $U(1)^{2d}$ gauge field coming from isometries of the $T^d$ and transformations of the B-field on the fibre.  It would be nice to show that these fields arise when one considers a connection in the doubled fibration.  However, in this case it becomes more subtle to find the constrained action akin to (\ref{Daction}).  It seems that the PST procedure would result in off-diagonal elements in both $\cM$ and $\cL$.  This would significantly increase the computational burden in calculating the Weyl divergences.

As with other treatments of the duality symmetric formalism we had to
specify that the fibre metric depended only on the base
coordinates. It would be nice to relax this assumption. The difficulty
with doing so is that  the chirality constraint would have to be
modified as would its implementation with a suitable generalisation
the PST action. Another interesting and perhaps more democratic 
generalisation along this line would be the doubling of all coordinates.  We will address this in the coming chapters when we consider Poisson-Lie T-duality which does involve internal coordinate dependance.

\chapter{Poisson-Lie T-duality Beyond the Classical Level}
\renewcommand{\G}{\Gamma}
\def\ba{\bea}
\def\ea{\eea}
\def\no{ }
\def\Om{\Omega}
\def\ha{\frac{1}{2}}
\def\inv{^{-1}}
\def\om{\omega}
\def\qq{\quad}
\def\D{\Delta}

\begin{quote}
In this chapter we examine a generalisation of T-duality known as Poisson--Lie T-duality.  We examine the renormalisation of the Poisson--Lie T-duality invariant sigma model using similar techniques to those in the previous chapter.  We then comment on the Lorentz Invariance properties of these duality invariant action.   Some of the results in this section have been published in  \cite{Sfetsos:2009vt} and some are as yet unpublished. 
\end{quote}

\section{Overview}

An interesting generalisation of the 
abelian T-duality \cite{Kikkawa:1984cp,Sakai:1985cs,Buscher:1987sk,Buscher:1987qj}
and non-abelian \cite{Fridling:1983ha,Fradkin:1984ai,delaossa:1992vc}  T-dualities is the Poisson-Lie T-duality proposed by Klimcik and Severa \cite{Klimcik:1995ux,Klimcik:1995jn,Klimcik:1995dy}.   Its most notable feature is that it does not
rely on the existence of isometries but rather on a rigid group-theoretical structure
known as the Drinfeld Double \cite{Drinfeld:1986in}.   Nevertheless, it shares some common features with ordinary T-duality.   For instance, it can be explicitly formulated as a canonical transformation between phase-space variables \cite{Sfetsos:1997pi}, similarly to ordinary T-duality
\cite{Alvarez:1994wj,Curtright:1994be}. 

 A further similarity to the abelian T-duality is that there exists a duality invariant `mother' theory
from which pairs of Poisson--Lie T-dual $\sigma$-models can be obtained.   This theory consists of a Wess-Zumino-Witten model on the Drinfeld Double supplemented with
a non-linear interaction term \cite{Klimcik:1995dy}.  Much like the duality invariant formalism of abelian T-duality  of the previous chapters this duality invariant theory involves a doubling of fields and a lack of manifest world sheet Lorentz invariance.   

 By choosing a parametrisation for a group element
of the Double and implementing the constraints that arise as equations of motion,
one can eliminate half of the fields and obtain a standard Lorentz invariant $\sigma$-model.
The T-dual partner to this is obtained by choosing a different parametrisation
for the group element and repeating the constraint procedure.

It is natural to ask whether such an equivalence holds beyond the classical level. Related to this is the question of whether  the procedure of quantisation before eliminating extra fields is equivalent to first eliminating fields and then quantising?

A first step in understanding whether Poisson--Lie T-duality holds beyond the classical
level was made in \cite{Sfetsos:1998kr} where it was shown,
for specific examples, that the system of
renormalisation group (RG) equations for couplings occurring in each of the T-dual theories
are equivalent at one-loop. The one-loop renormalisation of general
Poisson--Lie T-dual models was considered in \cite{Valent:2009nv} and a proof of
the equivalence of their RG equations was given in \cite{Sfetsos:2009dj}.

It is interesting to ask whether these RG equations can be recast in
a manifestly duality invariant form.  For this to be so we should hope
that the duality invariant theory also produces the same system of RG equations.
This need not be the case since the process of constraining used to arrive at the
pair of T-dual models may not commute with the process of quantisation.
This motivates our study of the one-loop renormalisation  of the PL T-duality
invariant theory.  

The organisation of this chapter is as follows; we begin by introducing the needed mathematical concept of the Drinfeld Double.  We then introduce the basics of Poisson-Lie T-duality and the duality invariant formalism.  Following this we shall consider the renormalisation of the PL duality invariant action by applying and extending the techniques established in the preceding chapter.   In this case the analysis is complicated by the non-trivial internal coordinate dependence of the group geometry.  As the duality invariant theory is not manifestly Lorentz invariant one might expect both a Lorentz and a Weyl anomaly to occur.  We shall see that the Lorentz anomaly vanishes through non-trivial calculations and that the counter terms for the Weyl divergence can be absorbed into a redefinition of the couplings.  Hence the model is one-loop renormalisable.    We will compare this renormalisation to that of the T-dual pairs of $\sigma$-models.   We finish the chapter with an investigation into the emergent on-shell Lorentz symmetry of sigma models that are similar to the PL T-duality invariant theory under consideration.

\section{Drinfeld Doubles}

A Drinfeld Double is a Lie algebra ${\cal D}$  which can be decomposed as the sum of two maximally isotropic subalgebras  ${\cal D} = {\cal G}\oplus \tilde{ \cal G}$ with respect to an inner product $ \langle \cdot \mid \cdot \rangle  $.
In terms of generators $T_a$ of $\cG$ and $\tilde{T}^a$ of $\tilde{\cG}$ the isotropy condition is expressed as
\bea
 \langle T_a \mid T_b \rangle  = \langle \tilde{T}^a \mid \tilde{T}^b \rangle=0\,, & \langle T_a \mid \tilde{T}^b \rangle = \delta_a^b\, .
\eea
We will often write $T_A = (T_a, \tilde{T}^a)$ for the complete set of generators of $\cD$ so that
\be
\langle T_A | T_B \rangle = \eta_{AB} =\left(\begin{array}{cc}
  \bf{0} & \mathbb{I}_{d_g\times d_{g}} \\
  \mathbb{I}_{d_g\times d_g} & \bf{0} \\
\end{array}\right)\, .
\label{fh23}  
\ee
  The structure constants of the double defined by  $[T_A, T_B]=i f_{AB}^{\ \ C} T_C$  must obey
\bea
[T_a, T_b] &=&i f^{\ \ c}_{ab} T_c\, , \\
\phantom{,}[ \tilde{T}^a , \tilde{T}^b]&=& i \tilde{f}^{ab}_{\ \ c} \tilde{T}^c\, ,\\
\phantom{,} [T_a , \tilde{T}^b] &=& i  \tilde{f}^{bc}_{\ \ a} T_c  - i f^{\ \ b} _{ac} \tilde{T}^c\, .
\eea
The Jacobi identity provides a compatibility condition on the two sets of structure constants:
\bea
\label{restr}
 \tilde{f}^{ac}_{\ \ k} f^{\  \ l}_{fa} -\tilde{f}^{al}_{\ \ k} f^{\ \ c}_{fa} - \tilde{f}^{ac}_{ \ \ f} f^{  \ \ l}_{ka} + \tilde{f}^{al}_{\ \ f} f^{\ \ c}_{ka} - \tilde{f}^{lc}_{\ \ a} f^{\ \ a}_{fk}=0\, .
\eea

For $g \in G$ the following definitions are extensively used:
\bea
\label{DEFabpi}
g^{-1} T_a g = a(g)_a^{\phantom{a} b} T_b\, , &  g^{-1} \tilde{T}^a g = b(g)^{ab} T_b + (a(g)^{-1})_b^{\phantom{b} a} \tilde{T}^b\, , &  \Pi^{ab} = b^{ca}a_c^{\phantom{c}b}\, .
\eea
Consistency demands that 
\bea
a(g^{-1})= a(g)^{-1}\,, & b(g)^T = b(g^{-1}) \,,& \Pi(g)^T = -\Pi(g)\,.  
\eea
We have similar definitions for the action of $\tilde{g} \in \tilde{G}$.

\subsection{Examples of Drinfeld Doubles}

The simplest example is where  $\cD= u(1)^{2d}$ and $\cG=\tilde{\cG} = u(1)^d$.  This is the abelian double.  In this case the structure constants all vanish and we have
\bea
a_a^{\phantom{a}b} = \tilde{a}^b_{\phantom{b}a} = \delta_\a^b\, & b=\tilde{b} =\Pi = \tilde{\Pi}=0\, .
  \eea
Poisson Lie T-duality for the case of this abelian double reduces to the familiar abelian T-duality and the duality  invariant formalism reduces to exactly that of the previous chapter.   

The next most simple example is the semi-abelian double where $\tilde{G}$ is abelian. In this case we parametrize  $\tilde{g} = e^{i \chi_a \tilde{T}^a} \in \tilde{G}$ and find
\bea
b^{ab}= \Pi^{ab} = 0\,, & \tilde{a}^b_{\phantom{b}a} = \delta^b_{\phantom{b}a}\,, & \tilde{b}_{ab} = f_{ab}^c \chi_c = - \tilde{\Pi}_{ab}\,. 
\eea
In fact, this semi-abelian double is relevant to `non-abelian' T-duality.  

We can also have doubles where both $\cG$ and $\tilde{\cG}$ are non-abelian.   For each decomposition of $\cD = \cG \oplus \tilde{\cG}$ there is a corresponding pair of dual sigma models. The fact that there may exist several such decompositions gives rise to the notion of Poisson-Lie plurality (this term was coined in \cite{VonUnge:2002ss}  though the notion dates back to the original work  \cite{Klimcik:1995ux}). The ordered set $(\cD, \cG, \tilde{\cG})$ is sometimes referred to as a Manin triple.

The simplest non-abelian double is four dimensional and is $\cD = gl(2,\mathbb{R})$  expressed as the sum of two Borelian subalgebras \cite{Klimcik:1995jn}.  In six dimensions there are a substantial number of doubles which have been completely classified in   \cite{Jafarizadeh:1999xv, Snobl:2002kq, Hlavaty:2002kp}.   An interesting example in six dimensions, which we shall use later, is the double  $so(3,1) = su(2) \oplus e_3$ where $e_3$ is a real three dimensional algebra whose non vanishing commutators are $[T_3 ,T_1] = T_1$ and $[T_3, T_2] = T_2$.  

\subsection{Subspaces of the Double}

A key step of the Poisson-Lie Duality construction is to choose two orthogonal subspaces of  ${\cal D}$ which we call ${\cal R}^\pm$.   We can do this by simply defining a constant matrix $E^+_0$ with no given symmetry. Then we write
\bea
{\cal R}^\pm = {\rm span} (T_a \pm E_{0\, ab}^{\pm} \tilde{T}^b)\,,
\eea 
where $E^+_{ab} = E^{-}_{ba}$.   In what follows, we shall use the notation that $\tilde{M}_s$ is the symmetric part of $E_{0\, ab}$ and $B$ the antisymmetric part. We shall also denote the matrix inverse $E_{0 }^{-1} = M$. 

If we write a basis vector
\bea
|R^{\pm}_a\rangle = |T_a \pm E_{0\, ab}^{\pm} \tilde{T}^b \rangle\,, 
\eea
we have
\bea
\langle R^{\pm}_a | R^{\mp}_b \rangle &=&0\\  
\langle R^{\pm}_a|R^{\pm}_b\rangle &=& \langle T_a |\pm E_{0\, bc}^{\pm} \tilde{T}^c\rangle + \langle \pm E_{0\, ac}^{\pm} \tilde{T}^c  | T_b \rangle  = \pm 2\tilde{M}_s{}_{ab}\,.
\eea
Then we may define an operator $R$ that takes value of $\pm 1$ on the subspaces ${\cal R}^\pm$ as:
\bea
R = \frac{1}{2} \left( |R^+_a\rangle (\tilde{M}_s^{-1})^{ab} \langle R^+_b | +  |R^-_a\rangle (\tilde{M}_s^{-1})^{ab} \langle R^-_b | \right)\,.  
\eea
It will be helpful to define a matrix given by the value of this operator on the generators of the double 
\bea
\label{rmatrix}
R_{AB} = \langle T_A | R | T_B \rangle  = \left( \begin{array}{cc}
                  ( \tilde{M}_s- BG^{-1} B)_{ab} & (B\tilde{M}_s^{-1})_{a}^{\ b}   \\
                  -(\tilde{M}_s^{-1}B)^{ a}_{\ b}  & (\tilde{M}_s^{-1})^{ab} 
                \end{array}\right) \, , 
\eea
which is, of course, a very familiar $O(d,d)/O(d)\times O(d)$ coset matrix.  

\section{Poisson Lie T-duality}

We now come to describing the actual sigma models that are related by PL T-duality.   Let $X^i$ and $\tilde{X}_i$ be local coordinates on $G$ and $\tilde G$ respectively.  For $g \in G$ a left invariant 1-form is given by
\be
L = L_i^a dX^i T_a = -i g^{-1} d g \, , 
\ee
so that 
\be
L_i^a = - i \langle  g^{-1} d g | \tilde{T}^a \rangle   \, , 
\ee
in which we have used the inner product on the  double. Right invariant forms are defined analogously $R= -i dg g^{-1}$. 

The two Poisson Lie T-dual models are (no spectator fields)  then given by 
\bea
\label{DEFgsimgamodel}
S &=& \int d^2 \sigma ((E_0^+)^{-1} + \Pi)_{ab}^{-1} L^a_i L^b_j \p_+ X^i \p_- X^j = \int  d^2 \sigma  Q^+_{ij}\p_+ X^i \p_- X^j \, , \\
\label{DEFtildedgsigmamodel}\tilde{S} &=& \int d^2 \sigma [(E_0^+ + \tilde{\Pi})^{-1}]^{ab} \tilde{L}_a^i \tilde{L}^j_b \p_+ \tilde{X}_i \p_- \tilde{X}_j = \int  d^2 \sigma  \tilde{Q}^{+ij}\p_+ \tilde{X}_i \p_- \tilde{X}_j \, .
\eea

The above sigma models may be supplemented by the inclusion of some extra `spectator' fields which we shall, at least for the time being, ignore.  This slightly simplified scenario, know as atomic duality, is sufficient to demonstrate all the salient features of Poisson--Lie T-duality.

In general, is not obvious that   $S$  is isometric under the (left action) action of $G$ since there is complicated coordinate 
dependence in  $Q^+_{ij}$.   However, the construction of $E$ is such that this coordinate dependence is highly constrained.   

The left action of the group given infinitesimally by $\delta g = \e g$  for $\e = \e^a T_a$ corresponds to the coordinate shift $\delta x^i = \epsilon^a R_a^i$ where the vector fields  $R_a$ are the duals to the right invariant forms.  The current associated to this transformation has components
\be
J_{+a} = Q_{ij} \partial_+ X^i R_a^j \, , \quad J_{-a} = Q_{ij} \partial_- X^j R_a^i \, . 
\ee
The result of performing the variation of the action is 
\be
\delta S = \int d^2\sigma  \partial_+ \e^a J_{a-} + \partial_{-} \e^a J_{a+} + \left( \e^a \cL_{R_a} Q_{ij} \right) \partial_+ X^i \partial_- X^j\,.
\ee
The special form of the matrices $Q_{ij}$ ensures that \cite{Klimcik:1995kw}
\be
\label{zxdc}
\cL_{R_a} Q_{ij} = Q_{ik}R^k_b \tilde{f}^{bc}{}_a R^l_c Q_{lj}  \, , 
\ee
and hence that 
\be
\delta S = \int d^2\sigma  \partial_+ \e^a J_{a-} + \partial_{-} \e^a J_{a+} + \e^a J_{+ b} J_{-c} \tilde{f}^{bc}{}_a \, .
\ee
Then demanding that $\delta S = 0$ means that the $\sigma$-model current $J_a$ associated to the action of $G$ is not closed. Instead it is covariantly closed  with respect to the dual gauge group $\tilde{G}$, in other words it satisfies non-commutative conservation property :
\be
d \ast J_a  + \frac{1}{2} \tilde{f}_{\ \ a}^{bc} \ast  J_b \wedge \ast J_c  = 0 \,.
\ee
For this to be a consistent construction it is necessary that $G$ and $\tilde G$ can form a Drinfeld Double \cite{Klimcik:1995ux}.   The reason for this is that one requires
\be
[\cL_{R_a} , \cL_{R_b} ] Q_{ij} = f_{ab}{}^c \cL_{R_c} Q_{ij}
\ee
and upon plugging in \eqn{zxdc} one finds a restriction on the structure constants which is exactly  that found in the Drinfeld double \eqn{restr}.

These sigma models given by \eqn{DEFgsimgamodel}  and \eqn{DEFtildedgsigmamodel} are dual, at least classically, since they can both be obtained from a single duality invariant starting point.  The duality invariant sigma model (no spectators again) is given by  the Lagrangian \cite{Klimcik:1995dy}:
\be
\label{DEFdualityinvariantwzw1}
S =  I_0[l] - \frac{1}{2}  \int_{\Sigma} d^2 \sigma \langle l^{-1} \partial_\sigma l | R|   l^{-1} \partial_\sigma
\rangle  \, ,
 \ee
 where $I_0[l]$ is the Wess-Zumino-Witten model on the Drinfeld double \cite{Witten:1983tw} but rotated so that $\sigma$ and $\tau$ play the role of light cone coordinates: 
 \be
  I_0[l]= \frac{1}{2} \int_{\Sigma} d^2 \sigma \langle l^{-1} \p_\sigma l   |  l^{-1} \p_\tau l  \rangle  + \frac{1}{12} \int_{\cB}  \langle  l^{-1} d l |[  l^{-1}dl, l^{-1}dl ] \rangle \, ,
  \ee
where $\cB$ is some three-dimensional manifold whose boundary is $\Sigma$.   Provided that the normalisation of the action is appropriately quantised the path integral is indifferent as to how the extension $\cB$ has been chosen \cite{Witten:1983tw,Witten:1983ar}.   The final term represents some deformation which certainly breaks manifest Lorentz invariance.\footnote{For the special case that $R_{AB} = \delta_{AB}$ this whole action is in fact the Chiral WZW model of Sonnenschein \cite{Sonnenschein:1988ug}.}

One can show  \cite{Klimcik:1995jn,Klimcik:1995dy} that the equations of motion that arise from the duality invariant theory implies those of the two T-dual models \eqn{DEFgsimgamodel}  and \eqn{DEFtildedgsigmamodel}.   By making use of the Polyakov-Wiegman identity \cite{Polyakov:1983tt} one can immediately see that \eqn{DEFdualityinvariantwzw1} is invariant under  $l(\tau, \sigma) \rightarrow \epsilon(\tau) l(\tau, \sigma)$. The equations of motion that follow from \eqn{DEFdualityinvariantwzw1} can be integrated and this invariance used to yield 
\bea
\label{EQMdouble}
\langle l^{-1} \p_\pm l  | {\cal R}^{\mp_a} \rangle = 0  \, .
\eea
Note that this relies on the splitting on ${\cal D}$ into orthogonal subspaces and therefore implicitly depends on the choice of $E_0$.   In the vicinity of the identity in  ${\cal D}$ we may  write $l = \tilde{h}g$ with $g\in G$ and $\tilde{h} \in \tilde{G}$.  Substitution of this into (\ref{EQMdouble}) shows that\footnote{Our conventions are aligned with \cite{Sfetsos:1996xj} and differ slightly from  \cite{Klimcik:1995dy} since we are using left invariant forms to express the duality invariant theory.}:
\be
A_{\pm \, a} \equiv  (  \tilde{h} \partial_{\pm} h)_a  = \pm i a_{a}^{\, b} E^{\mp}_{bc} L_{\pm}^c\, ,
\ee
where 
\be
E^{\pm}_{bc} = \left( (E^\pm_{0})^{-1} \pm \Pi \right)^{-1}_{bc} \, . 
\ee
These gauge fields are pure gauge and have a vanishing field strength.  When this is expressed as a  condition on $g$ one recovers the equation of motion coming from the sigma model (\ref{DEFgsimgamodel}).  Alternatively, if one applies the same decomposition and makes use of the PW formula one finds that \eqn{DEFdualityinvariantwzw1} depends on  $A_\sigma$ and not $A_\tau$ .   Eliminating $A_\sigma$ by means of the above equation of motion yields the action (\ref{DEFgsimgamodel}) \cite{Klimcik:1995dy,Sfetsos:1996xj}.  The same procedure but with decomposition $l =g \tilde{h}$ yields the field equations for the dual sigma model (\ref{DEFtildedgsigmamodel}).  This essentially demonstrates classical equivalence.   

A slightly different philosophy to the duality invariant model is provided in \cite{Tyurin:1995bu}, wherein the duality invariant action has manifest Lorentz invariance but must be supplemented with a constraint.   This is reminiscent of the difference between the Abelian T-duality symmetric formulations   of Hull \cite{Hull:2004in} and Tseytlin \cite{Tseytlin:1990va}. 
\subsection{Geometry of the Double}
At this stage it is helpful to recap some geometrical concepts that allow us to describe the Drinfeld Double.   We introduce local coordinates $X^I$, $I = 1\dots 2d_g$ on the Drinfeld Double group manifold so that a group element is given by $l(X) \in D$.   Then  left invariant one-forms can be defined as
\be
L = L^A T_A = -i l^{-1} d l = L^A_I T_A dX^I\, , 
\ee
and which obey the Maurer-Cartan equation
\be
dL^A = \frac{1}{2} f_{BC}^{\phantom{BC} A} L^B \wedge L^C\, .
\ee
The standard (bi-invariant) metric on the group manifold is given by
\be
\eta_{IJ}  =  L_I^AL_J^A \eta_{AB}\, .
\ee
The components of the left invariant forms can be viewed as vielbeins converting flat indices (signified by early Roman letters) to curved space indices (late Roman letters).  In terms of covariant derivatives built from $\eta_{IJ}$ we have 
\be
D_I L^A_J = \frac{1}{2} f_{BC}^{ \phantom{BC} A} L^B_I L^C_J\, .
\ee
A natural object to define is the torsion given by 
\be
\label{torsion}
S_{IJK} = - f_{ABC} L_I^AL_J^B L_K^C
\ee
in which $\eta_{AB}$ has been used to raise and lower indices.  We may  define a generalised connection
\be
\label{gcon}
\hat{\Gamma}^I_{JK} = \Gamma^I_{JK}  - \frac{1}{2} S^I_{JK}\, , 
\ee
with respect to which the  vielbeins are covariantly constant 
\be
\hat{D}_I L^A_J = D_I L^A_K + \frac{1}{2} S^M_{IJ} L_M^A = 0 \, ,
\ee
and thus, according to the first vielbein postulate, the spin connection and torsion are related according to 
\be
\frac{1}{2} S_{IJK} = - \omega_{I \, BC} L^B_J L^C_K\, ,
\ee
or, equivalently,
\be
\label{torsionrel}
\frac{1}{2} f_{ABC} L^A_I = \omega_{I \, BC}\, . 
\ee
A further useful identity is to note that the Riemann curvature is given by
\be
R_{IJKL} = \frac{1}{4} L^A_I L^B_J L^C_K L^D_L f_{ABE} f_{CD}^{\phantom{CD} E}\, , 
\ee
and 
\be
\hat{R}_{IJKL}= 0
\ee
where $\hat{R}$ is the curvature that arises from using derivatives that are covariant for the generalised connection (\ref{gcon}).\footnote{Expression for the generalised curvature can be found in the appendix.}  It is interesting to note that the generalised curvature also vanishes when the torsion (\ref{torsion}) is defined with an opposite sign.  However we found that the relative sign of the relation in (\ref{torsionrel}) is crucial to our calculations;  with the wrong sign one finds that at quantum level there will be a Lorentz anomaly and that even at the classical level there is a loss of emergent on-shell Lorentz invariance. 

With these geometrical concepts we may write the duality invariant action in the form 
\be
\label{PLTaction}
S= \frac{1}{2} \int d^2 \sigma (\eta_{IJ} + B_{IJ})\partial_0 X^I \partial_1 X^J - R_{IJ} \partial_1 X^I \partial_1 X^J \, ,
\ee
where $B_{IJ}$ is an antisymmetric potential\footnote{The reader should keep in mind that we use $B_{IJ}$ to represent this potential and $B_{ab}$ for the antisymmetric part of $E_{0}$.  Also in what follows $R_{IJ}$ refers to the matrix defined in \eqn{Rmat} and not the Ricci tensor which makes no appearance. } whose field strength is given by the torsion:
\be
B_{IJ,K} + B_{KI,J} + B_{JK, I} = S_{IJK}\, ,
\ee 
and where 
\be
\label{Rmat}
R_{IJ} = R_{IJ}(X) = L_I^A L_J^B R_{AB} \, . 
\ee
In this form, it is quite clear that the Poisson--Lie duality invariant action shares many of the features of the Doubled formalism,  with chirality constraint implemented as in (\ref{Daction}).   We still have a metric $\eta_{IJ}$ of signature $(d_g, d_g)$ and a matrix that depends on $d_g^2$ free constant parameters in $R_{IJ}$ that obeys the compatibility
\be
\eta^{JK} R_{KL} = (R^{-1})^{JK} \eta_{KL}\, .
\ee
Additionally, we have a similar structure where Lorentz invariance is not manifest, a point to which we shall return shortly.    

However, there are two key differences with the action considered in the previous chapter. First is that the matrices $\eta_{IJ}$ and $R_{IJ}$,  which should be compared to $L_{IJ}$ and $\cH_{IJ}$ in (\ref{doubleact}) and (\ref{constrainthew}), are no longer constant since they include vielbein dressing factors which contain implicit coordinate dependance.  The second alteration is the inclusion of the torsion term in (\ref{PLTaction}) .  

\section{Renormalisation of the Poisson-Lie Duality Invariant Theory} 

We now wish to calculate the quantum effective action defined as 
\be
e^{iS_{eff}(X)} = \int [D\xi] e^{iS[X+\xi] -iS[X]} 
\ee
which to one-loop is given by expanding the action (\ref{PLTaction})  to quadratic order in fluctuations and considering the interaction terms $S_{int}$.  As with the previous chapter we shall adopt the geometric background field expansion method \cite{Mukhi:1985vy} to perform this calculation.  The first point to make is we again find an ambiguity as to which is the best metric to use in the geometric expansion; the two choices are the group invariant metric $\eta_{IJ}$ and the symmetric $R_{IJ}$.   Because we wish to take advantage of the simplifications that the group geometry allows we shall choose to perform a geometric expansion using $\eta_{IJ}$ as the metric.   

We now consider the expansion of (\ref{PLTaction}) which we shall separate into two parts;   the  Wess-Zumino-Witten terms given by
\be
S_{WZW} = \frac{1}{2} \int d^2 \sigma (\eta_{IJ} + B_{IJ})\partial_0 X^I \partial_1 X^J\, ,
\ee
and the remaining part which we call the non-Lorentz (NL) term
\be
S_{NL} = - \frac{1}{2} \int d^2 \sigma  R_{IJ} \partial_1 X^I \partial_1 X^J \, .
\ee
First we consider the expansion of the WZW term for which we find the following second order piece
\bea
S_{WZW}^{(2)} &=& \frac{1}{2} \int d^2 \sigma\, \eta_{IJ} D_0\xi^I D_1 \xi^J + \left( R_{IKLJ} + \frac{1}{2} D_LS_{IJK} \right) \xi^K \xi^L \partial_0 X^I \partial_1 X^J \nonumber\\
&& \quad \quad \quad +\frac{1}{2} S_{IJK} \xi^K \left(D_0 \xi^I \partial_1 X^J + \partial_0 X^I D_1 \xi^J  \right) \, . 
\eea
We define generalised derivatives
\bea
\hat{D}_0 \xi^I  &=& D_0 \xi^I - \frac{1}{2} S^I_{\ LK} \partial_0 X^L  \xi^K \, ,  \label{xxx1}\\ 
\hat{D}_1 \xi^I  &=& D_1 \xi^I +  \frac{1}{2} S^I_{\ LK} \partial_1 X^L  \xi^K \,   \label{xxx2},
\eea
and note that the sign swap in the torsion terms of these expressions indicate that these are not simply the covariant derivatives with generalised connection pulled back to the world sheet in a naive way.   In terms of these derivatives we have 
\be
S_{WZW}^{(2)} = \frac{1}{2} \int d^2 \sigma\, \eta_{IJ} \hat{D}_0\xi^I \hat{D}_1 \xi^J + \left( R_{IKLJ} + \frac{1}{2} D_LS_{IJK} - \frac{1}{4} S_{MIK} S^M_{\ JL} \right) \xi^K \xi^L \partial_0 X^I \partial_1 X^J\, .
\ee
The combination of terms in parentheses in this above expression can be recognised as the parts of the generalised Ricci tensor $\hat{R}_{IKLJ}$ that are symmetric under the interchange of the indices $K,L$ which are contracted with the fluctuations.   Because of the group geometry this term  vanishes leaving
\be
S_{WZW}^{(2)} = \frac{1}{2} \int d^2 \sigma\, \eta_{IJ} \hat{D}_0\xi^I \hat{D}_1 \xi^J \, .
\ee
As in the previous chapter we shall require a conventional kinetic term for the propagators and thus we wish to introduce vielbeins to absorb all the coordinate dependance from $\eta_{IJ}$.  For a standard sigma-model this is done, as in the appendix, by pulling the vielbeins through the derivatives, the result of which is to introduce `gauge fields' which minimally couple to the fluctuations $\xi$.   Normally the contributions from these gauge fields produce divergences that cancel and one arrives at the standard result of non-renormalisation of the WZW term.  However, in our case the kinetic term for the fluctuations is non-standard and there will be contributions from the gauge fields. 

To proceed we expand out the derivatives and note that 
\bea
L^A_I \hat{D}_0 \xi^I &=& L_A^I D_0 \xi^I - \frac{1}{2} S^I_{LK} \partial_0 X^L \xi^K L^A_I \nonumber \\
&=& \partial_0 \xi^A + \omega_0{}^A{}_B \xi^B  - \frac{1}{2} S^I_{LK} \partial_0 X^L \xi^K L^A_I\nonumber \\
&=& \partial_0 \xi^A  + \frac{1}{2} f_E{}^A{}_B L_0^E\xi^B -\frac{1}{2}\eta^{IJ}\left( -f_{EFG} L^E_J L^F_L L^G_K \right) \xi^K \partial_0X^L L^A_I \nonumber  \\
&=& \partial_0 \xi^A + \frac{1}{2}f_E{}^A{}_B L_0^E \xi^B + \frac{1}{2} f^{A}{}_{EB}\xi^B L_0^E \nonumber \\
&=&  L^I_A  \partial_0  \xi^A 
\eea
and, due to the sign difference between   (\ref{xxx1}) and (\ref{xxx2}), that
\be
\hat{D}_1 \xi^I = L_A^I \left( \partial_1 \xi^A +  f_{BC}^{\phantom{BC} A} L_1^B\xi^C  \right)\, .
\ee
Finally we have 
\be
\label{xxx3}
S_{WZW}^{(2)} = \frac{1}{2} \int d^2 \sigma\, \eta_{AB}\partial_0\xi^A \partial_1 \xi^B - f_{ABC} \xi^A \partial_0 \xi^B  L^C_I \partial_1 X^I \, .
\ee
To evaluate the NL piece of the action we need to evaluate certain derivatives of the matrix $R_{IJ}$.  We find
\be
D_K R_{IJ } = \frac{1}{2} f^{\ \ A}_{CD} L^C_K L^D_I R_{AB} L^B_J + \frac{1}{2} f^{ \ \ A}_{CD} L^C_K L^D_J R_{AB} L^B_I = -\frac{1}{2} S^L_{KI}R_{LJ} -\frac{1}{2}S^L_{KJ}R_{LI}\,,
\ee
and
\bea
D_KD_LR_{IJ } &=& \frac{1}{4} S^M_{KI}R_{MN}S^N_{LJ} + \frac{1}{4} S^N_{LM}R_{NJ} S^M_{KI} + I\leftrightarrow J \nonumber \\
&=& \frac{1}{4} L^A_IL^B_JL^C_K L^D_L \left(f^{\ \ E}_{DF}R_{EA}f^{\ \ F}_{CB}  +  f^{\ \ E}_{CA}R_{EF}f^{\ \ F}_{DB}  + A\leftrightarrow B\right)\, . \quad
\eea
We will find the following tensor useful:
\bea
\label{TIJ}
T_{IJ} &=& \xi^K\xi^L \left(D_K D_L  R_{IJ} + R_{IM}R^M_{\ KLJ} +   R_{JM}R^M_{\ KLI}\right)\nonumber \\
&=&\frac{1}{4} \xi^K \xi^L\left( 2 S^M_{KJ} S^N_{LM} R_{NI} + S^N_{LI} S^M_{KJ} R_{NM}  + I \leftrightarrow J \right)  \quad  \nonumber \\ 
&=& \frac{1}{4}L^A_IL^B_JL^C_K L^D_L  \xi^K \xi^L\left(2f^{\ \ E}_{CF}R_{EB}f^{\ \ F}_{DA}  + f^{\ \ E}_{CA}R_{EF}f^{\ \ F}_{DB} + A\leftrightarrow B \right)\, . \nonumber \\
&& 
\eea
Then one finds that the second order term is given by
\be
S^{(2)}_{NL} = -\frac{1}{2} \int d^2 \sigma \, R_{IJ} D_1 \xi^I D_1 \xi^J + T_{IJ} \partial_1 X^I \partial_1 X^J\,  .
\ee
If we combine this piece with (\ref{xxx3}) and expand out the covariant derivatives we find that the complete second order expansion of (\ref{PLTaction}) can be written as a sum of a kinetic and interaction terms
\be
\label{A-21a}
S^{(2)}=S_{\rm kin}+S_{\rm int}\ ,
\ee
where
\be
S_{\rm kin}=\ha\int d\s d\tau\left(
\eta_{AB}\del_0\xi^A\del_1\xi^B-R_{AB}\del_1\xi^A\del_1\xi^B\right)\ ,
\ee
and
\be
S_{\rm int}=\ha\int d\s d\tau\left(
I_{AB}\xi^A\xi^B+J_{AB}\xi^A\del_1\xi^B+K_{AB}\xi^A\del_0\xi^B\right) \ ,
\ee
with
\ba
&&I_{AB}=-\ha L^C_1 L^D_1 \left[
f_{AC}{}^E f_{BD}{}^FR_{EF}+
(2f_{AF}{}^ER_{EC}+f_{AC}{}^ER_{EF})f_{BD}{}^F\right]\ ,
\nonumber \\
&&J_{AB}=(f_{BA}{}^CR_{CE}+2f_{EA}{}^CR_{CB})L_1^E \ ,
\label{A-22}\\
&&K_{AB}=-f_{ABC}L_1^C \ .
\nonumber
\ea
We note that the terms labeled by $I_{AB}$ and $J_{AB}$ come from the ${\cal S}^{(2)}_{NL}$ action, while the term $K_{AB}$ comes from the ${\cal S}^{(2)}_{WZW}$ action.

The effective action $S_{\rm eff}(X)$ at one-loop is given in terms of
the interacting Lagrangian as
\ba
\label{A-23}
S_{\rm eff}(X)=\langle S_{\rm int}\rangle +
\frac{i}{2} \langle S_{\rm int}^2 \rangle\ .
\ea
Substituting \eqn{A-21a} in \eqn{A-23} we find that the effective Lagrangian is given by
\ba
\label{A-24}
S_{\rm eff}(\chi)&=&\ha\int d\s d\tau(q_1+q_2+q_3+q_4)
\nonumber\\
&=&\ha\int d\s d\tau\left(I_{AB} \langle\xi^A\xi^B \rangle
+ \frac{i}{4}J_{AB}J_{CD} \langle \xi^A\del_1\xi^B\xi^C\del_1\xi^D \rangle
\right.
\nonumber\\
&& +\frac{i}{4}K_{AB}K_{CD} \langle \xi^A\del_0\xi^B\xi^C\del_0\xi^D \rangle +
\\
&&+ \left.\frac{i}{4}(J_{AB}K_{CD}+K_{AB}J_{CD}) \langle \xi^A\del_1\xi^B\xi^C\del_0\xi^D \rangle \right)\ ,
\nonumber
\ea
where $q_i, i=1,2,3,4$ are respectively the first, second, third and fourth term in the above expression.
The $\xi^2$ contractions and $(\xi\del\xi)^2$ are evaluated identically to the those in the abelian case (details can again be found in the appendix), with the result, reprinted here for convenience:
\ba
\langle \xi^A\xi^B \rangle & = & R^{AB}\Delta(0)+\eta^{AB}\Theta(0)\ ,
\nonumber\\
i\langle \xi^A\del_1\xi^B\xi^C\del_1\xi^D \rangle & \simeq & \ha(R^{A[C}R^{D]B}-\eta^{A[C}\eta^{D]B})\Delta(0)\ ,
\nonumber\\
i\langle \xi^A\del_0\xi^B\xi^C\del_0\xi^D \rangle & \simeq & -\ha(R^{A[C}R^{D]B}+3\eta^{A[C}\eta^{D]B})\Delta(0)
\label{A-25}\\
&& \phantom{}
-(R^{A[C}\eta^{D]B}+\eta^{A[C}R^{D]B})\Theta(0)\ ,
\nonumber\\
i\langle \xi^A\del_1\xi^B\xi^C\del_0\xi^D \rangle &  \simeq & -\ha(R^{A[C}\eta^{D]B}+
\eta^{A[C}R^{D]B})\Delta(0)-\eta^{A[C}\eta^{D]B}\Theta(0) \ ,
\nonumber
\ea
where $\Delta(0)$ and $\Theta(0)$ are the Weyl and Lorentz anomalies, respectively and $\simeq$ denotes the fact that we have kept only the terms relevant to the Lorentz and the
 Weyl divergence.\footnote{In these expressions, $\langle \xi^A\partial \xi^B \xi^C \partial \xi^D \rangle$
is really short hand for  $\int d^2 \sigma' \langle \xi^A(\sigma)\partial \xi^B(\sigma) \xi^C(\sigma^\prime) \partial \xi^D (\sigma^\prime)  \rangle $.} 

\subsection{Weyl Anomaly}

We shall now compute the Weyl anomaly of the effective action
by plugging \eqn{A-22} and \eqn{A-25} in \eqn{A-24}. The result can be written as
\ba
q_1^W&=&-\ha L^C_1 L^D_1 \left(
f_{AC}{}^E f_{BD}{}^F R_{EF}R^{AB}+\right.
\nonumber \\
&&\left.(\ 2f_{AF}{}^E R_{EC}+f_{AC}{}^E R_{EF})f_{BD}{}^F R^{AB}\right)\ , \\
q_2^W&=&\frac{1}{4} f_{AB}{}^C f_{DE}{}^F R_{CE}R_{FK}(R^{AD}R^{BE}-\eta^{AD}\eta^{BE})
 L^E_1 L^K_1\ ,
\label{A-26}\\
q_3^W&=&-\frac{1}{4}\left(f_{ABC}f_{DEF}R^{AD}R^{BE}+3f_{ABC}f^{AB}{}_F\right)
 L^C_1 L^F_1\ ,\nonumber \\
q_4^W&=&\left(f_{ABC}f^{AB}{}_D+f_{CE}{}^F R_{FB} f_{DA}{}^B R^{EA}+f_{ABC}f^{AF}{}_E R_{FD}R^{EB}\right)
L^C_1 L^D_1\ . \nonumber \\
&&
\ea
In the calculation of $q_2$ we make some heavy use of the compatibility condition
\bea
R^{IJ}\eta_{JM} = \eta^{IJ}R_{JM}
\eea
to show that
\bea
&&\left( R^{JQ} R^{IP} - R^{IQ} R^{JP}  - \eta^{JQ}\eta^{IP} +  \eta^{IQ} \eta^{JP}\right)R_{IM} S^M_{KJ} R_{PN} S^N_{RQ} = 0 \, , \nonumber  \\
&&\left( R^{JQ} R^{IP} - R^{IQ} R^{JP}  - \eta^{JQ}\eta^{IP} +  \eta^{IQ} \eta^{JP}\right)R_{MK} S^M_{IJ}R_{PN} S^N_{RQ} = 0 \, , \nonumber  \\
&& \left( R^{JQ} R^{IP} - R^{IQ} R^{JP}  - \eta^{JQ}\eta^{IP} +  \eta^{IQ} \eta^{JP}\right)R_{IM}S^M_{KJ} S^N_{PQ}R_{NR} = 0 \, . 
\eea

Adding up the contributions we find that the Weyl anomaly reads
\ba
\label{A-27}
{\rm Weyl}&=&\frac{1}{4} f_{AB}{}^Cf_{DE}{}^F\left(R_{CK}R_{FH}R^{AD}R^{BE}-
R_{CK}R_{FH}\eta^{AD}\eta^{BE}\right.
+\nonumber\\
&&\left.\eta_{CK}\eta_{FH}\eta^{AD}\eta^{BE}-\eta_{CK}\eta_{FH}R^{AD}R^{BE}\right)
L^K_1 L^H_1 
\nonumber\\
&=&\frac{1}{4}
(R_{AC}R_{BF}-\eta_{AC}\eta_{BF})
(R^{KD}R^{HE}-\eta^{KD}\eta^{HE})
f_{KH}{}^Cf_{DE}{}^F \times
\\
&&L^A_I L^B_J \del_1X^I \del_1X^J \ .
\nonumber
\ea
However, the zeroth order in the expansion of ${\cal S}_{\rm NL}$ has the Lagrangian density
\ba
\label{A-28}
R_{AB}L^A_I L^B_J \del_1X^I \del_1X^J \ .
\ea
Since the Weyl anomaly counter terms can be absorbed as a redefinition of the $R_{AB}$
we find that the action \eqn{act1} is renormalisable at one loop.
Moreover the RG flow of  $R_{AB}$ can be directly read off from \eqn{A-27}
and it is equal to
\ba
\label{A-29}\boxed{
\frac{dR_{AB}}{dt}=\frac{1}{4}
(R_{AC}R_{BF}-\eta_{AC}\eta_{BF})
(R^{KD}R^{HE}-\eta^{KD}\eta^{HE})
f_{KH}{}^Cf_{DE}{}^F}\ ,
\ea
where $t=\ln \m$, with $\m$ being the energy scale.
This is a quite simple formula and constitutes one of the main
results of the present chapter.\footnote{This result appeared in a paper by the author of this thesis together with collaborators K. Sfetsos and K. Siampos \cite{Sfetsos:2009vt}.  During the final days of the preparation of that preprint, we learnt of another group \cite{Avramis:2009xi} who had arrived at the same conclusion albeit with different motivation. }

\subsection{Lorentz Anomaly}

We shall now compute the Lorentz anomaly of the effective action
by plugging \eqn{A-22} and \eqn{A-25} in \eqn{A-24}. The result can be written as
\ba
q_1^L&=&-\ha  L^C_1 L^D_1 \left(
f_{AC}{}^E f_{BD}{}^F R_{EF}\eta^{AB}+\right.
\nonumber
\\
&&\left.(\ 2f_{AF}{}^E R_{EC}+f_{AC}{}^E R_{EF})f_{BD}{}^F \eta^{AB}\right)\ ,\nonumber \\
q_2^L&=&0\ ,
\label{A-30}
\\
q_3^L&=&f_{ABC}f_{DE}{}^B R^{AD}L^C_I L^E_J\del_1X^I\del_1X^J\ ,\nonumber\\
q_4^L&=&(-f_{AB}{}^C R_{CE}+2f_{EA}{}^C R_{CB})f^{AB}{}_D L^E_I L^D_J\del_1X^I\del_1X^J\ .\nonumber
\ea
Adding up the contributions we find that the Lorentz anomaly is zero. Thus, the system
is Lorentz invariant at one loop and this represents an important and non-trivial demonstration of consistency.

\section{Some Examples and a general proof of equivalence}

We now wish to compare the result (\ref{A-29}) with the renormalisation of T-dual pairs of sigma models.  We first give some examples in which the known renormalisation of the T-dual pairs match that of the duality invariant theory.  We then present an algebraic proof that this is equivalence holds in general.  

\subsection{Renormalisation of T-dual pairs}
Since the pairs of dual models, 
\bea
\label{DEFgsimgamodel2}
S &=& \int d^2 \sigma ((E_0^+)^{-1} + \Pi)_{ab}^{-1} L^a_i L^b_j \p_+ X^i \p_- X^j = \int  d^2 \sigma  Q^+_{ij}\p_+ X^i \p_- X^j \, , \\
\label{DEFtildedgsigmamodel2}\tilde{S} &=& \int d^2 \sigma [(E_0^+ + \tilde{\Pi})^{-1}]^{ab} \tilde{L}_a^i \tilde{L}^j_b \p_+ \tilde{X}_i \p_- \tilde{X}_j = \int  d^2 \sigma  \tilde{Q}^{+ij}\p_+ \tilde{X}_i \p_- \tilde{X}_j \, ,
\eea
are of a standard form, the one-loop counter terms are given by the generalised Ricci tensor constructed out of $Q^+_{ij}$  \cite{Braaten:1985is,Fridling:1985hc,Osborn:1989bu}.    One can then deduce the RG equations for the $d^2$ couplings we call $a^\mu$   that are hidden inside $Q^+_{ij}$ and provided by the parameter matrix $E_{0, ab}$.   This is expressed as 
\ba
\frac{da^\mu}{dt}= \frac{ 1 }{ \pi} a^\mu_1\ ,
\ea
where the $a^i_1$'s are chosen such that
\be
\ha \hat R_{ij } = \del_{a^\mu} Q^+_{ij} a^\mu_1 \ . 
\label{1loop}
\ee
In this treatment we have also omitted possible field renormalisations as they are not needed in our case. 

 Since the metrics that appear in \eqn{DEFgsimgamodel2} and \eqn{DEFtildedgsigmamodel2} are, in general, very complicated it may seem very hard to algebraically evaluate the RG equations described by \eqn{1loop}.  However, an algebraic treatment was provided in \cite{Valent:2009nv} which allowed a demonstration \cite{Sfetsos:2009dj}  of the equivalence of the system  \eqn{1loop} for \eqn{DEFgsimgamodel2} and \eqn{DEFtildedgsigmamodel2}.

To present the result for the beta function equations
we recall the notation introduced in \cite{Sfetsos:2009dj}.
We first define the matrices
\be
A^{ab}{}_{c} = \tilde f^{ab}{}_c - f_{cd}{}^a M^{db}\ ,\qq
B^{ab}{}_{c} = \tilde f^{ab}{}_c + M^{ad}f_{dc}{}^b \ ,
\label{fhh11}
\ee
as well as their duals
\be
\tilde A_{ab}{}^{c} =  f_{ab}{}^c - \tilde f^{cd}{}_a (M^{-1})_{db}\ ,\qq
\tilde B_{ab}{}^{c} = f_{ab}{}^c + (M^{-1})_{ad}\tilde f^{dc}{}_b \ .
\ee
Using these we also construct
\ba
L^{ab}{}_c & = &  \ha (M_s^{-1})_{cd}\left( B^{ab}{}_e M^{ed} + A^{db}{}_e M^{ae}- A^{ad}{}_eM^{eb}
  \right) \ ,
\nonumber\\
R^{ab}{}_c & = & \ha (M_s^{-1})_{cd}\left( A^{ab}{}_e M^{de} + B^{ad}{}_eM^{eb} - B^{db}{}_e M^{ae}
\right) \, ,
\label{kkll1a}
\ea
and
\ba
\tilde L_{ab}{}^c & = &  \ha (\tilde M_s^{-1})^{cd}\left(
\tilde B_{ab}{}^e (M^{-1})_{ed} + \tilde A_{db}{}^e (M^{-1})_{ae}- \tilde A_{ad}{}^e (M^{-1})_{eb}
  \right) \ ,
\nonumber\\
\tilde R_{ab}{}^c & = & \ha (\tilde M_s^{-1})^{cd}\left( \tilde A_{ab}{}^e (M^{-1})_{de}
+ \tilde B_{ad}{}^e (M^{-1})_{eb} - \tilde B_{db}{}^e (M^{-1})_{ae}
\right) \ ,
\label{kkll1}
\ea
where
\be
M_s = \ha (M+M^T) \ ,\qq \tilde M_s = \ha \left[(M^{-1}) + (M^{-1})^T\right]\ .
\ee

\no
The one-loop RG equations corresponding to \eqn{DEFgsimgamodel2} were caluclated in  \cite{Valent:2009nv} and are given by
  (in the notation of \cite{Sfetsos:2009dj})
\be
\frac{d M^{ab}}{ dt}
= R^{ac}{}_d L^{db}{}_c\ .
\label{rg1}
\ee
Similarly, for its dual \eqn{DEFtildedgsigmamodel2} we have
\be
\frac{d (M^{-1})_{ab}}{ dt}
= \tilde R_{ac}{}^d \tilde L_{db}{}^c\ .
\label{rg2}
\ee
Developing certain identities among the various quantities defined above it has been shown
in \cite{Sfetsos:2009dj} that the two systems \eqn{rg1} and \eqn{rg2} are in fact equivalent. Therefore, at one-loop in perturbation theory, general $\s$-models related
by Poisson--Lie T-duality are equivalent under the renormalisation group flow. Perhaps the reason for this can be traced to that fact that the $\s$-model actions \eqn{DEFgsimgamodel2} and \eqn{DEFtildedgsigmamodel2} are formally canonically equivalent in phase space \cite{Sfetsos:1997pi,Sfetsos:1996xj}.  The equivalence of the systems \eqn{rg1} and \eqn{rg2} does not necessarily mean that the beta functions one would have computed for the
original action \eqn{PLTaction} using \eqn{A-29} would have been the same. The reason is that
as we have mentioned above, certain constraints were solved in order to obtain \eqn{DEFgsimgamodel2} and
\eqn{DEFtildedgsigmamodel2}, and quantisation before and after solving them might not be commuting operations.

At this stage we cannot help but remark at the compactness of the duality invariant form of the RG equation \eqn{A-29} compared with the rather complicated equations in \eqn{rg1} and \eqn{rg2}.

We will present three non-trivial examples in which we will explicitly compute the beta functions equations using \eqn{A-29} and recover the same equations that were previously computed using \eqn{rg1}.

\subsubsection{ Semi-Abelian Doubles }
As a first example we consider the case when the commutator relations for the double are given as
\ba
&& [T_a,T_b]  =  i f_{ab}{}^c T_c \ ,
\nonumber\\
&& [\tilde T^a, \tilde T^b]  =  0 \ ,
\\
&& [T_a,\tilde T^b ] = - i f_{ac}{}^b \tilde T^c\ .
\nonumber
\ea
This is known as the semi-abelian double since $\tilde{ {\cal G}} = U(1)^d$
and coincides with the regular notion non-abelian T-duality.
We leave the group ${\cal G}$ general but to keep the
problem simple we consider a point in the coupling space where
\be
M^{ab}  = \kappa \delta^{ab}\ .
\ee
Then from \eqn{kkll1a} we find that $\displaystyle R^{ab}{}_c = - L^{ab}{}_c= \frac{\kappa}{ 2} f^{ab}{}_c $. Then,
from the general RG equations \eqn{rg1} we find the result
\be
\frac{ d \kappa} {dt} \delta^{ab} = -\frac{\kappa^2}{  4} f^{ac}{}_d f^{db}{}_c = -\frac{C_2}{4}\kappa^2 \delta^{ab}\ ,\ee
where $C_2$ is the quadratic Casimir in the adjoint representation.
In the duality invariant expressions we have that
$R_{AB} = {\rm diag} ( \frac{1}{\kappa} \delta_{ab}, \kappa \delta^{ab} )$
 and can simply calculate the RG equation for $\kappa$ using \eqn{A-29}.
 We find that the above result is recovered.

\subsubsection{ A six-dimensional Drinfeld Double}

In this example we use a six-dimensional Drinfeld Double based on the
three-dimensional Lie algebras, IX for $G$ and V for $\tilde{G}$ in the Bianchi classification.  In fact, this Drinfeld double is ismorphic to $so(3,1)$ represented as a sum $su(2) \oplus e_3$.  

The corresponding generators are
$T_a$ and $\tilde{T}^a$, where $a=1,2,3$. It is convenient to
split the index
$a=(3,\a)$, with $\a=1,2$. The non-vanishing commutation relations are
\ba
&& [T_a,T_b]= i \e_{abc} T_c \ ,\qq [\tilde T_3,\tilde T_\a]= i \tilde T_\a \ ,
\nonumber
\quad [T_\a,\tilde{T}_\b]=i \e_{\a \b}\tilde{T}_3-i \delta_{\a \b}T_3\ ,\\ 
&& \quad [T_3,\tilde{T}_\a]=i\e_{\a \b}\tilde{T}_\b\ ,
\quad [\tilde{T}_3,T_\a]=i \e_{\a \b}\tilde{T}_\b-i T_{\a} \ ,
\label{commu1}
\ea
where $\delta_{\a\b}, \e_{\a\b}$ are the Kronecker delta and the antisymmetric
symbol in two dimensions, and $\e_{abc}$ is the totally antisymmetric tensor in three dimensions,
$\e_{123}=1$. 

To see the equivalence of this Drinfeld double with the $so(3,1)$ algebra given by
\be
-i [\cM_{ij} \cM_{kl}] =   \eta_{ik} \cM_{jl} +   \eta_{jl} \cM_{ik} - \eta_{il} \cM_{jk} - \eta_{jk} \cM_{il}\, , 
\ee 
one splits the generators into rotations $J_a = \frac{1}{2} \e_{abc} \cM_{bc}$ for $a,b,c=1,2,3$ and boosts $K_a = \cM_{a4}$ to obtain the contraction
\be
[J_a, J_b] = i \e_{abc} J_c\, , \quad  [K_a, J_b] = i \e_{abc} J_c\, ,   \quad [K_a, K_b] = - i \e_{abc} J_c\, . 
\ee
These commutators are realised by the algebra \eqn{commu1} by defining 
\be
J_a = T_a , \quad K_1= \tilde{T}^1 + T_2\, , \quad K_2 =  \tilde{T}^2 - T_1\, , \quad K_3=  \tilde{T}^3  \, .
\ee

An explicit representation of this algebra is given by the following generators.  For the $su(2)$ factor we have the standard
\be
T_i = {\rm diag} \left( \frac{\sigma_i}{2},   \frac{\sigma_i}{2} \right) \, , \quad i = 1\dots 3
\ee
with
\ba
\sigma_1 = 
\left( \begin{array}{cc}
  0 & 1 \\ 
  1 & 0  
 \end{array}	
\right) & 
 \sigma_2= 
\left( \begin{array}{cc}
  0 & - i  \\ 
  i & 0
 \end{array}	
\right) & 
 \sigma_3= 
\left( \begin{array}{cc}
  1 & 0 \\ 
  0 & -1
 \end{array}	
\right)\, . 
\ea
For the $e_3$ factor we have 
\be
\tilde{T}^1 = {\rm diag} \left(\frac{i \sigma_+}{2} , \frac{-i \sigma_-}{2} \right)  \ , \quad   \tilde{T}^2 = {\rm diag} \left( \frac{\sigma_+}{2} ,\frac{ \sigma_-}{2} \right) \ , \quad  \tilde{T}^3 ={\rm diag} \left(\frac{i \sigma_3}{2} ,\frac{ - i \sigma_3}{2} \right) \, , 
\ee
where $\sigma_\pm =  \sigma_1 \pm i \sigma_2$.  An inner product with respect to which $su(2)$ and $e_3$ are maximally isotropic is given by
\be
< A , B > = Tr \left(  P_+ A P_ + B -  P_- A P_- B  \right) \ ,
\ee 
in which we have defined projectors $P_+ = {\rm diag}\left( 1 , 0 \right)$ and  $P_- = {\rm diag} \left( 0, 1 \right)$ in the relevant 4-dimensional matrix representation.  If we define a set of six generators $T_A = \{ T_a , \tilde{T}^a \} $ then the inner product is given in components as 
\be
\eta_{AB} = < T_A , T_B > = \left( \begin{array}{cc}
   0 &  1_3  \\ 
  1_3    &  0
 \end{array}	
\right)\, . 
\ee

We now wish to test the RG equations for this six-dimensional double. We consider the matrix 
 $M$ to be given by
\be
\label{Rmatrixb}
M=\left( \begin{array}{ccc}
     a & b-1 & 0 \\
     1-b & a & 0 \\
     0 & 0 & \frac{a}{1+g} \\
   \end{array}
\right) \, .
\ee
The reason for this restricted choice of parameter is that in this case the geometries of the two sigma models are much more tractable than in the general case. 
Then from \eqn{rmatrix} we find that
\be
R=\left(\begin{array}{cccccc}
    \frac{1}{a} & 0 & 0 & 0 & \frac{1-b}{a} & 0 \\
    0 & \frac{1}{a} & 0 & \frac{b-1}{a} & 0 & 0 \\
    0 & 0 & \frac{1+g}{a}& 0 & 0 & 0 \\
    0 & \frac{b-1}{a} & 0 & \frac{a^2+(b-1)^2}{a} & 0 & 0 \\
    \frac{1-b}{a} & 0 & 0 & 0 & \frac{a^2+(b-1)^2}{a} & 0 \\
    0 & 0 & 0 & 0 & 0 & \frac{a}{1+g} \\
  \end{array}\right)\ .
\label{Rmatrix}
\ee
By plugging the structure constants of the double and this parameter matrix into the doubled RG equation \eqn{A-29}
we can calculate the running of the parameters $a, b$ and $g$. We find (using the Mathematica computer programme): 
\ba
\label{6dim}
&&\frac{da}{ dt}=\frac{1+a^2-b^2}{2a^2}((g-1)a^2+(g+1)(b^2-1))\ ,
\nonumber \\
&&\frac{db}{ dt}=\frac{b}{ a}(a^2(g-1)+(g+1)(b^2-1))\ ,\\
&&\frac{dg}{dt}=\frac{1+g }{ a}(g(1+a^2)+(g+2)b^2)\ .
\nonumber
\ea
This is precisely the same system one derives using \eqn{rg1} or \eqn{rg2} as shown in \cite{Sfetsos:1999zm,Sfetsos:2009dj}.

\subsubsection{ A sixteen-dimensional Drinfeld double}

In this final example we study a sixteen-dimensional Drinfeld double group
based on an $SU(3)$ group with generators $T_a$ and an abelian
eight-dimensional group
with generators $\tilde{T}^a$, where $a=1,2,\dots,8$.
For the $SU(3)$ group we use the structure constants in the Gell-Mann
basis (see for instance, eq.(5.2) of \cite{MuellerHoissen:1987cq})
\be
\label{strSU(3)}
f^{12}{}_3=2,\quad
f^{14}{}_7=-f^{15}{}_6=f^{24}{}_6=f^{25}{}_7=f^{34}{}_5=-f^{36}{}_7=1,\quad
f^{45}{}_8=f^{67}{}_8=\sqrt{3}\ ,
\ee
where the rest are obtained by antisymmetrisation in the three indices.
Whilst the structure constants between the $\tilde{T}$ generators vanish,  the full set of structure constants for the double includes those coming from commuting $\tilde{T}$ and $T$ generators and these may be read from the defining equations of the double.  Unlike the previous case, finding a matrix representation of this double may be tricky.

The matrix $R_{AB}$ is built
in terms of the symmetric and antisymmetric part of $M^{-1}$,
where $M$ is taken for simplicity to be a diagonal eight-dimensional matrix
\ba
M=\left(\frac{1}{g}\mathbb{I}_{3\times 3}\right)\oplus
\left(a\ \mathbb{I}_{4\times 4}\right)\oplus b\ .
\ea
The result is a diagonal sixteen-dimensional matrix given by
\ba
\label{Rmatrix2}
R=
\left(g\mathbb{I}_{3\times 3}\right)\oplus
\left(\frac{1}{a}\mathbb{I}_{4\times 4}\right)\oplus\frac{1}{b}\oplus
\left(\frac{1}{g}\mathbb{I}_{3\times 3}\right)\oplus
\left(a\mathbb{I}_{4\times 4}\right)\oplus b\ .
\ea
Plugging the structure constants and this parameter matrix into the doubled RG equation \eqn{A-29} we find
\ba
\frac{da}{dt}=\frac{3a^2}{2}\frac{a(bg+1)-4b}{b}\ ,\quad
\frac{db}{dt}=-3a^2\ ,\quad
\frac{dg}{dt}=g^2a^2+2\ ,
\ea
which were precisely the expressions in eq.(5.9) of \cite{Sfetsos:2009dj}.

\subsection{General Equivalence}

Encouraged by the above non-trivial checks, we now prove the general equivalence of the RG equations for the parameters contained in the matrix $R_{AB}$. 
It will be useful to state the running of $\tilde{M}_s$ and $B$ explicitly by expanding out the compact expression \eqn{rg2}.  Whilst this calculation is straightforward it is exceedingly tedious. The result is that we can write:
\be
\frac{d (M^{-1})_{ab}}{ dt}
=  \frac{1}{4} \sum_{i, j} \Phi^{(i,j)}_{ab}
\label{RGM}
\ee
with $\Phi^{(i,j)}_{ab}$ consisting of terms of order $i$ in $B_{ab}$ and order $j$ in the structure constant $\tilde{f}^{ab}_{\ \ c}$.   When $i+j$ is even  $\Phi^{(i,j)}_{ab}$ is symmetric and hence contributes to the running of $\tilde{M}_s$ whereas when $i+j$ is odd, $\Phi^{(i,j)}_{ab}$ gives the running of $B$.  We define $K_{ab} = B_{ac}(\tilde{M}_s^{-1})^{cd} B_{d b}$ and introduce the notation that hatted indices have been raised or lowered from their natural position using $\tilde{M}_s$ and its inverse.  For future reference we provide the exact expressions for  $\Phi^{(i,j)}_{ab}$:
\ba
\Phi^{(0,0)}_{ab} &=& 2 f_{ad}^{\ \ e}f_{e b}^{ \ \ d} + 2 f_{a \ \hat{e}}^{\ \hat{d}} f_{db}^{ \  \ e} - f^{\hat{d} \hat{e}}_{\ \ \hat{a}} f_{e d \hat{b}}  \\
\Phi^{(0,1)}_{ab} &=&-2 f_{ac}^{\ \ d} \tf_{\hat{d}\hat{b}}^{\ \ \hat{c}} -2 f_{ac}^{\ \ d} \tf^{\hat{c}}_{\ b \hat{d}}+ f^{\hat{d}}_{\ c \hat{a}} \tf^c_{\hat{d} b} -(a \leftrightarrow b)  \\
\Phi^{(0,2)}_{ab} &=&  2 \tf^e_{\ \hat{a} d} \tf_{\ \hat{b} e}^d - 2 \tf_{\hat{a}}^{\ e \hat{d}} \tf_{\hat{e}\hat{b}d} + \tf_{\hat{e} \hat{d} a} \tf^{de}_{\ \ b} \\
\Phi^{(1,0)}_{ab} &=& 2B_{de} f_{ac}^{\ \ e} f^{\hat{c}\hat{d}}_{\ \ \hat{b}} - B_{af}f^{\hat{d}\hat{c} f} f_{cd \hat{b}} - (a \leftrightarrow b)  \\
\nonumber \Phi^{(1,1)}_{ab} &=& -2 B_{af} \tf^{df}_{\ \ e} f_{db}^{ \ \ e} - 2 B_{af} \tf_{\hat{e}}^{\ f \hat{d} } f_{db}^{ \ \ e}  + 2 B_{dk} \tf^{e k}_{\ \ a}f^{\hat{d}}_{\ b e} \\
&&  +2 B_{eh} \tf^{ h \ \hat{d}}_{\ \hat{a}} f^{\hat{e}}_{\ d \hat{b}} + B_{af} \tf^{d e }_{ \ \ b} f_{e d }^{\ \ f} + (a \leftrightarrow b) \\
\Phi^{(1,2)}_{ab} &=& B_{cf}[-2 \tf^{df}_{\ \ a} \tf^c_{\ \hat{b} d} - \tf^{df}_{\ \ a}  \tf_{\hat{d}\hat{b}}^{\ \ \hat{c}} - 2 \tf^f_{\ \hat{a} d} \tf^{cd}_{\ \ b} + \tf^{fd}_{\ \ a} \tf_{\hat{d} \ b}^{\ c}  ] \nonumber \\
&& +2B_{af}[ \tf^{df}_{\ \ c} \tf_{\hat{d} \hat{b}}^{ \ \ \hat{c}} + \tf^{df}_{\ \ c} \tf^c_{\ \hat{b} d}  ]   - (a \leftrightarrow b)\\
\nonumber\Phi^{(2,0)}_{ab} &=& B_{f b} B_{a c} f^{\hat{d} \hat{e} c } f_{de}^{\ \ f} -  2B_{af} B_{e h } f_{db}^{\ \ e} f^{\hat{d} \hat{h} f}  -  2B_{bf} B_{e h } f_{da}^{\ \ e} f^{\hat{d} \hat{h} f}\\
&&  -2 B_{eh} B_{dk} f_a^{\ \hat{k} h} f_b^{\ \hat{e} d} - 2 K_{he} f_{a}^{\ \hat{d} h} f_{d b}^{ \ \ e }\\
\nonumber \Phi^{(2,1)}_{ab} &=& 2 K_{gf} f_{db}^{ \ \ g} \tf^{f \ \hat{d}}_{\ \hat{a}} -2 B_{cf}B_{gb}f_d^{\ \hat{c} g} \tf^{f \ \hat{d}}_{\ \hat{a}} -2 B_{de}B_{gc} f^{\hat{d} \ g}_{\ b} \tf^{e \ \hat{c}}_{\ \hat{a}}\\
&& -2B_{af}B_{de} f^{\hat{c}\hat{d}}_{\ \ \hat{b}} \tf^{ef}_{\ \ c} - B_{de} B_{gc} f^{\hat{c} \hat{d}}_{\ \ \hat{b}} \tf^{eg}_{ \ \ a} - (a \leftrightarrow b) \\
\nonumber \Phi^{(2,2)}_{ab} &=& -2B_{af}B_{gc} \tf^{f d \hat{c} } \tf^g_{\ \hat{d} b} - 2B_{bf}B_{gc} \tf^{f d \hat{c} } \tf^g_{\ \hat{d} a} -2 B_{ac} B_{bf} [ \tf^{ce}_{ \ \ d} \tf^{f d}_{\ \ e} +  \tf^{ce}_{\ \ d}  \tf^{f \ \hat{d}}_{\ \hat{e}}  ] \\
&& -2B_{he}B_{dk}\tf^{h\ \hat{k}}_{\ \hat{a}}\tf^{d \ \hat{e}}_{\ \hat{b}} - 2 K_{gf} [\tf^{f \ \hat{d}}_{\ \hat{a}} \tf^g_{\ \hat{b} d}-  \tf^{fc}_{\ \ a} \tf^{g}_{\ \hat{c} b} ] \\
\nonumber\Phi^{(3,1)}_{ab} &=& -2 B_{af} B_{eh} B_{dk} \tf^{h f \hat{k}} f^{\hat{e} \ d}_{\ b} + 2 B_{af} K_{eh} \tf^{h f}_{\ \ d} f^{\hat{d} \ e}_{\ b} + 2B_{ac} B_{fb} B_{eh} \tf^{hc \hat{d}} f^{\hat{e} \ f}_{\ d} \\
&& + 2 K_{eh} B_{dk}\tf^{kh}_{\ \ a} f^{\hat{d}\ e}_{\ b}  +   B_{dk}B_{he}B_{fb}\tf^{kh}_{\ \ a} f^{\hat{d}\hat{e} f} + (a \leftrightarrow b)  \\
\Phi^{(3,2)}_{ab} &=& - 2K_{eh} B_{af} \tf^{ef}_{\ \ c} \tf^{h \ \hat{c}}_{\ \hat{b}} + 2 B_{af}B_{de}B_{hc} \tf^{ef\hat{c}} \tf^{h \ \hat{d}}_{\ \hat{b}} \nonumber \\ && - 2 K_{eh}B_{gc} \tf^{eg}_{\ \ a} \tf^{h \  \hat{c}}_{\ \hat{b}} - (a \leftrightarrow b) \\
\nonumber \Phi^{(4,2)}_{ab} &=& -K_{kd} K_{eh} \tf^{kh}_{\ \ a} \tf^{de}_{\ \ b} + 2 B_{ac} B_{bf} B_{eh} B_{dk} \tf^{ hc \hat{k} } \tf^{fd \hat{e} } + 2 B_{ac} B_{bf} K_{he} \tf^{ hc \hat{d} } \tf^{e f }_{ \  \ d}  \\
&&    + 2 B_{af} B_{dk} K_{he} \tf^{ hf \hat{k} }\tf^{e d }_{ \  \ b} + 2 B_{bf} B_{dk} K_{he} \tf^{ hf \hat{k} } \tf^{e d }_{ \  \ a} \\
\Phi^{(4,0)}_{ab} &=& \Phi^{(4,1)}_{ab}= \Phi^{(3,0)}_{ab} =  0 \, .
\ea

We now turn to the  duality invariant RG equation \eqn{A-29} and from it infer the running of the parameters in $B$ and $\tilde{M}_s$.  If we write the RG equation for $R_{AB}$ as
 \be
 \frac{d R_{AB}}{dt} = \left(\begin{array}{cc}W_{ab} & W_a^{\ b} \\
 (W_a^{\ b})^T & W^{ab} \end{array} \right)\ ,
 \label{RG2}
\ee
then it is clear from the $O(d,d)/O(d)\times O(d)$ coset structure of the matrix $R_{AB}$  that the bottom right hand block entry of \eqn{RG2}  gives
\be
\frac{d (\tilde{M}_s)_{ab} }{d t} = -(\tilde{M}_s)_{ac}(\tilde{M}_s)_{bd}W^{cd} \ .
\ee
Expanding out the duality invariant RG equation \eqn{A-29} yields
\be
\frac{d (\tilde{M}_s)_{ab} }{d t} = \frac{1}{4} \left( \Phi^{(0,0)}+\Phi^{(0,2)}+\Phi^{(2,0)}+ \Phi^{(1,1)}+\Phi^{(2,2)} + \Phi^{(3,1)} + \Phi^{(4,2)}\right)_{ab} \ .
\label{RGG}
\ee
Similarly, from the the top right block of \eqn{RG2} we have
\be
\left(\frac{dB}{dt}\tilde{M}_s^{-1} - B \tilde{M}_s^{-1}\frac{d\tilde{M}_s}{dt} \tilde{M}_s^{-1} \right){}_a^{\ b} = W_a^{\ b} .
\ee
With a little work one can show that
\ba
\nonumber 4 W_a^{\ b} &=& - B_{ac}(\tilde{M}_s^{-1})^{cd}\left(  \Phi^{(0,0)}+\Phi^{(0,2)}+\Phi^{(2,0)}+ \Phi^{(1,1)}+\Phi^{(2,2)} + \Phi^{(3,1)} + \Phi^{(4,2)}      \right)_{de}(\tilde{M}_s^{-1})^{eb}\\
&&+ \left( \Phi^{(0,1)}+\Phi^{(1,0)}+\Phi^{(1,2)}+ \Phi^{(2,1)}+\Phi^{(3,1)}\right)_{ac}(\tilde{M}_s^{-1})^{cb}\ ,
\ea
and hence
\be
\frac{dB_{ab}}{dt} = \frac{1}{4} \left( \Phi^{(0,1)}+\Phi^{(1,0)}+\Phi^{(1,2)}+ \Phi^{(2,1)}+\Phi^{(3,1)}\right)_{ab}\,.
\label{RGB}
\ee
We can now see that the renormalisation group equations calculated in the duality invariant way in \eqn{RGG} and \eqn{RGB} are equivalent to the  renormalisation group equations obtained previously in \eqn{RGM}.   

We remark that whilst it is a small chore to obtain the expressions above in the duality invariant formalism it is a significantly more lengthy calculation to obtain the form of the $\Phi^{(i,j)}$ from expanding out the compact expression \eqn{rg2}.  It is somewhat interesting that there are  two very compact ways of writing the same result but demonstrating their equivalence is an extremely lengthy exercise.  It would be desirable to have a better understanding of the underlying geometrical reason for this equivalence.

\section{On-shell Lorentz invariant actions}

We now return to discuss the Lorentz invariance properties of the Poisson-Lie duality invariant action.   Instead of doing this directly we will consider the  most general form of a bosonic sigma model which has the Lorentz structure of  the duality invariant formalism.  We will examine the conditions under which such a theory has: a) first order equations of motion; b) emergent on-shell Lorentz invariance and c) an off-shell invariance under modified Lorentz transformations.  

Consider a general non-Lorentz invariant two-dimensional bosonic $\s$-model action
\be
S= \frac{1}{2}\int d\s d\tau \left( C_{MN} \del_0 X^M \del_1 X^N + M_{MN} \del_1 X^M
\del_1 X^N \right) ~ ,
\label{act1}
\ee
where the general matrix $C_{MN}$ and the
symmetric matrix $M_{MN}$ depend on the $X^M$'s.
It is straightforward to show that the Lorentz transformations
\be
\delta X^M = -\s \del_\tau X^M -\tau \del_\s X^M \ ,
\label{Lllor}
\ee
do not leave \eqn{act1} invariant, resulting instead in (discarding total derivatives)
\be
\delta_{\rm Lorentz} S = \int d\s d\tau \left( \ha S_{MN}(\del_1 X^M \del_1 X^N + \del_0 X^M \del_0 X^N)
+ M_{MN} \del_0 X^M \del_1 X^N \right)\ ,
\label{tralor}
\ee
where $S_{MN}$ is the symmetric part of $C_{MN}$, i.e. $S_{MN}=\ha (C_{MN}+
C_{NM})$.  

We would like to find the conditions under which \eqn{act1}
is on-shell Lorentz invariant, generalising similar considerations have been made in \cite{Tseytlin:1990nb} for constant matrices $C$ and $M$.   A word of caution is in order; obviously the variation of an action with regard to any arbitrary $\delta X$ is by definition zero when the equations of motion are satisfied.  What we are investigating is not trivial in this regard. Firstly we will not need to employ all the equations of motion to make the variation vanish and the way in which equations of motion are employed is very special.  Typically, equations of motion are second order, here we shall find conditions underwhich the equations of motion can be integrated to give first order equations.  A crucial feature is that the equations of motion themselves enjoy Lorentz covariance.  Finally, we shall see that the action \eqn{act1} is completely invariant under a set of modified transformations which, on shell, reduce to the standard invariance \eqn{Lllor}. 

We will split the bosons into $X^M=(X^\m,Y^i)$ and investigate conditions under which this anomalous variation can be removed using the 
motion corresponding to the $X^\m$'s.  Since the equations of motion
for the $Y^i$'s will not be used, we may add to the
Lagrangian corresponding to \eqn{act1} any Lorentz invariant term $Q_{ij}
\del_+ Y^i \del_- Y^j$ for some matrix $Q_{ij}$ that may depend on the
$Y^i$ spectator fields.

The variation of \eqn{act1} with respect to the $X^\mu$ is given by
\be
\delta S = \int d\s d\tau \ \delta X^\m \left(-\del_1 E_\m + \ha \del_\m M_{NP}
\del_1 X^N \del_1 X^P + \hat \G_{N;\m P} \del_0 X^N \del_1 X^P \right) ~ ,
\label{varact1}
\ee
where
\be
E_\m = S_{\m N} \del_0 X^N + M_{\m N} \del_1 X^N \ ~ ,
\label{eqEm}
\ee
and
\be
\hat\G_{N;\m \L} = \ha (\del_\m C_{N P} + \del_P C_{\m N} -\del_N C_{\m P})~ ,
\label{gmnl}
\ee
which as we will soon see, when it is appropriately restricted, plays the role
of connection.

Since \eqn{eqEm} and the equations of motions
are first and second order in worldsheet derivatives, respectively,
we should require that the latter can be written in the form $\del_1(\dots)$ in order to get
conditions that can be used to make \eqn{tralor} vanish.
However, the last two terms in \eqn{varact1} cannot be written immediately
in this form.  To proceed we require that the last two terms in \eqn{eqEm} can be cast into the form
\be
-\del_1 \L^\n_A \L^A_\m E_\n\ ,
\label{assup}
\ee
for some $X^\m$ dependent square matrix $\L^A_\m$ and its inverse $\L_A^\m$.
This is a very stringent condition with
severe consequences that restrict the type of backgrounds in the $\s$-models
that can be finally admitted by requiring
that Lorentz invariance emerges on-shell.
Then, assuming that \eqn{assup} holds, we find that \eqn{varact1} becomes
\be
\delta S = -\int d\s d\tau  \ \delta X^\m\left (\del_1 E_\m +\del_1 \L^\n_A \L^A_\m E_\n\right) =
-\int d\s d\tau  \ \delta X^\m \L^A_\m \del_1 (\L^\n_A E_\n)\ .
\label{jsg2}
\ee
Hence, the equations of motion can be integrated once and read
\be
\L^\n_A E_\n = f_A(\tau)\ ,
\label{invars}
\ee
where the $f_A(\tau)$ are otherwise arbitrary. However, \eqn{jsg2} shows that
\eqn{act1} has a local symmetry under the transformation
\be
\delta X^\m = \L^\m_A \e^A(\tau)\ ,
\ee
where $\e^A(\tau)$ are some $\tau$-dependent parameters. This can be used
to set $f^A(\tau)=0$ showing that the equations of motion are
first order and simply read
\be
E_\m = S_{\m N} \del_0 X^N + M_{\m N} \del_1 X^N  =0  ~ .
\label{eqEm2}
\ee

\subsection{Recovering on-shell Lorentz invariance}

Using \eqn{eqEm2}, we may solve for $\del_0 X^\m$ and after some algebraic manipulations recast the
anomalous Lorentz variation \eqn{tralor} into the form
\ba
&& \delta_{\rm Lorentz} S\Big |_{\rm on \ shell}  =  \int d\s d\tau \Big[\ha (S_{MN} - M_{M\a}S^{\a\b} M_{\b N}) \del_1 X^M \del_1 X^N
\nonumber\\
&& + (M_{iN} -S_{i\a} S^{\a\b}M_{\b N})\del_0 Y^i \del_1 X^N + \ha (S_{ij} -S_{i\a} S^{\a\b} S_{\b j})
\del_0 Y^i \del_0 Y^j  \Big]\ , \nonumber \\
&&
\ea
where $S^{\a\b}$ is the inverse matrix to $S_{\a\b}$.  Thus, to ensure that the Lornetz variation can be set to zero we find additional algebraic conditions 
\ba
S_{MN} & = & M_{M\a} S^{\a\b} M_{\b N} ~ ,
\nonumber\\
S_{ij} & = & S_{i\a} S^{\a\b} S_{\b j} ~ ,
\label{con2} \nonumber\\
M_{iN} & = & S_{i\a} S^{\a\b} M_{\b N} ~ .
\ea

Having established the on-shell Lorentz invariance of the action it remains to show the Lorentz invariance of the equations of motion  \eqn{eqEm2}. In order to do
that we define a set of projection operators as
\be
(P_\pm)^\m{}_\n = \ha (\delta^\m{}_\n \mp S^{\m\l} M_{\l\n})
\label{proojj}
\ee
and their invariant subspaces as
\be
(Q_\pm)^\m{}_i = \ha S^{\m\n} (S_{\n i} \mp M_{\nu i}) ~ .
\label{prjj}
\ee
Indeed, using \eqn{con2}, one can readily verify that they satisfy the required properties
\be
P_\pm^2 = P_\pm ~ ,~~~~~ P_\pm P_\mp = 0 ~ , ~~~~~ P_\pm Q_\pm = Q_\pm ~ ,
~~~~~ P_\pm Q_\mp =0 ~ .
\label{proop}
\ee
Then \eqn{eqEm2} with the definition \eqn{eqEm} can be written in the
form
\be
E_\m = S_{\m\n} (E_+^\n + E_-^\n) ~ ,
\label{empm}
\ee
where
\be
E_\pm^\m = (P_\mp)^\m{}_\n \del_\pm X^\n + (Q_\mp)^\m{}_i \del_\pm Y^i ~ .
\label{loreq}
\ee
Then using the properties \eqn{proop} the equations of motion \eqn{eqEm2}
can be easily shown to be equivalent to
\be
E_\pm^\m = 0~ ,
\label{epm}
\ee
which has the required form, since they remain invariant under Lorentz transformations.  Indeed, if we ignore the spectator $Y$ fields, the equations of motion for the $X^\mu$ are no more than the chirality condition
\be
d X^\mu = \ast S^{\mu \nu} M_{\mu \nu} d X^\nu \, , 
\ee
which is clearly Lorentz covariant and which we recognise as a slight generalisation of the chirality constraint obtained from the doubled formalism to the case of non-constant matrices $S \sim L $ and $M \sim \cR$.  

\subsubsection{The Off-shell Symmetry}
In addition to the on-shell Lorentz-symmetry \eqn{Lllor} we can construct some
modified Lorentz transformations under which the action is invariant off-shell.  These transformations are similar to those appearing in a simpler setting of \cite{Floreanini:1987as,Tseytlin:1990va,Tseytlin:1990nb}. 

These transformations are given by
\ba
\delta X^\m & =  & -\s \del_\tau X^\m - \tau \del_\s X^\m -\s (E^\m_+ + E^\m_-)~ ,
\nonumber \\
\delta Y^i & =  & -\s \del_\tau Y^i - \tau \del_\s Y^i ~ .
\label{lornew}
\ea
We see that the $Y^i$'s have the usual global Lorentz transformation rules.
On-shell the same is of course true for the $X^\m$'s as well.  Again this is best illustrated with the spectator fields turned off.  Then using \eqn{tralor} and \eqn{jsg2} we find that the variation of the action under these modified transformations is given by 
\be
\delta_S = \int d^2 \sigma \frac{1}{2} E_{\m} S^{\m \n} E_{\n} + \sigma E_{\a}S^{\a \b} \Lambda_{\b}^A \partial_1 \left( \Lambda_A^\gamma E_\gamma   \right)\, . 
\ee
Provided one identifies $\Lambda$ with the veilbein of the metric $S_{\a\b}$, which we shall shortly see is the correct thing to do, the second term cancels the first after an integration by parts. 

\subsection{Solving the Conditions}

We first solve the algebraic constraints \eqn{con2} before turing to the requirement that the equation of motion is first order.   We introduce a set of vielbeins $e_\m^A$ and their inverses $e^\m_A$ such that 
\be
S_{\m\n} = e^A_\m e^B_\n \eta_{AB}\ ,
\label{smn}
\ee
where $\eta_{AB}$ is the constant tangent space metric.  Similarly, let's choose matrices $M_{MN}$ such that
\be
M_{\m\n}  =  - R_{AB} e^A_\m e^B_\n\ ,
\label{smn1}
\ee
for some constant symmetric matrix $R_{AB}$.  The first of the constraints is solved \eqn{con2} providing that $R_{AB}$ obeys the compatibility condition
\be
R_{AC} \eta^{CD} R_{DB} = \eta_{AB} \ .
\label{rac}
\ee

Then, the remaining conditions in \eqn{con2} are solved by introducing a set of matrices $F^a_i$ that may depend on the spectator field $Y^i$ (but not on the internal coordinates $X^\mu$) and defining
\ba
S_{\m i} & = & \eta_{AB} e^A_\m F^B_i ~ ,
\nonumber\\
S_{ij} & =  & \eta_{AB} F^A_i F^B_j ~ ,
\nonumber
\\
M_{ij} & = & - R_{AB} F^A_i F^B_j ~ .
\\
M_{\m i} & = & - R_{AB} e^A_\m  F^B_i ~ .
\nonumber
\ea

We may choose for the matrix $C_{ij}$ to equal $S_{ij}$ since its antisymmetric part is Lorentz invariant by itself.

We now turn to the requirement that the equation of motion is first order namely
\be
\ha \del_\m M_{NP}\del_1 X^N \del_1 X^P + \hat \G_{N;\m P} \del_0 X^N \del_1 X^P  = -\partial_1 \Lambda^\nu_A \Lambda^A_\mu E_\nu\, .
\label{xx1}
\ee

\no
To proceed further, we found necessary to impose for the matrix $M_{MN}$ the conditions
\ba
\hat \nabla_\m M_{\n\l} & = & 0 ~ ,
\nonumber
\\
\hat \nabla_\m M_{\n i} & = & 0 ~ ,
\label{con41} \\
\del_\m M_{ij} &  = & 0 ~ ,
\nonumber
\ea
where the derivatives are covariant with respect to a connection defined as 
\be
\hat{\G}^\m_{\n \l } = S^{\m\r}\hat \G_{\r;\n\l}\ .
\ee
 In the covariant differentiation the index $i$ is assumed not to transform.  We also demand that 
 \be
 \hat \G_{\m;\n i} = \hat \G_{j;\n i} =0\, , \quad \hat{\Gamma}^{\tau}_{\m \n} S_{\tau i} = \hat{\Gamma}_{i ; \m \n}\, , \quad \partial_i C_{\mu\nu} = 0\, . 
 \label{con5}
 \ee
 An immediate consequence of these constraints is that $\partial_{[\mu} C_{\nu] i} = 0$ which we shall solve by simply assuming that $C_{\mu i }=0$.  We then have 
\be
\frac{1}{2}\del_\m C_{i\l} = S_{\n i} \L^\n_A \del_\l \L^A_\m\ .
\label{jsdh}
\ee
 
When the conditions (\ref{con41}, \ref{con5}) are satisfied we find that \eqn{xx1} becomes
\be
\hat{\Gamma}^\tau_{\mu \nu} E_\tau \p_1 X^\nu =  -\partial_1 \Lambda^\nu_A \Lambda^A_\mu E_\nu\, ,
\ee
and hence
\be
\label{xx2}
\hat{\Gamma}^\tau_{\mu \nu}  = \Lambda^\tau_A \partial_\n \Lambda_\mu^A\, .
\ee
 We can evaluate the connection to find
 \be
 \hat{\G}^\tau_{\mu \nu} =  \G^\tau_{\mu \nu} - \frac{1}{2}S^{\tau\rho} H_{\rho \mu \nu}
 \ee
 where  $\G^\tau_{\mu \nu}$ are the components of the usual symmetric Levi--Cevita connection built from the $S_{\m \n}$ and $H_{\rho \mu \nu} = \partial_\rho B_{\m \n} + \partial_\n B_{\rho \m }+\partial_\m B_{\n \rho}$ with $B_{\m \n}$ the antisymmetric part of $S_{\m \n}$.  We then find that the relation \eqn{xx2} gives 
  \be
 \G^\tau_{\mu \nu} = \Lambda^\tau_A \partial_\mu \Lambda_\nu^A -  \frac{1}{2}S^{\tau\rho} H_{\rho \mu \nu}\, ,
 \ee
and therefore  
\be
\boxed{ \del_\m \L_\l^A - \del_\l \L_\m^A = \L^A_\n H^\n{}_{\m\l}\ .}
\label{rra}
\ee

 Using the form of $M_{\mu\nu} =-e_\mu^A R_{AB}e_\nu^B$ we find that the first constraint of  \eqn{con41} is solved when the spin connection
 \be
\omega_\mu{}^A{}_B = \Gamma_{\mu \nu}^{\tau} e_\tau^A e_B^\nu - e^\tau_B \partial_\mu e_\tau^A
\ee
is related to the torsion via 
\be
\boxed{ \omega_\m{}^{AB} +\ha H_\m{}^{\n\l} e^A_\n e^B_\l = 0 \ . } 
\label{omh}
\ee

These two boxed equations remind us of the Maurer-Cartan equation and spin connection of a group manifold with the coordinates $X^\mu$ parameterising a group element $g\in G$.  Let us see this precisely by introducing generators, an inner product and left and right invariant forms on the group:
\bea
[T_A, T_B] = if_{AB}{}^C T_C\, , &&   \langle T_A | T_B \rangle = \eta_{AB} \, , \nonumber\\
L^A_\mu = -i \langle g^{-1}dg | T_B \rangle \eta^{AB} \, , & ~~ &  R^A_\mu = -i \langle dg g^{-1} | T_B \rangle \eta^{AB} \, . 
\eea
The left and right invariant forms obey the Maurer-Cartan equations
\bea
\partial_{\mu} L^A_{\nu} - \partial_{\nu} L^A_{\mu} &=& f_{BC}{}^A L_\mu^B L_\nu^C \, , \\ 
\partial_{\mu} R^A_{\nu} - \partial_{\nu} R^A_{\mu} &=& - f_{BC}{}^AR_\mu^B R_\nu^C \, ,
\eea
and can be related by
\be
R^A_\mu = \eta^{AB}\langle T_B | g T_C g^{-1} \rangle L^C_\mu = \eta^{AB} N_{BC}  L^C_\mu \, .
\ee
A couple of helpful identities are 
\be
\eta^{EC}N_{DC} \eta^{DA} N_{AB} = \delta_B^E\, , \quad \partial_\mu N_{AB} =-L^C_\mu f_{CB}{}^D N_{AD}\, .
\ee

We can now choose whether we wish to associate the veilbein $e_\mu^A$ with the left or the right Maurer-Cartan form.  For each case the relationships above determine the torsion, the spin connection and the matrix $\Lambda$ uniquely. Choosing the left invariant form we have 
\bea
 e_\mu^A = L^A_\mu\, , &\quad &\omega_\mu{}^{AB} = \frac{1}{2} f_E{}^{AB}L^E_\mu\, , \nonumber \\ H_{\mu\nu\lambda}= -f_{ABC} L^A_\mu L^B_\nu L^C_\lambda \, ,& \quad &  \Lambda_\mu^A = R_\mu^A  \, , 
\eea
whereas for the right invariant form 
\bea
e_\mu^A = R^A_\mu\, , &\quad& \omega_\mu{}^{AB} = - \frac{1}{2} f_E{}^{AB}R^E_\mu\, , \nonumber \\ H_{\mu\nu\lambda}= + f_{ABC} R^A_\mu R^B_\nu R^C_\lambda \, , &\quad &   \Lambda_\mu^A = L_\mu^A \, . 
\eea
 The relative signs in the above equations are crucial in order that the equations of motion can be written in first order form.   In performing our background field expansion for the Poisson Lie duality invariant action we found that if the wrong sign choice for torsion was used the contributions to the Lorentz anomaly do not  cancel.   So whilst the above is somewhat of a technicality it turns out to be an important one.  When dealing with a standard WZW model one does not need to take so much care; one can define the torsion with either sign relative to the spin connection and still find the same result since the generalised curvature vanishes in both cases.   

For concreteness we choose the first case $e_\mu^A= L_\mu^A$.  Using the identities for the matrix $N_{AB}$ one finds that 
\bea
C_{i \m } & = & 2 \eta_{AB} F^A_i e^B_\m  ~  ,
\label{cim}
\eea
 solves the requirement   \eqn{jsdh}. 

Putting everything together we find that the on-shell Lorentz invariance conditions (first order equation of motion) are solved by an action with Lagrangian:
\ba
 \cL & = &   \cL_{\rm WZW}  + \ha  \eta_{AB}\left(
F^A_i F^B_j \del_0 Y^i \del_1 Y^j + 2 F^A_i L^B_\m  \del_0 Y^i \del_1 X^\m\right)
\nonumber\\
&&
+ \ha  R_{AB}\left(- L^A_\m L^B_\n \del_1 X^\m \del_1 X^\n
- F^A_i F^B_j \del_1 Y^i \del_1 Y^j - 2 L^A_\m F^B_i \del_1 X^\m \del_1 Y^i
\right)\ , \qquad \
\label{actwzw}
\ea
where $\cL_{\rm WZW}$ corresponds to a WZW model action rotated such that $\sigma$ and $\tau$ play the role of light cone coordinates.  

When the group is a Drinfeld Double this action is exactly that of the duality invariant action for Poisson--Lie T-duality with $Y^i$ `spectator' fields as described in e.g. \cite{Sfetsos:1996xj}.   
However, we emphasize that, unlike the discussions related to Poisson-Lie T-duality, the group $D$ need not necessarily be a Drinfeld Double.   We remark that our results for the one-loop Weyl anomaly and Lorentz anomaly did not seem to make use of the fact that the group was a Drinfeld Double and we anticipate that they hold in more generality. For example, the application to generalised T-duality and doubled twisted tori \cite{Hull:2009sg,ReidEdwards:2010vp} has been explored in \cite{Avramis:2009xi}.

\section{Discussion and Conclusions}
We showed that demanding on-shell Lorentz invariance highly
constrains the structure of a general Lorentz non-invariant action.
The resulting theories have an underlying group structure and consist
of a WZW term together with some interaction term.   In the case that
the group is a Drinfeld Double these theories are \PL\ invariant $\s$-models.
 
 By using a background field method we calculated the one-loop effective
 action of these models and found that they were renormalisable and that
a possible quantum Lorentz vanished.  This is an important consistency
condition of such models.   We also obtained the renormalisation group
equations for the couplings of the interaction term in our model.
For the \PL\ invariant theories this provides a duality invariant
description of the RG equations, a key motivator for this work.
 We also verified that for specific examples of the Drinfeld Double
that these duality invariant RG equations agreed with those obtained
from either of the T-dual standard $\s$-models.  This agreement was extended algebraically to prove equivalence in all generality. 

An observation that hints at the utility of
a duality invariant framework is that duality invariant
RG equations can be computed using simple contractions of structure constants
with constant matrices.
When dealing with the standard $\s$-model one obtains these equations
only by calculating the generalised curvature of what are, in general, extremely complicated
target space backgrounds.

\chapter{Conclusions and outlook}
T-duality is a striking and fundamental property of String Theory which distinguishes it from other quantum field theories. Among its many facets, T-duality can be realised as a canonical transformation in phase space.  We saw that this canonical approach to the cases of bosonic T-duality in Ramond backgrounds and also to Fermionic T-duality. 

 There is a growing body of evidence which supports the idea of using T-duality to construct consistent non-geometric or T-fold backgrounds which challenge our conventional notions of geometry.   
String Theory can be reformulated in a way which promotes T-duality to the level of a manifest symmetry of a classical action.   Such an approach could be vital to gain a full understanding of these novel non-geometric string backgrounds.  Moreover, the philosophy of making symmetries manifest in an action seems extremely natural and a potential pathway to uncovering some deeper understanding of String Theory. 

In this thesis we have examined in some detail one such approach: the Doubled Formalism.   In this thesis we have added some substantial support to the full quantum validity of such approaches.

We have been able to calculate the background field equations related to the Weyl divergence  for the Doubled Formalism.  This computation was made complex  by the need to supplement the extra degrees of freedom in the Doubled Formalism with a constraint.  We used the PST procedure to implement the constraint at the level of the Lagrangian.  Upon gauge fixing we found an action where Lorentz invariance had been lost, a feature characteristic of many duality invariant approaches.   Working without manifest Lorentz invariance required being careful of both Lorentz and Weyl anomalies in the effective action.  For the Doubled Formalism we found two crucial results:
\begin{enumerate}
\item Global Lorentz anomalies cancel non-trivially at one-loop
\item The background field equations that need to be satisfied for the vanishing of the Weyl anomaly are equivalent to those of the conventional String Theory in a toroidally compactified background. 
\end{enumerate}
These two results are important indicators of the consistency of the Doubled Formalism.  

There are many ways to directly extend this research, for example,  the inclusion of higher loop effects and additional `spectator' fields. Within the calculation approach used in this thesis such generalisations may be  rather time consuming and it does seem, at least in this author's opinion, rather probable that the validity will continue to hold.  

Generalising T-duality beyond the Abelian situation is notoriously difficult due to complications with global topological effects. Nevertheless, Poisson--Lie T-duality represents an import and very interesting step in that direction.   We investigated the effective action of the duality invariant formalism for this case of Poisson--Lie T-duality. Once again we found that the formulation appears consistent: Lorentz anomalies cancel and Weyl divergences match those of the conventional T-dual related sigma models. 

It is somewhat remarkable that the results for the beta functions could, in both cases, be expressed compactly in terms of the duality invariant theory.  To show the equivalence with the conventional sigma models requires a large amount of manipulation.   It would be highly desirable to have a more profound conceptual way of understanding how expressions without manifest duality invariance can be shown to match those with duality invariance. 
To this end perhaps one needs to adopt a different geometric perspective for example the doubled geometry of Hitchin \cite{Hitchin:2004ut,Gualtieri:2003dx}.

We close with a few remarks about possible future directions of research in the subject. One avenue of research would be the construction of explicit non-geometric  string backgrounds and understanding of their associated phenomenology.  A different direction is to consider applications of the duality invariant philosophy to other branches of theoretical physics.  For example, perhaps the formulation of a Doubled Formalism for fermionic T-duality might help shed further light on the Wilson-loop/amplitude connection in supersymmetric gauge theory.  Indeed one might even hope to reformulate the actual gauge theory so that both conformal and dual superconformal symmetries are manifest.  It is interesting to ask whether such an approach can be used to constrain and determine amplitudes.   Another possibility is the extension of duality invariance to M-theory -- a duality invariant approach may provide insights into fundamental structure of M-theory.

\begin{appendix}
\chapter{Conventions}
\section{String $\sigma$-model Conventions}
The worldsheet action for a string is given, in conformal gauge, by
\be
S = \frac{1}{4\pi \alp} \int d^2 \sigma \eta^{\a \b} \partial_\a X^I \partial_\b X^I\, ,
\ee
in which the constant $\alp$  is proportional to the inverse string tension.  The world-sheet is defined by coordinates $(x^0, x^1) \equiv(\tau, \sigma)$ and the metric is given by $\eta_{00}=-\eta_{11}=1$. 

 We define light cone coordinates as $\s^\pm= \frac{1}{2} (\tau\pm \s)$ and hence $\del_\pm = \del_0 \pm \del_1 $, where $\del_0 = \del_\tau$ and $\del_1= \del_\s$.  Occasionally we use the complex world sheet coordinates obtained by performing a Wick rotation $i \tau =  t$ and defining 
$z=\sigma + i t$ and $\partial =  \frac{1}{2} \left( \partial_\sigma - i \partial_t \right)$. The reader should note that these conventions do differ in parts from the main text books but have been chosen so as to be in maximal agreement with the literature on T-duality and in particular with \cite{Hull:2004in,Berman:2007yf,Berman:2007xn,Tseytlin:1990nb,Tseytlin:1990va}.

The two-dimensional $\sigma$-model describing a bosonic string with curved worldsheet coupled to background fields is given by 
\be
S=\frac{1}{4\pi \alp} \int d^2 \sigma \left[\sqrt{g} g^{\a\b} G_{IJ} \p_\a X^I \p_\b X^J - \e^{\a\b}B_{IJ}\p_\a X^I \p_\b X^J  + \frac{\alp}{2} \sqrt{g} \phi {\cal R}^{(2)}  \right]\, ,
\ee
where $(G,B,\Phi)$ are the target space metric, 2-form Kalb-Ramond potential and dilaton respectively.  The antisymmetric tensor density $\e^{\a\b}$ is defined by $ \e^{01}= - \e^{10} = 1$.  Note that the Dilaton normalisation contains a factor of two, again this is a common normalisation in T-duality literature used in e.g. \cite{Giveon:1994fu}. 

For applications to T-duality it is quite helpful to rescale fields so that they become dimensionless.  For example, consider the theory of a single compact boson describing an $S^1$ target of radius $R$. Then the worldsheet theory is given by
\be
S=  \frac{1}{4\pi \alp} \int d^2 \sigma  \partial_\a X \partial^\a X\, ,
\ee
where $X$ is a dimensional coordinate with periodicity $X\simeq X+2\pi R n$  (where $n \in \mathbb{Z}$ describes the number of times the string is wound round the $S^1$).   By performing the rescaling  $X' = X/R$ we can work with a dimensionless field whose periodicity is $2\pi$. Then the action becomes
\be
S=  \frac{1}{4\pi \alp} \int d^2 \sigma  R^2 \partial_\a X' \partial^\a X'\, ,
\ee
but now the target space metric is dimensional. A final rescaling is to work with the dimensionless radius $R' = R/\sqrt{\alp}$ yielding the final result
\be
S = \frac{1}{4\pi} \int d^2 \sigma  R'^2 \partial_\a X' \partial^\a X'\, .
\ee 
T-duality is often described in terms of these rescaled fields since the duality transformation becomes simply $R' \rightarrow 1/R'$.   When factors of $\alp$ are not explicit it is because the above type of redefinitions have been used. Also,  in the treatment of the Doubled Formalism a factor of $2\pi$ is suppressed throughout for convenience.  

With these conventions and these normalisations, the correct T-duality transformation for the dilaton is given by
\be 
\phi^\prime \rightarrow \phi - 2 \ln R\, ,\\
\phi^\prime \rightarrow \phi - \frac{1}{2} \ln \frac{\det G}{\det G^\prime} \, .
\ee

\newpage 
\section{Notation}
\begin{table}[htdp]
\caption{Notation used in this thesis and the chapters to which it refers}
\begin{center}
\begin{tabular}{|c|c|c|}
\hline
 & Description & Chapters\\
 \hline
$\sigma^\a\, , \partial_\b$ & string world-sheet coordinates and derivatives  & all\\
$X^I$ &  generic target space coordinates   & 2\\
$Z^M$ & generic target superspace coordinates & 2 \\
$X^i$ & coordinates on a $T^d$ fibre  & 3,4\\
$\tilde{X}_i$ & T-dual coordinates of $T^d$ fibre  & 3,4\\
$Y^a$ &  coordinates on the base of a fibre-bundle  & 4 \\
$\X^I = ( X^i, \tilde{X}_i )$ &  coordinates of doubled torus $T^{2d}$   & 3,4\\
$X^\a = (\X^I , Y^a)  $ & coordinates of doubled torus bundle    &  4 \\ 
$\X^{\bar{A}} $ & chiral tangent frame coordinates of doubled torus $T^{2d}$  & 3,4\\ 
$Y^{\bar{a}} $ & tangent frame  coordinates of base of fibre-bundle  & 3,4\\ 
$X^{\bar{\a}} $ & chiral tangent frame coordinates of doubled torus bundle & 3,4\\ 
$\phi$ & standard dilaton field & 2,3,4  \\
$\Phi$ & duality invariant dilaton & 3,4 \\
$\cM_{\a\b}, \cR_{\a\b}, \cD_{\a}$ & Geometric quantities on doubled bundle & 4  \\
$g_{ab}, \hat{R}_{ab}, \hat{\nabla}_a$ & Geometric quantities on base only & 4\\
$T_A$ & Generators of Drinfeld Double ${\cal D} = \cG \oplus \tilde{\cG} $ & 5 \\
$T_a\, , \tilde{T}^a$   & Generators of  $\cG\, ,  \tilde{\cG}  \subset  {\cal D} $  & 5 \\
$L_I^A$ & Left invariant forms on double  Drinfeld Double group manifold & 5 \\
$L^a_i\, , \tilde{L}^i_a$   & Left invariant forms  of subgroups $G, \tilde{G}$ & 5 \\
\hline
\end{tabular}
\end{center}
\label{default}
\end{table}%

\chapter{Chiral Fields}
Chiral bosonic fields which are defined in $d=2p+2$ dimensions as p-form potentials whose field strength is self-dual,  have been encountered in many areas of theoretical physics including the five-form Ramond field strength  of IIB supergravity (and Superstring Theory) and the self-dual two-form contained in the $\cN= (2,0)$ tensor multiplet living on an M-theory fivebrane worldvolume.  Chiral fields are also very important in duality-symmetric theories, not just the T-duality invariant models considered in this thesis but also in four dimensional electromagnetic duality symmetric models. 

In this appendix we provide some detail of the treatment of chiral bosons in two-dimensions used in this thesis.  We start with a theory of a free boson 
\be
\cL_0  = \frac{1}{2} \partial_\mu \phi \partial^\mu \phi
\ee
 and wish to supplement this with a constraint that
\be
C= \partial_- \phi =  \dot{\phi}- \phi^\prime = 0 \, . 
\ee
From a Hamiltonian perspective the treatment of this constraint is slightly subtle.  In Dirac terminology this is a second class constraint\footnote{A complete presentation of constrained systems can be found in \cite{henneauxbook}.} since under the Poisson bracket\footnote{Equal time dependance is implicit in the equations (\ref{X1}, \ref{X2}).}
\be
\label{X1}
[\pi(x), \phi(y)  ]_{P.B.}= \delta(x-y) 
\ee
the constraint does not commute (even weakly):
\be
\label{X2}
[C(x), C(y) ]_{P.B.} = 2\delta^\prime(x-y) \, . 
\ee
At the Hamiltonian level one can proceed by quantisation using Dirac brackets and for the Doubled Formalism this is the approach taken in \cite{HackettJones:2006bp}.   To incorporate such a second class constraint into a Lagrangian description is some what tricky; an immediate mechanical problem is that the constraint is first order where Euler-Lagrange equations are typically second order in derivatives.  An early approach was proposed by Siegel \cite{Siegel:1983es} in which the constraint is squared and then implemented via a Lagrange multiplier
\be
\cL_{Siegel}  =  \frac{1}{2} \partial_+ \phi \partial_- \phi  - \lambda_{++} (\partial_- \phi)^2\, , 
\ee
however, there are some difficulties in quantising this system (see e.g.  \cite{Chiralprob1, Labastida:1987zy}).   A different approach was proposed by Floreanini and Jackiw \cite{Floreanini:1987as},  and further developed in \cite{Sonnenschein:1988ug}, with the Lagrangian density\footnote{In the presentation of \cite{Floreanini:1987as} the action (\ref{X3}) is actually derived as a non-local description of another chiral theory, however it serves equally well to describe chiral dynamics.}
\be
\label{X3}
\cL_{FJ} =  \phi^\prime \dot{\phi} - (\phi^\prime)^2 =  \partial_- \phi \partial_1 \phi\, . 
\ee
The action corresponding to (\ref{X3}) has both time translation and spatial translational symmetries with Noether charges related as $H=-P$ which is to be expected for chiral dynamics.  Although (\ref{X3}) is not manifestly Lorentz invariant there exists a modified Lorentz transformation
\be
\delta_M \phi = (t+x)\phi^\prime\, . 
\ee
The local invariance 
\be
\delta_\alpha \phi = \alpha(t) 
\ee
means that the equations of motion that follow from (\ref{X3}), namely 
\be
\partial_1 \partial_- \phi = 0 
\ee
may be integrated to give solutions which gauge equivalent to the desired chirality constraint $\partial_- \phi = 0$.   It is worth noting that the modified Lorentz invariance is, on-shell, equivalent to the regular Lorentz transformation
\be
\delta_{\tilde{M}} =  t  \phi^\prime + x \dot{\phi}
\ee
and so Lorentz transformations can be viewed as an emergent, on-shell symmetry (or solution generating symmetry) of (\ref{X3}). 

A formulation of chiral fields which has manifest Lorentz invariance was introduced by Pasti, Sorokin and Tonin \cite{Pasti:1996vs}.  Lorentz invariance is made possible by the introduction of auxiliary fields.   In the case of two dimensions, one introduces an auxiliary field $u$ which is a closed one form written locally as $u_\alpha= \partial_\alpha a$ for some scalar $a$.   The action is given by
\bea
\cL_{PST} &=& \frac{1}{2} \partial_+ \phi \partial_- \phi - \frac{1}{2}\frac{ \partial_+ a}{\partial_- a} (\partial_- \phi)^2\\
&=&   \frac{1}{2} \partial_+ \phi \partial_- \phi - \frac{1}{2 u_\a u^\a} ( u^\b \cP_\b )^2
\eea
where, to explicitly show the manifest Lorentz covariance in the second line,  we have introduced a one-form $\cP$ whose components are $\cP_0 = \p_- \phi= - \cP_1$ which vanish on the desired constraint.  The non-polynomial form of the PST action means that the action is actually different to the Siegel form.     The canonical momenta of this system obey a first class constraint which is related to the PST local symmetry 
\be
\delta a = \zeta \, , \quad \delta \phi = \frac{ \partial_- \phi}{\partial_- a} \zeta \, . 
\ee
By fixing the PST symmetry $u_\a = \delta_\a^0$ one recovers the FJ action (\ref{X3}). One must be a little careful in defining the field space of $u_\a$ to ensure that $u^2 \neq 0$ and to only consider gauge fixing choices that respect this.  Also, when asking questions of a topological nature one must remember that $u$ need only be closed and need not be exact.   The equation of motion that comes from varying $\phi$ is 
\be
\partial_{+-}\phi - \partial_-\left( \frac{\partial_+ a }{\partial_- a} \partial_- \phi \right) \propto  \e^{\a\b}\partial_\a \left(\frac{1}{u^2} u_\b u^\gamma \cP_\gamma \right)=0
\ee
and the equation of motion that comes from variation of $a$ is proportional to this result. It is shown in \cite{Pasti:1996vs} how the desired chirality constraint is obtained as a consquence of this equation. 
\chapter{Background Field Method}
In this appendix we provide some details of the calculation of the effective action obtained from an $\a^\prime$ expansion of String Theory sigma-models using the geometric background field method.  We introduce the necessary geometrical equations and  we show some details of the regularisation and renormalisation which are assumed in the thesis. 

We begin by establishing some geometrical concepts that are relevant for string sigma models.  When the string is coupled to background fields by the action 
\be
\label{C1}
S=\frac{1}{4\pi \alp} \int d^2 \sigma \left[\sqrt{g} g^{\a\b} G_{IJ}(X) \p_\a X^I \p_b X^J - \e^{\a\b}B_{IJ}(X)\p_\a X^I \p_b X^J   \right]\, ,
\ee
one can observe that in the equations of motion, $B$ enters only through its  field strength 
\be
H_{IJK} = \partial_I B_{JK} + \partial_J B_{KI} +\partial_K B_{IJ} \, . 
\ee
Geometrically $H_{IJK}$ can be interpreted as torsion in a Riemann-Cartan space by defining the non-symmetric connection
\be
\hat{\Gamma}^{I}_{\ JK} = \Gamma^{I}_{\ JK} - \frac{1}{2}G^{IL}H_{LJK} \, . 
\ee
The covariant derivatives constructed from this connection yield the algebra
\be
[\hat{D}_{I}, \hat{D}_{J}] V_{K} = \hat{R}^L_{IJK}V_L + G^{LM}H_{LIJ}\hat{D}_M V_K
\ee
with the curvatures given by 
\bea
 \hat{R}_{KLIJ}&=& R_{KLIJ} - \frac{1}{2} D_{I} H_{KJL} + \frac{1}{2} D_J H_{KIL} + \frac{1}{4} H_{KIM}H^M_{\ JL}-\frac{1}{4} H_{KJM}H^M_{\ IL} \, , \quad \quad \\ 
 \hat{R}_{JL} &=&R_{JL} + \frac{1}{2} D^{K} H_{KJL}  -\frac{1}{4}H^{MK}_{\phantom{MK}J}H_{MKL} \, . 
\eea
The action of these modified derivatives on a contravariant vector $\xi^I$ can be pulled back to the worldsheet by defining\footnote{Note the definition of this derivative is slightly altered in Chapter 5  due to the Lorentz index structure of the sigma model (\ref{act1}).}   
\be
\hat{D}_\a \xi^I = \partial_\mu \xi^I + \Gamma^{I}_{\ JK} \partial_\mu  X^J \xi^K - \e_{\a\b} \frac{1}{2}H^{I}_{\ JK}\partial^\beta X^J \xi^K\, .
\ee
These formulae are heavily used in the derivation below and also in the renormalisation of the Poisson-Lie T-duality model.

We now wish to calculate the effective action of this sigma model obtained by expanding quantum fluctuations around a classical solution: 
\be
e^{i\Gamma[X_0]} = \langle e^{iS[X_0+ \pi] }\rangle \, . 
\ee
We shall calculate as a perturbation series in $\alpha^\prime$ and in the following analysis consider only 1-loop diagrams to obtain the leading order result  \cite{AlvarezGaume:1981hn,Mukhi:1985vy,Braaten:1985is}.  Instead of expanding in the flucuation $\pi^I$, a more geometric approach is to expand in the tangent $\xi^I |_{X_0}$ to the geodesic connecting $X^I_0$ and $X^I_0 + \pi^I$ as indicated in the figure.   

\begin{figure}[!ht]
\label{geometricexpansion}
\begin{center}
\scalebox{0.75}{\includegraphics{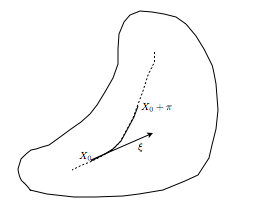}}
\end{center}
\caption{\em Geoemetric interpretation of quantum flucuations}
\end{figure}
 
The geodesic is given by $\rho(s)$ such that $\rho(0) = X_0$, $\rho(1) = X_0 + \pi$ and obeying the geodesic equation
\be
 \ddot{\rho}^I + \Gamma^I_{JK} \dot{\rho}^J \dot{\rho}^K = 0 \, . 
\ee
Making multiple use of the geodesic equation gives
\be
\rho(s)^I = X_0^I + \xi^I s -\frac{1}{2} \Gamma^I_{JK} \xi^J \xi^K s^2 + \frac{1}{3!} \Gamma_{JKL}^I   \xi^J \xi^K \xi^L s^3 + \dots
\ee
where $\Gamma^I_{JKL}$ is the covariant derivative of the Christoffel symbol acting only on lower indices.  Evaluated at $s=1$ gives
\be
\pi^i =  \xi^I  -\frac{1}{2} \Gamma^I_{JK} \xi^J \xi^K + \dots .
\ee
One can use a choice of coordinates for which $\pi^I = \xi^I$ to perform the expansion (these are known as Riemann Normal Coordinates) and then translate the results of the expansion to hold in arbitrary coordinates.   Expanding in this way allows one to write $S[X_0 + \pi]  = \sum_{n} S^{(n)}[X_0, \xi] $ where the expansion is in powers of the fluctuation.   For the sigma model above (\ref{C1}) we have that 
\bea
S^{(0)}[X_0] &=& S[X_0]\, , \quad  S^{(1)}[X_0] \simeq 0\, , \\
S^{(2)}[X_0] &=& \frac{1}{4\pi \alp} \int d^2 \sigma\, G_{IJ}(X_0) \hat{D}_\a \xi^I \hat{D}_\b \xi^J + \hat{R}_{IJKL}\xi^J\xi^K \left( \eta^{\a\b} - \e^{\a\b}  \right) \partial_\a X_0^I \partial_\b X_0^L \, , \nonumber \label{secondorder}\\   
\eea   
where the linear term vanishes since $X_0$ obeys the classical equations of motion.   In order to  calculate Feymann diagrams it is helpful to put the kinetic term into a conventional form which is done by introducing vielbeins  $V_{I}^{\ \bar{A}}$  and gauge fields $B_\a^{\ \bar{A}\bar{B}}= -B_\a^{\ \bar{B}\bar{A}} $ such that 
\bea
G_{IJ}(X) &=& V_{I}^{\ \bar{A}} V_{J}^{\ \bar{B}} \delta_{\bar{A}\bar{B}} \, .\\
 V_{I}^{\ \bar{A}} \hat{D}_\a \xi^I &=& \partial_\a \xi^{\hat{A}} + B_\a^{\ \bar{A}\bar{B}}\xi^{\bar{B}} \, .
\eea
The quadratic term in the expansion (\ref{secondorder}) is then
\ba
S^{(2)}[X_0] &=& \frac{1}{4\pi \alp} \int d^2\sigma\,  \partial_\a \xi^{\bar{A}} \partial^\a \xi^{\bar{B}} + 2 B_\a^{\ \bar{A}\bar{B}} \xi^{\bar{A}}\partial^\a \xi^{\bar{B}} + B_\a^{\ \bar{A}\bar{C}}B^{\a  \bar{B}\bar{C}} \xi^{\bar{A}}\xi^{\bar{B}}\, \nn  
\\&& \quad \quad \quad  + \hat{R}_{I\bar{A}\bar{B}J}\xi^{\bar{A}}\xi^{\bar{B}} \left( \eta^{\a\b} - \e^{\a\b}  \right) \partial_\a X_0^I \partial_\b X_0^J\, . 
\ea
The propagator is now standard and given by
\be
\langle \xi^{\bar{A}}(\sigma)  \xi^{\bar{B}}(\sigma^\prime)  \rangle = \frac{i \mu^{-\epsilon} \alpha^\prime \delta^{\bar{A}\bar{B} }}{(2 \pi)^{2+\e} } \int d^{2+\e} p\,  \frac{e^{i p(\sigma - \sigma^\prime)}}{p^2 - m^2}\, , 
\ee 
in which we have continued to $d=2+\epsilon$ dimensions to regulate the UV behaviour of integrals and introduced a mass parameter which is taken to zero at the end of calculations to regulate the IR divergences.  At one loop there are three diagrams that contribute to the effective action illustrated schematically in the figure below.
\begin{figure}[!ht]
\label{geometricexpansion}
\begin{center}
\scalebox{0.5}{\includegraphics{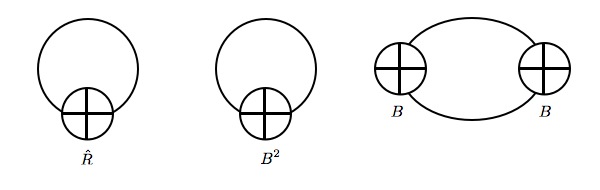}}
\end{center}
\caption{\em Contributing one-loop diagrams}
\end{figure}

In fact, it turns out that divergent contributions cancel out between the two diagrams involving the gauge field $B$ \cite{Mukhi:1985vy,Braaten:1985is}.  The first diagram contains the integral 
\bea
\langle \xi^{\bar{A}}(\sigma)  \xi^{\bar{B}}(\sigma)  \rangle & = &\frac{i \mu^{-\epsilon} \alpha^\prime \delta^{\bar{A}\bar{B} }}{(2 \pi)^{2+\e} } \int d^{2+\e} p\,  \frac{1}{p^2 - m^2}\, ,  \nn 
\\
&=& \frac{i \alp  \delta^{\bar{A}\bar{B} }}{4\pi} \left( \frac{1}{\epsilon} - \frac{1}{2} \gamma_E + ln \frac{m}{\mu} \right)
\eea
and provides a contribution
\bea
&& \frac{1}{4\pi \alp} \int d^2 \sigma  \hat{R}_{I\bar{A}\bar{B}J} \left( \eta^{\a\b} - \e^{\a\b}  \right)\partial_\a X_0^I \partial_\b X_0^J \langle \xi^{\bar{A}}(\sigma)  \xi^{\bar{B}}(\sigma)  \rangle\\
&=& \frac{1}{4 \pi  \epsilon } \int d^2 \sigma \hat{R}^{\phantom{I} \bar{A}} _{I\phantom{\bar{A}}\bar{A}J} \left( \eta^{\a\b} - \e^{\a\b}  \right) \partial_\a X_0^I \partial_\b X_0^J\, .
\eea   
These divergences can be absorbed through counter terms into the couplings $G_{IJ}$ and $B_{IJ}$. 

One can define renormalised couplings (or matrices of couplings) given to this order in perturbation theory as
\bea
G^{R}_{IJ }(\mu) = G_{IJ} + \alp \ln (\mu) \hat{R}_{(IJ)} + \dots \, , \\ 
B^{R}_{IJ }(\mu) = B_{IJ} + \alp \ln (\mu) \hat{R}_{[IJ]} + \dots \, ,
\eea  
 and corresponding beta-functions (or rather beta-functionals to signify the implicit $\sigma$ dependance) \cite{Mukhi:1985vy,Braaten:1985is}:
 \bea
  \beta[G_{IJ}] &=&\frac{ \partial G^{R}_{IJ}}{\partial \ln \mu} = \alp \hat{R}_{(IJ)} = \alp\left( R_{IJ} -\frac{1}{4}H^{MK}_{\phantom{MK}I}H_{MKJ}\right) \, , \\
 \beta[B_{IJ}]&=& \frac{ \partial G^{R}_{IJ}}{\partial \ln \mu} = \alp \hat{R}_{[IJ]} =- \frac{\alp}{2} D^{K} H_{KIJ}\, .
 \eea
Practically one simply reads off the beta-functions directly from the simple poles in the dimensional regularisation parameter.   The non-triviality of the beta-functions show that whilst the sigma model (\ref{C1}) is classically conformal it is not conformal at the quantum level.  In string theory we therefore demand that these beta functions identically vanish.   
 
 Including a dilaton term
 \be
 S_{dil} = \frac{1}{8\pi} \int d^2 \sigma \phi(X) R^{(2)}\, ,
 \ee
 into the action produces extra contributions to the above one-loop beta functions and also a dilaton beta function (which requires a two-loop calculation to get the leading order result in $\alp$) \cite{Callan:1985ia}:
 \bea
  \beta[G_{IJ}] & =&\alp\left( R_{IJ}-\frac{1}{4}H^{MK}_{\phantom{MK}I}H_{MKJ} + D_{I}D_{J}\phi\right) \,  , \\
 \beta[B_{IJ}]&=& - \frac{\alp}{2}\left(  D^{K} H_{KIJ} - D^K \phi H_{KIJ} \right) \, ,\\
 \beta[\phi] &=& - \frac{\alp}{2} \left(  D^2\phi - \frac{1}{2} D_I \phi D^I \phi +\frac{1}{2}R -\frac{1}{24} H^2  \right)\, .
 \eea
 These beta functions vanish when the background fields obey the equations of a gravity theory with action
\be 
S= \frac{1}{2\kappa^2} \int d^{26} X \sqrt{-G}e^{-\phi} \left(  R+  D_I \phi D^I \phi -  \frac{1}{12}H_{IJK} H^{IJK}  \right)\, . 
\ee 
 The schematic equivalence between equations of motion and beta functions is
 \be
 \frac{\alp\kappa^2 e^{\phi}}{ \sqrt{-G} }\frac{\delta S}{\delta \phi} \sim  \beta[\phi]\, ,
  \quad  \frac{\alp \kappa^2 e^{\phi}}{ \sqrt{-G}} \frac{\delta S}{\delta B_{IJ} }\sim \beta[B_{IJ}]\, ,  
  \quad  \frac{\alp \kappa^2 e^{\phi}}{\ \sqrt{-G}} \frac{\delta S}{\delta G^{IJ}} \sim \beta[G_{IJ}] +2 G_{IJ} \beta[\phi]\, . 
 \ee
 There are several additional classic papers which provide further perspective on these techniques and results including \cite{Callan:1985ia,Callan:1986jb,CallanThor,Callan:1986bc,Hull:1985rc}.
 
\chapter{Propagators and Wick Contractions}
In this appendix we provide some details as to the propagators and Wick contractions used in evaluation of the one-loop effective actions for duality symmetric theories. 

\section*{Propagators for Chiral Fields}

For a normal two-dimensional boson with kinetic term
\be
{\cal L}_0 = \frac{1}{2} \partial_\a \phi \partial^\a \phi 
\ee
the propagator is given by
\be
\langle \phi(\sigma) \phi(\sigma^\prime) \rangle =  \frac{i}{(2\pi)^2} \int d^2p \frac{e^{i p\cdot(\sigma - \sigma^\prime)}}{p^2} = \Delta_0(\sigma - \sigma^\prime) \, .
\ee
Actually this has problems in the IR so in practice on should consider adding a mass parameter and then taking the massless limit at the end of calculations.  Since we are interested in  UV divergences we will not concern ourselves with this.

For (anti)chiral fields with Floreanini-Jackiw style kinetic terms 
\be
{\cal L}_\pm = \pm \frac{1}{2} \partial_1 \phi_\pm \partial_\mp \phi_\pm
\ee
the propagator is then 
\be
\langle \phi_\pm(\sigma) \phi_\pm(\sigma^\prime) \rangle = \pm \frac{i}{(2\pi)^2} \int d^2p \frac{e^{i p\cdot(\sigma - \sigma^\prime)}}{p_1 p_\mp} = \Delta_\pm(\sigma - \sigma^\prime) \, .
\ee
We will normally write $\Delta_\pm$ to indicate the $\s\rightarrow\s'$ limit. Simply by examining the integrals we see
\bea
\Delta_+ + \Delta_-& =& 2\Delta_0, \label{D00}\\
\Delta_+ - \Delta_-& =& 2\t\, ,
\eea
 and we take this as the definition of $\theta$. Of course these propagator integrals are divergent and must regularised for example by dimensional regularisation as described in appendix C.

The propagators can be calculated in $z$-space after Wick rotation \cite{Tseytlin:1990va,Tseytlin:1990nb}
 with $z=\s+it=\s+\tau$ and $\d_\s=\d+\bar{\d},\d_\tau=\d-\bar{\d}$. Using a $z\rightarrow0$ regularisation such that $\bar{\d}z^{-1}=\pi\delta^{(2)}(z)$ one finds
\bea
\Delta_{+}(z , z^\prime ) &=& -\frac{1}{2\pi}\ln ( z- z^\prime) , \\
\Delta_{-}(z , z^\prime ) &=& -\frac{1}{2\pi}\ln ( \bar{z}- \bar{z}^\prime) ,\\
\Delta_+(z , z^\prime ) +  \Delta_-(z , z^\prime ) &=& -\frac{1}{2\pi}\ln | z- z^\prime|^2 = 2\Delta_0( z , z^\prime), \label{D0}\\
\Delta_+(z , z^\prime ) - \Delta_-(z , z^\prime ) &=& -\frac{1}{2\pi}\ln \frac{ z- z^\prime}{\bar{z}- \bar{z}^\prime}=-\frac{i}{\pi} \mbox{arg}(z-z')=2\t \label{theta} \, ,\eea
where in (\ref{D0}) we have noted after regularisation we have the same relation as (\ref{D00}) to the standard two-dimensional scalar propagator in this regularisation scheme. 

Terms in the path integral of the effective action that are proportional to $\Delta_0$ will be those related to a breakdown in Weyl invariance whereas terms proportional to $\t$ will correspond to a breakdown in worldsheet Lorentz invariance.  This can be understood from (\ref{D0}) and (\ref{theta}) by noticing that a scaling of $z$ shifts $\Delta_0$ and not $\t$ and rotation by a phase shifts $\t$ and not $\Delta_0$. 

\section*{Propagators in the Duality Symmetric Formalisms} 

In both the Poisson-Lie model and the Doubled Formalism we have a similar kinetic term for fluctuations
\bea
S= \frac{1}{2}\int \left[-\cH_{\Ab\bar{B}} \d_1\xi^{\Ab} \d_1\xi^{\bar{B}}  + L_{\Ab\bar{B}} \d_1\xi^{\Ab} \d_0\xi^{\bar{B}}\right],
\eea
where in the chiral frame the matrices are constant and take the form
\bea
\H = \left(\begin{array}{cc}
\openone & 0 \\ 0 & \openone\end{array}
\right), &  L = \left(\begin{array}{cc}
\openone & 0 \\ 0 & -\openone\end{array}
\right).
\eea
The projectors $\frac{1}{2}(\H \pm L)$ then project onto chiral and anti chiral sectors of the theory, hence the general propagator for $\xi^\Ab$ is given by 
\bea
\langle \xi^{\Ab}(z) \xi^{\bar{B}}(z) \rangle &=&\left(\begin{array}{cc}
\openone & 0 \\ 0 & 0\end{array}
\right)\Delta_{+}+ \left(\begin{array}{cc}
0 & 0 \\ 0 & \openone\end{array}
\right) \Delta_{-}\\&=& \frac{1}{2}(\H + L) \Delta_{+} + \frac{1}{2}(\H - L) \Delta_{-}\\
&=&  \frac{1}{2}\H^{\Ab\bar{B}}(\Delta_+ + \Delta_- ) +  \frac{1}{2}L^{\Ab\bar{B}}(\Delta_+ - \Delta_- ) \\
&=& \Delta_0\H^{\Ab\bar{B}}  +\theta L^{\Ab\bar{B}}\, .
\eea
We can use this result and a Wick contraction procedure, described for the ordinary string in \cite{Braaten:1985is},
 to determine the divergent behaviour of higher-order propagator contractions which appear in the expansion of the exponential of our action. 

For example
\bea
&&i\int d^2\s'\langle \xi(\s)^{\bar{A}} \d_1 \xi(\s)^{\bar{B}} \xi(\s')^{\Cb} \d_0 \xi(\s')^{\Db} \rangle\nn\\
&=&\frac{i}{4} \int d^2\s'\left( (\H + L)^{\bar{A}\Cb} \Delta_{+}(p) + (\H - L)^{\bar{A}\Cb}  \Delta_{-}(p)\right)e^{ip.(\s-\s')}\nn\\
&&\times\left( (\H +L)^{\bar{B}\Db} \Delta_{+}(q) + (\H - L)^{\bar{B}\Db}  \Delta_{-}(q)\right)q_1q_0e^{iq.(\s-\s')}\nn\\
&&+\frac{i}{4} \int d^2\s'\left( (\H + L)^{\bar{A}\Db} \Delta_{+}(p) + (\H - L)^{\bar{A}\Db}  \Delta_{-}(p)\right)p_1e^{ip.(\s-\s')}\nn\\
&&\times\left((\H +L)^{\bar{B}\Cb} \Delta_{+}(q) +(\H - L)^{\bar{B}\Cb}  \Delta_{-}(q)\right)q_0e^{iq.(\s-\s')}.
\eea
where for shorthand $\Delta_\pm = \pm (p_1 p_\mp)^{-1}$.   Performing the $\s'$ integral gives a delta function which allows one to simply carry out one of the momenta integrals. 
We make use of identities like 
\be \frac{p_0}{p_1}\frac{1}{p_\pm^2} = \frac{1}{p_1 p_\pm} \mp \frac{1}{p_\pm^2} \ee
and discard integrals of the form
\be 
\int d^2 p \frac{1}{p_\pm^2}
\ee
which though divergent in both IR and UV do not produce logarithmic divergences which we are interested in,  to  find that 
\bea
\nonumber
&&i \int d^2\s'\langle \xi(\s)^{\bar{A}} \d_1 \xi(\s)^{\bar{B}} \xi(\s')^{\Cb} \d_0 \xi(\s')^{\Db} \rangle\\\nonumber
&& \simeq \frac{i}{4} \int \dtp   (\H+L)^{\bar{B}\bar{C}}(\H + L)^{\bar{A}\bar{D}} \frac{1}{p_1 p_-}   +  (\H-L)^{\bar{B}\bar{C}}(\H - L)^{\bar{A}\bar{D}} \frac{1}{p_1 p_+} \\ \nonumber
&& \quad \qquad \quad \quad     -(\H+L)^{\bar{B}\bar{C}}(\H - L)^{\bar{A}\bar{D}} \frac{p_0}{p_1 p_- p_+ }   -  (\H-L)^{\bar{B}\bar{C}}(\H + L)^{\bar{A}\bar{D}} \frac{p_0}{p_1 p_- p_+}    -  ({\bar{C}} \leftrightarrow \bar{D})  \\
&&\qquad\qquad\qquad\qquad\qquad\qquad
\\
&&= -\frac{1}{2} \left(\H^{{\bar{A}}[\Cb}L^{\Db]\bar{B}} + L^{{\bar{A}}[\Cb}\H^{\Db]\bar{B}}   \right)\Delta_0    -  L^{{\bar{A}}[ \Cb}L^{\Db ] \bar{B}}  \t \, ,  
\eea
where $\simeq$ indicates equality up to convergent terms which are irrelevant for our purpose.  

A similar procedure can be used to calculate the two-propagator contractions with any combination of worldsheet derivatives.  One finds that\footnote{We will simplify notation by using $\langle \xi^{\Ab} \d_1 \xi^{\bar{B}}\xi^{\Cb} \d_0 \xi^{\Db} \rangle =i\int d^2\s'\langle \xi(\s)^{\Ab} \d_1 \xi(\s)^{\bar{B}} \xi(\s')^{\Cb} \d_0 \xi(\s')^{\Db} \rangle$.} 
\bea
\langle \xi(\s)^{\bar{A}} \d_1 \xi(\s)^{\bar{B}} \xi(\s')^{\Cb} \d_1 \xi(\s')^{\Db} \rangle &=& \frac{1}{2} \left(\H^{{\bar{A}}[\Cb}\H^{\Db]\bar{B}} - L^{{\bar{A}}[\Cb}L^{\Db]\bar{B}}   \right)\Delta_0 \, , \\
\langle \xi(\s)^{\bar{A}} \d_1 \xi(\s)^{\bar{B}} \xi(\s')^{\Cb} \d_0 \xi(\s')^{\Db} \rangle &=&-\frac{1}{2} \left(\H^{{\bar{A}}[\Cb}L^{\Db]\bar{B}} + L^{{\bar{A}}[\Cb}\H^{\Db]\bar{B}}   \right)\Delta_0  \nonumber \\ && \qquad  -  L^{{\bar{A}}[ \Cb}L^{\Db ] \bar{B}}  \t \, , \\
\langle \xi(\s)^{\bar{A}} \d_0 \xi(\s)^{\bar{B}} \xi(\s')^{\Cb} \d_0 \xi(\s')^{\Db} \rangle &=& 
-  \frac{1}{2} \left(\H^{{\bar{A}}[\Cb}\H^{\Db]\bar{B}} + 3 L^{{\bar{A}}[\Cb}L^{\Db]\bar{B}}   \right)\Delta_0 \nonumber \quad\\
  && \qquad-   \left(\H^{{\bar{A}}[\Cb}L^{\Db]\bar{B}} + L^{{\bar{A}}[\Cb}\H^{\Db]\bar{B}}  \right)    \t \, ,  
\eea
 For the Doubled Formalism one can also include indices on the base.  The results can be compactly summarised in terms of the total space metric $\cM$ and  $\cL$ as
\bea
 \langle \xi^{\ab} \d_1 \xi^{\bar{\beta}} \d_1 \xi^{\cb} \xi^{\db} \rangle &=& \frac{1}{2} \Delta_0\left(\cM \cM -\cL\cL\right)^{(\ab\cb\bar{\beta}\db - \ab\db\bar{\beta}\cb)},\\
\langle {\xi^\ab \d_1 \xi^{\bar{\beta}} \d_0 \xi^{\cb} \xi^{\dlb}} \rangle &=& -\frac{1}{2} \Delta_0\left(\cM\cL+\cL\cM\right)^{(\leftrightarrow)} -  \t \cL\cL^{(\leftrightarrow)},   \\
 \langle {\xi^{\ab} \d_0 \xi^{\bar{\beta}} \d_0 \xi^{\cb} \xi^{\dlb}}\rangle &=& -\frac{1}{2} \Delta_0\left(\cM\cM+3\cL\cL\right)^{(\leftrightarrow)} -  \t \left( \cL\cM + \cM\cL\right)^{(\leftrightarrow)},   
\eea
where $\cM \cM^{(\ab\cb\bar{\beta}\db - \ab\db\bar{\beta}\cb)} $ represents $ \cM^{\ab\cb} \cM^{\bar{\beta}\db}  - \cM^{\ab\db} \cM^{\bar{\beta}\cb}$ and  $(\leftrightarrow)$ is understood in the same way. 

\chapter{Dimensional Reduction}
\begin{quote} In this appendix we provide some details of the calculation of dimensional reduction of target space effective actions.  \end{quote}

We begin with the $D=26$ dimensional target space effective action for the bosonic string given by
\be 
S= \frac{1}{2\kappa^2} \int d^{26} X \sqrt{-G}e^{-\phi} \left(  R+  D_I \phi D^I \phi -  \frac{1}{12}H_{IJK} H^{IJk}  \right)\, . 
\ee
For reference the equations of motion that follow from this action are 
\bea
0 &=& R_{IJ} + D_I D_J \phi - \frac{1}{4} H^2_{IJ} - G_{IJ}\left(\frac{R}{2} + D^2\phi - \frac{1}{2} (D\phi)^2 - \frac{1}{24}H^2   \right) \,,\\
 0 &=& \frac{1}{2} D^I \left( e^{-\phi} H_{IJK} \right) \, ,\\ 
 0 &=& -2\left(\frac{R}{2} + D^2\phi - \frac{1}{2} (D\phi)^2 - \frac{1}{24}H^2 \right) \, .
\eea
To establish the equation of motion for the metric variation it is most efficient to observe that 
\bea
\delta R_{IJ} &=& D_K \delta \Gamma^K_{IJ} -  D_I \delta \Gamma^K_{KJ}\, , \\ 
 G^{IJ} \delta R_{IJ} &=& D_K \bigl( \left( G^{KL}G^{MN} - G^{KM}G^{LN} \right) D_L \delta G_{MN}  \bigr) \, . 
\eea
In pure Einstein-Hilbert gravity the total derivative produced by varying the curvature results in surface term whereas in the string effective action above one must be careful to appropriately pull the derivatives through the dilaton term.   

We now reduce with the ansatz $X^I = (X^i, Y^a)$, $G_{IJ} = {\rm  diag}\left( h_{ij}(Y), g_{ab}(Y)\right) $ and $B_{IJ} = {\rm diag}\left( b_{ij}(Y), 0\right)$.   The only non-vanishing Christofel symbols and components of field strength are 
\be
\Gamma^i_{ja} = \frac{1}{2} (h^{-1} \partial_a h )^i_{\ j}\, , \quad \Gamma^a_{ij} = -\frac{1}{2}\partial^a h_{ij}\, , \quad \Gamma^a_{bc} = \hat{\Gamma}^a_{bc} \, , \quad H_{aij} = \partial_a b_{ij} \, . 
\ee
Using this one finds 
\bea
R_{ij} &=& -\frac{1}{2}\hat{\nabla}^2 h_{ij} -  \frac{1}{4} \Lambda_a \partial^a h_{ij} + \frac{1}{2}( \partial_a h  h^{-1} \partial^a h )_{ij} \, , \\
R_{ab} &=& \hat{R}_{ab} -\frac{1}{2}  \hat{\nabla}_b \Lambda_a - \frac{1}{4} {\rm Tr}(h^{-1} \partial_a h  h^{-1} \partial_b h h^{-1} ) \, ,\\
R &=& \hat{R} -  \hat{\nabla}_a \Lambda^a + \frac{1}{4} \Lambda_a \Lambda^a - \frac{1}{4} {\rm Tr}(h^{-1} \partial_a h  h^{-1} \partial^a h h^{-1})\, ,
\eea
in which $\Lambda_a = {\rm Tr} (h^{-1} \partial_a h )$.  Recalling that the fibre moduli fields are naturally packaged in to the $O(d,d)/O(d)\times O(d)$ coset form $\cH_{AB}$ we have the identity 
\be
{\rm Tr}\left( \partial_a \cH^{-1} \partial^a \cH  \right) = 2 {\rm Tr}\left( \partial_a h^{-1} \partial^a h  + h^{-1} \partial_ab  h^{-1} \partial^a b \right) \, . 
\ee
Armed with the above it is easy to see that the dimensional reduction produces the action
\be
S= \frac{vol}{2\kappa^2} \int d^{26-d}y  \sqrt{-g}e^{-\Phi} \left( \hat{ R}+  \hat{\nabla}_a \Phi \hat{\nabla}^a \Phi  +   \frac{1}{8} {\rm Tr}  \left( \partial_a \cH^{-1} \partial^a \cH  \right)     \right)\, , 
\ee
where we have defined $\Phi = \phi - \frac{1}{2} \ln \det h$. The equations of motion that follow are
\bea
 0& =& \hat{\nabla}^2 \Phi - \frac{1}{2} (\hat{\nabla} \Phi)^2 + \frac{1}{2} \hat{R} + \frac{1}{16}  {\rm Tr}  \left( \partial_a \cH^{-1} \partial^a \cH  \right)\, ,\\ 
 0&=&-\frac{1}{4} \hat{\nabla}^2 \cH_{IJ} + \frac{1}{4}  \left( \partial_a \cH  \cH^{-1} \partial^a \cH  \right)_{IJ} + \frac{1}{4} \hat{\nabla}^a \Phi \hat{\nabla}_a \cH_{IJ} \, , \\ 
 0 &=& \hat{R}_{ab} + \hat{\nabla}_a\hat{\nabla}_b \Phi + \frac{1}{8}  {\rm Tr}  \left( \partial_a \cH^{-1} \partial^a \cH  \right) \nonumber   \\
 && \quad \quad  - g_{ab} \left(  \hat{\nabla}^2 \Phi - \frac{1}{2} (\hat{\nabla} \Phi)^2 + \frac{1}{2} \hat{R} + \frac{1}{16}  {\rm Tr}  \left( \partial_a \cH^{-1} \partial^a \cH  \right)    \right)\, . 
\eea

\end{appendix}

\providecommand{\href}[2]{#2}\begingroup\raggedright\endgroup

\end{document}